\documentclass[a4paper,11pt]{article}
\usepackage{jheppub}
\usepackage[T1]{fontenc}
\usepackage{lmodern}

\usepackage{amstext}
\usepackage{babel}
\usepackage{verbatim}
\usepackage{amsfonts}
\usepackage{mathrsfs}
\usepackage{amsmath}
\usepackage{mathtools}
\usepackage{amssymb}
\usepackage{bm}
\usepackage{extarrows}
\usepackage{slashed}
\usepackage{isodateo}
\usepackage{xcolor}
\usepackage{graphicx}
\usepackage{array}
\usepackage[skins]{tcolorbox}
\usepackage[low-sup]{subdepth}
\usepackage{float}
\usepackage{isodateo}
\usepackage{tensor}
\usepackage{multirow}
\usepackage{enumitem}
\usepackage{ytableau}
\usepackage{framed}
\usepackage{scalerel}
\usepackage{stackengine,wasysym}
\usepackage{tikz-feynhand}
\usepackage{hyperref}
\usepackage{stackengine}
\usepackage{subcaption}
\usepackage{xparse}
\usepackage{makecell}
\usepackage{fontawesome5}
\hypersetup{
colorlinks=true,
citecolor=blue,
linkcolor=blue,
urlcolor=blue,
pdfauthor={},
pdftitle={},
pdfsubject={}}
\makeatletter
\def\simgt{\mathrel{\lower2.5pt\vbox{\lineskip=0pt\baselineskip=0pt
\hbox{$>$}\hbox{$\sim$}}}}
\def\simlt{\mathrel{\lower2.5pt\vbox{\lineskip=0pt\baselineskip=0pt
\hbox{$<$}\hbox{$\sim$}}}}
\numberwithin{equation}{section}
\definecolor{shadecolor}{gray}{0.925}
\definecolor{lblu}{RGB}{15,125,255}
\definecolor{lgre}{RGB}{40,170,40}
\definecolor{magg}{RGB}{157,0,255}
\usepackage{amsthm}
\theoremstyle{definition}
\newtheorem{definition}{Definition}[section]
\theoremstyle{plain}
\newtheorem{theorem}[definition]{Theorem}
\newtheorem{corollary}[definition]{Corollary}
\theoremstyle{remark}
\newtheorem{remark}[definition]{Remark}
\newcommand{\red}{\color{red}}

\newcommand{\lgray}{\color{gray}}
\newcommand{\lblu}{\color{lblu}}
\newcommand{\lgre}{\color{lgre}}

\newtcolorbox{shadowequation}{
colback=gray!18,
colframe=gray!18,
boxrule=0.7pt,
arc=0mm,
left=0pt,
right=2.5pt,
boxsep=0pt,
top=3mm,
bottom=3mm
}
\newcommand{\eqrefe}{eq.~\eqref}
\newcommand{\beqs}{\begin{subequations}}
\newcommand{\eeqs}{\end{subequations}}

\renewcommand{\rm}{\mathrm}

\graphicspath{{fig/}}       
\DeclareGraphicsExtensions{.pdf}

\newcommand{\letterfig}[5][0]{%
\ensuremath{%
\vcenter{\hbox{%
\raisebox{\if\relax\detokenize{#2}\relax0pt\else#2\fi}[\height][\depth]{%
\makebox[#5][l]{%
\includegraphics[scale=#4,angle=#1,origin=c]{#3.pdf}}}}}}}

\newcommand{\cubeHarpoons}[2]{%
\ensuremath{\xrightleftharpoons[\scalebox{1.15}{$\scriptstyle\mkern14mu #2\mkern14mu$}]{\scalebox{1.15}{$\scriptstyle\mkern14mu #1\mkern14mu$}}}%
}
\newcommand{\onetubeL}{%
\mathord{\vcenter{\hbox{%
\begin{tikzpicture}[scale=0.18,baseline=-0.5ex]
\draw[thick] (0,0) circle (0.8);
\fill (0,0) circle (0.12);
\fill (1.7,0) circle (0.12);
\end{tikzpicture}}}}%
}
\newcommand{\onetubeR}{%
\mathord{\vcenter{\hbox{%
\begin{tikzpicture}[scale=0.18,baseline=-0.5ex]
\fill (0,0) circle (0.12);
\draw[thick] (1.7,0) circle (0.8);
\fill (1.7,0) circle (0.12);
\end{tikzpicture}}}}%
}
\newcommand{\twositetube}{%
\mathord{\vcenter{\hbox{%
\begin{tikzpicture}[scale=0.18,baseline=-0.5ex]
\draw[thick] (0.85,0) ellipse (1.55 and 0.85);
\fill (0,0) circle (0.12);
\fill (1.7,0) circle (0.12);
\end{tikzpicture}}}}%
}
\def\leqq{\leqslant}
\def\geqq{\geqslant}
\def\({\left(}
\def\){\right)}
\def\[{\left[\,}
\def\]{\,\right]}
\def\LB{\left\{}
\def\RB{\right\}}
\def\nn{\nonumber}
\def\pd{\partial}
\def\pp{\prime}
\def\to{\rightarrow}

\def\vs{\vspace*{1mm}}
\def\hs{\hspace*{0.3mm}}
\def\hsm{\hspace*{-0.3mm}}
\def\A{\mathcal{A}}
\def\B{\mathcal{B}}
\def\tC{\widetilde{C}}
\def\td{\text{d}}
\def\ii{\text{i}}
\def\bk{\boldsymbol{k}}
\def\LL{\mathcal{L}}
\def\L{\text{L}}
\def\PP{\mathcal{P}}
\def\R{\text{R}}
\def\SS{\mathfrak{S}}
\def\TT{\mathscr{T}}
\def\hTT{\widehat{\mathscr{T}}}
\def\bT{\boldsymbol{T}}
\def\W{\mathcal{W}}

\def\YY{\boldsymbol{Y}}
\def\mY{\mathcal{Y}}
\def\tmY{\widetilde{\mathcal{Y}}}
\def\bmY{\boldsymbol{\mathcal{Y}}}
\def\al{\alpha}
\def\be{\beta}
\def\ga{\gamma}
\def\lam{\lambda}
\def\blam{\boldsymbol{\lambda}}
\def\re{\mathrm{Re}}
\def\op{\mathrm{OP}}
\def\hop{\widehat{\mathrm{OP}}}
\def\tot{\mathrm{tot}}
\def\emp{\varnothing}
\def\id{\mathrm{id}}
\def\Vol{\mathrm{Vol}}
\def\Li{\mathrm{Li}_2}
\def\frw{\mathrm{FRW}}
\def\ds{\mathrm{dS}}
\def\ch{\operatorname{Ch}}
\allowdisplaybreaks
\title{From Trees to All Loops: The Combinatorial Geometry of Two-Site Cosmological Wavefunctions}

\author[a]{Yanfeng Hang,}
\emailAdd{yhang24@wisc.edu}

\author[b]{Cong Shen}
\emailAdd{CongShen2028@u.northwestern.edu}

\affiliation[a]{Department of Physics,\\ 
University of Wisconsin--Madison, Madison, WI 53706, USA}

\affiliation[b]{Department of Physics and Astronomy,\\ 
Northwestern University, Evanston, IL 60208, USA}

\abstract{We prove, at arbitrary loop order and including arbitrary tadpole decorations, the shifted-tree decomposition of two-site cosmological wavefunction coefficients proposed in our previous work, and uncover the combinatorial and geometric structure underlying it.
For conformally coupled scalars with polynomial interactions, this decomposition expresses each loop coefficient as a finite signed sum of tree coefficients evaluated at shifted energies.
For bubble-like graphs, the alternating sum over a Boolean lattice can be written in terms of finite differences or maximal chains.
We prove the maximal-chain identity algebraically by induction on loop order and geometrically by decomposing a cubical Laplace integral into simplices.
The maximal chains correspond to permutohedron vertices, whose incident facets encode the chain-dependent divisors.
For tadpole-decorated graphs, the finite-difference Leibniz rule gives an explicit expansion indexed by ordered partitions and local maximal chains, which glue into global maximal chains and organize the coefficient as a combinatorial ``path integral''.
A natural ordering of the divisors yields a term-by-term match with graph tubings and completes the all-loop proof, while other orderings give alternative tubing representations of the same coefficient.
These identities hold before spatial loop-momentum integration and extend to power-law FRW backgrounds by linearity of the twisted integral.
Two-site loop coefficients can therefore be obtained directly from known tree coefficients without solving new loop-level differential systems.}

\begin{document}
\maketitle
\flushbottom

\section{Introduction}
\label{sec:1}

Cosmological wavefunction coefficients provide a perturbative description of late-time spatial correlation functions in an expanding universe.\ 
For conformally coupled scalars with polynomial interactions in power-law Friedmann-Robertson-Walker (FRW) backgrounds, the time dependence can be encoded via a Mellin transform, expressing the wavefunction coefficients as twisted integrals of their flat-space counterparts, which are rational functions of the external and internal energies associated with a graph.\ 
The singularity and combinatorial structures of these rational functions are closely related to cosmological polytopes, twisted cohomology, and the differential equations satisfied by the corresponding twisted integrals \cite{Arkani-Hamed:2017fdk,Arkani-Hamed:2018bjr,De:2023xue,Arkani-Hamed:2023bsv,Arkani-Hamed:2023kig,Benincasa:2024ptf,Hang:2024xas,Arkani-Hamed:2024jbp,Chen:2024glu}.\ 
Related developments include direct approaches to correlation functions \cite{Qin:2024gtr,Pimentel:2025rds,Zhang:2025nzd,Chowdhury:2026dwm}, approaches based on scattering amplitudes \cite{Cespedes:2025dnq,Glew:2025arc,Glew:2025ugf,Chowdhury:2025ohm}, and cluster-algebra methods \cite{Capuano:2025myy,Mazloumi:2025pmx,Capuano:2026pgq}.

At loop level, the rational coefficients are usually organized by graph tubings, while the corresponding twisted integrals are obtained by solving systems of differential equations whose size depends on the graph topology \cite{Hang:2024xas,De:2024zic,Fevola:2024acq,Capuano:2025ehm,Chen:2026dqp,Fu:2026dqb}.\
Kinematic-flow methods provide a systematic framework at and beyond tree level \cite{Arkani-Hamed:2023bsv,Baumann:2025qjx,Glew:2025ypb,Ke:2026laa,Hang:2024xas,Westerdijk:2026msm}, but the combinatorial and analytic complexity generally increases with the number of sites and loops.\
For the two-site graphs considered here, ref.~\cite{Hang:2024xas} proposed a useful alternative and verified it through two loops: a loop coefficient can be written as a finite signed sum of tree coefficients evaluated at shifted energies,
\begin{equation}
\label{eq:loop-decom}
\psi_{\mathrm{loop}}^{} \,=\,\sum_\al\,\mathrm{sgn}_\al\,\psi_{\mathrm{tree}}^{\al} \,,
\end{equation}
where $\al$ labels a shift pattern and $\mathrm{sgn}_\al=\pm1$ is its sign.\
The relation provides a separate loop-level construction from shifted tree-level data.\
In this paper, we prove this relation for the two-site wavefunction coefficients of arbitrary loop order and topology, including the bubble-like family and graphs with arbitrary tadpole decorations.\
In addition, we uncover a novel all-loop combinatorial and geometric structure behind this hidden simplicity.\

The first part of the analysis concerns the bubble-like family with $\ell+1$ internal lines, where $\ell$ is the loop order.\
After the local tree divisors are evaluated with the total internal energy, the dependence on the shifted subset of internal lines is carried by the total-energy divisors $D_V$, with $V\subseteq\YY$.\
The shifted-tree relation can therefore be reorganized as a finite difference acting on the inverse divisor $1/D$.\
This gives a vertex expansion on the Boolean lattice $\PP(\YY)$, whose vertices are the subsets of internal lines and whose signs are the corresponding M\"obius weights.\
The same finite difference also admits a maximal-chain expansion.\
Its terms are labeled by permutations of the internal-line labels and reproduce the products of inverse divisors associated with the graph-tubing representation, a graphical bookkeeping of the all-loop recursion derived from the Feynman rules.

We prove the maximal-chain identity in two complementary ways.\
The algebraic proof uses the finite-difference Leibniz rule and induction in the number of internal lines; inserting the new line at each possible position generates every enlarged maximal chain exactly once.\
The geometric proof uses Laplace representations of the products of inverse divisors.\
For each chain, a change to cumulative variables maps the integration region to an ordered simplex, and the sum over permutations fills the corresponding Boolean cube.\
This makes the relation between the alternating subset sum and the chain expansion explicit. We also illustrate the associated permutohedral organization in low-loop examples.\
It complements the two-dimensional hyperplane arrangements in the plane of twist variables of ref.~\cite{Hang:2024xas} by providing a higher-dimensional realization of the same combinatorics.

We then incorporate tadpole decorations, whose energies we collect in the set $\bT$.\
Left and right tadpoles act as commuting finite-difference operators on the two external energies.\
Unlike the pure bubble case, a single tadpole shift can modify several divisors within one central-chain contribution, and iterating the asymmetric finite-difference Leibniz rule distributes these shifts among the divisor factors.\
The resulting terms are indexed by an ordered partition of $\bT$, together with a local maximal chain for each nonempty block.\
Gluing these local chains consistently yields an explicit all-loop expansion of an arbitrary two-site wavefunction in terms of global maximal chains: a combinatorial ``path integral'' over central maximal chains in $\PP(\YY)$ and global maximal chains in $\PP(\bT)$ compatible with the partition.\
For a natural ordering of the divisor product, we match this expansion to graph tubings by organizing each tubing according to its smallest 2-site tube, thereby proving the tadpole-decorated shifted-tree relation.\
Other divisor orderings redistribute the cumulative shifts among the individual factors, giving term-wise different but equivalent expansions of the same wavefunction coefficient.

Our proof is carried out for the flat-space rational coefficients.\
The corresponding statements in power-law FRW backgrounds follow from the linearity of the twisted integral and its commutation with the relevant energy shifts, with each shifted tree integrated with the twist measure of the target graph.\
The construction relies on the divisor structure of two-site graphs: for multi-site topologies, several overlapping shifted divisors can occur, and the associated combinatorial organization need not reduce to a single Boolean lattice.\
The two-site case thus provides both an all-loop proof and a controlled setting in which the relations among shifted trees, finite differences, and graph tubings can be made explicit.

The rest of the paper is organized as follows.\
Section~\ref{sec:2} reviews the two-site tree cosmological wavefunction coefficient, reduces the shifted-tree relation for the bubble-like family to an alternating sum over shifted total-energy divisors, and explains how it carries over to FRW backgrounds.\
Section~\ref{sec:3} formulates this sum on the Boolean lattice as a maximal-chain representation and illustrates the associated permutohedra at one and two loops.\
Section~\ref{sec:4} proves the maximal-chain identity algebraically, using the finite-difference Leibniz rule, and geometrically, through a simplex decomposition of the Boolean cube.\
Section~\ref{sec:5} extends the construction to tadpole-decorated graphs, yielding an expansion in ordered partitions and local chains that we interpret as a combinatorial ``path integral'' over maximal chains.\
Section~\ref{sec:6} matches this expansion to graph tubings, completing the proof of the shifted-tree decomposition.\
Section~\ref{sec:7} presents explicit examples of the general construction, including the chain--tubing correspondence and alternative divisor orderings. Section~\ref{sec:8} concludes with possible extensions.\
Appendix~\ref{app:A} contains additional examples, while appendix~\ref{sec:B} collects the lattice-theoretic preliminaries.

\subsection*{Summary of the Main Results}

The main result of this paper is an all-loop shifted-tree decomposition of two-site cosmological wavefunction coefficients.\
The flat-space identities below are stated before spatial loop-momentum integration, with the normalization of section~\ref{sec:2}.\
The shifted-tree decomposition carries over to power-law FRW backgrounds by integrating each shifted tree with the twist measure of the target graph, as discussed above.\
For the bubble-like family with internal-energy set $\YY=\{Y_1,\ldots,Y_{\ell+1}\}$, the loop coefficient can be written as a finite signed sum of tree coefficients shifted in the divisor basis \eqref{eq:TLD-chain} as follows:
\begin{equation}
\psi^{(2;\ell)}(X_1,X_2;\YY) \,=\, \sum_{V\subseteq\YY}(-1)^{|V|}\,\psi^{(2;0)}(X_1,X_2;Y_\tot)\big|_{D\to D_V}\,,\qquad
\ell\geqq0\,.
\end{equation}
This relation is equivalent to the compact finite-difference form
\begin{equation}
\label{eq:intro-summary-bubble}
\psi^{(2;\ell)} 
\,=\, \frac{D_{\YY}}{L_1L_2}\sum_{V\subseteq\YY}\frac{(-1)^{|V|}}{D_V}
\,=\, \frac{L_1+L_2}{L_1L_2}\Delta_{2\YY}\bigg(\frac1D\bigg)\,, \qquad 
\ell\geqq0\,.
\end{equation}
The first equality exhibits the shifted-tree decomposition as a M\"obius-weighted vertex sum on the Boolean lattice $\PP(\YY)$, while the second rewrites the same sum as a product of commuting finite-difference operators. The central structural identity of the paper is the maximal-chain decomposition
\begin{equation}
\label{eq:intro-summary-chain}
\Delta_{2\YY}\bigg(\frac1D\bigg)
= \Bigg(\prod_{y\in\YY}2y\Bigg)
\sum_{\pi\in\SS_{|\YY|}}
\prod_{r=0}^{|\YY|}\frac1{D_{C[\pi]_r}}\,.
\end{equation}
It converts the Boolean-lattice vertex sum into a sum over maximal chains in $\PP(\YY)$, equivalently over permutations of the internal-line labels. We prove \eqrefe{eq:intro-summary-chain} algebraically by induction on the operator product and geometrically by a standard simplex triangulation of the Boolean cube. Substituting this identity into eq.~\eqref{eq:intro-summary-bubble} gives the explicit all-loop maximal-chain representation in eq.~\eqref{eq:wfc-loop-max-chain}, which satisfies the all-loop recursion relation. Thus, for the bubble-like family, the proof map may be summarized schematically as
\begin{equation}
\label{eq:intro-summary-proof-map}
\text{shifted trees}
~\Longleftrightarrow~
\text{finite differences}
~\Longleftrightarrow~
\text{maximal chains}~\Longleftrightarrow~
\text{all-loop recursion}\,.
\end{equation}
The first three forms are different representations of the same combinatorial/geometric object.\ 
One can view the $(\ell+1)$-dimensional Boolean cube as the polytope of the shifted trees, and regard the associated $\ell$-dimensional permutohedron as the polytope of the maximal chains.\ 
The last equivalence is achieved by a technical term-by-term matching with the graph-tubing representation, connecting the previous mathematical objects to physical wavefunction coefficients.

We then extend the construction to a generic two-site graph with $\ell$ central bubble loops and arbitrary left and right tadpole loops $(\ell_\L,\ell_\R)$.
For the left and right tadpole energy sets $\bmY_\L$ and $\bmY_\R$, with $\bT\equiv\bmY_\L\sqcup\bmY_\R$, the wavefunction coefficient in flat space or power-law FRW obeys the shift relation
\begin{equation}
\label{eq:intro-summary-tadpole}
\psi^{(2;\ell,\ell_\L,\ell_\R)}\,=\,\sum_{U_\L\subseteq\bmY_\L}\sum_{U_\R\subseteq\bmY_\R} (-1)^{|U_\L|+|U_\R|}\, \psi^{(2;\ell)}\,\Big|{\substack{\\[0mm]
X_1\to X_{1}+\sigma[U_\L]
\\[1mm]
X_2\to X_{2}+\sigma[U_\R]}}\,.
\end{equation}
Combining this with the bubble-like shifted-tree formula reduces every two-site coefficient to a finite signed sum of shifted two-site trees, written in the divisor basis with the shifted divisors defined in \eqrefe{eq:shifted-divisors-tadpole},
\begin{equation}
\label{eq:intro-summary-completely}
\psi^{(2;\ell,\ell_\L,\ell_\R)}_{\rm{Flat/FRW}}=\!\sum_{U_\L\subseteq\bmY_\L}\sum_{U_\R\subseteq\bmY_\R} \sum_{ V\subseteq\YY}(-1)^{|U_\L|+|U_\R|+|V|}\,
\psi^{(2;0)}_{\rm{Flat/FRW}}(L_{1,U_\L},L_{2,U_\R};D_{V,U_\L,U_\R})\,,
\end{equation}
which allows us to read loop wavefunction coefficients directly from the known tree coefficients, without solving new loop-level differential systems.

The corresponding subset sum is naturally organized by the product Boolean lattice $\PP(\YY)\times\PP(\bmY_\L)\times\PP(\bmY_\R)$, while its chain expansion, in addition to chains in $\PP(\YY)$, is organized by partition-associated global maximal chains in $\PP(\bT)\cong\PP(\bmY_\L)\times\PP(\bmY_\R)$, obtained by gluing together local chains in each block.\
This yields the explicit all-loop all-topology chain formula summarized in eqs.~\eqref{eq:all-loop-wfc}--\eqref{eq:all-loop-wfc-weight}.\ 
For a natural divisor ordering, a similar term-wise matching with the graph-tubing representation completes the all-loop proof. As a byproduct, different divisor orderings generate different but equivalent graph-tubing rules, showing that the particular term-wise matching is a technical choice of construction rather than an intrinsic property of the wavefunction.

\section{Wavefunction Coefficient from Tree to Bubble Loops}
\label{sec:2}

We consider a conformally coupled scalar field with polynomial self-interactions in an FRW background, as studied in ref.~\cite{Hang:2024xas}
\begin{equation}
\label{eq:FRW-action}
S_{\frw} = -\int \td^d x \int_{-\infty}^0 \td \eta \sqrt{-g}\[\frac{1}{2} g^{\al\be} \pd_\al \phi \pd_\be \phi+\frac{d-1}{8d} R \phi^2+\sum_{n\geqq3} \frac{\lam_n}{n!} \phi^n\],
\end{equation}
where $\eta$ denotes the conformal time, $R$ represents the Ricci scalar and $\lam_n$ is the coupling constant. The metric reads
\begin{equation}
\td s^2=g_{\al\be} \td x^\al \td x^\be = [a(\eta)]^2\(-\td \eta^2+\delta_{ij} \td x^i \td x^j\).
\end{equation}
Under a Weyl transformation with the corresponding field redefinition, the theory is equivalent to a massless scalar field in flat space with time-dependent couplings
\begin{equation}
\label{eq:time-dep-flat-action}
S_{\rm{Flat}}=-\int \td^d x \int_{-\infty}^0 \td \eta\[\frac{1}{2}(\pd\phi)^2+\sum_{n \geqq 3} \frac{\lam_n(\eta)}{n!} \phi^n\],
\end{equation}
where $\lam_n(\eta) \equiv \lam_n[a(\eta)]^{2-(n-2)(d-1) / 2}$. We will focus on the wavefunction of the universe for the system \eqref{eq:FRW-action} in terms of its fundamental field $\phi$.

The probability distribution underlying late-time correlations in an expanding universe is encoded in the wavefunction of the universe,
\begin{equation}
\log\Psi[\phi] \,=\, \!-\!\sum_{m\geqq2} \frac{1}{m!}\int\prod_{i=1}^m\Bigg[\frac{\td^d\bk_i}{(2\pi)^{d}}\,\phi_{\bk_i}^{}\Bigg]
\psi_m^{}(\bk_1,\ldots,\bk_m)\,(2\pi)^d\,\delta^{(d)}\Bigg(\sum_{j=1}^{m}\bk_j^{}\Bigg)\,,
\end{equation}
where $\psi_m(\bk_1,\ldots,\bk_m)$ denotes the $m$-point cosmological wavefunction coefficient, with $m$ the number of external late-time legs. 
It characterizes the transition from the Bunch--Davies vacuum \cite{Bunch:1978yq} to a specified late-time field configuration. 
In perturbation theory, these coefficients are organized as sums of graph contributions analogous to Feynman diagrams. 
Since the number of external legs will not be essential below, we instead label them by their bulk topology, with $N_{\rm s}$ denoting the number of bulk sites and $\ell_{\tot}$ the total loop order, and write them as $\psi^{(N_{\rm s};\ell_{\tot})}$. 
In this work, we focus on the two-site family where $N_{\rm s}=2$, which is reviewed in ref.~\cite{Hang:2024xas} up to $\ell_\tot=2$. For a generic two-site graph, the bulk topology is fully specified by a tuple of loop orders $ (\ell, \ell_\L,\ell_\R)$ denoting respectively the number of bubble loops, left-tadpole loops, and right-tadpole loops, subject to $\ell+\ell_\L+\ell_\R=\ell_\tot$. We refer to the tadpole-less graph where $\ell_\L=\ell_\R=0$ as a two-site bubble-like graph which has two bulk vertices connected by internal propagators, as illustrated in figure~\ref{fig:2siteloop}.  
The graph with one internal line is the two-site tree; adding more internal lines between the two vertices gives the bubble-like loop-level two-site family.  
We write $\ell+1$ for the number of internal lines, so that $\ell=0$ is the tree and $\ell\geqq 1$ is the loop case. 
A generic two-site graph with possible tadpoles is illustrated in figure~\ref{fig:tadpole-two-site}. Accordingly, we denote the two-site coefficient studied in this work by $\psi^{(2;\ell)}$ for a bubble-like graph, and $\psi^{(2;\ell,\ell_\L,\ell_\R)}$ for a generic two-site graph with arbitrary topology.

For the action \eqref{eq:FRW-action}, the cosmological wavefunction coefficients in power-law FRW backgrounds can be expressed as twisted integrals over their flat-space counterparts, with the external energies shifted by the twist variables $x_i$ as $X_i\to X_i+x_i$ \cite{Arkani-Hamed:2017fdk,Arkani-Hamed:2023kig}. Specializing to the generic two-site family, we have
\begin{equation}
\label{eq:psi-FRW}
\psi^{(2;\ell_\tot)}_{\frw}(X_1,X_2;\YY,\bT) \,=
\int_0^{\infty} \prod_{a=1}^2(\td x_a x_a^{\ga_a})\, \psi^{(2;\ell_\tot)}(X_1\!+\!x_1,X_2\!+\!x_2;\YY,\bT)\,. 
\end{equation}
Here $\YY=\{Y_1,\ldots,Y_{\ell+1}\}$ denotes the internal energies and similarly $\bT$ denotes the tadpole ones when they are present, while the exponents $\ga_a$ of the twist measure depend on the valence of vertex $a$ and on the background; their explicit values do not enter the analysis below. Throughout this paper, wavefunction coefficients without the $\frw$ subscript, including $\psi^{(2;\ell_\tot)}$, $\psi^{(2;\ell)}$, and $\psi^{(2;\ell,\ell_\L,\ell_\R)}$, denote those of the flat-space theory in \eqrefe{eq:time-dep-flat-action} evaluated with \emph{time-independent} couplings.\
This formula enables us to pass from simpler flat-space wavefunction coefficients to ones on the cosmological background that probe the quantum nature of the early universe. However, in most cases, the FRW wavefunction coefficients on the left-hand side (lhs) are too difficult to calculate via straightforward twisted integrals, and are instead obtained by solving finite-dimensional PDE systems generated from the IBP relations of a certain class of twisted integrands which descend from the shifted flat-space coefficient on the rhs. These systems can be constructed systematically, but they couple the physical coefficient to auxiliary functions and can involve a large number of basis integrals. At tree level, related IBP relations and PDE systems are studied, for example, in refs.~\cite{De:2023xue,Arkani-Hamed:2023kig}. At loop level, even for two-site graphs, the number of auxiliary basis functions can grow rapidly with loop order, increasing the size of the differential system. Examples of such systems for FRW loop integrands can be found in refs.~\cite{De:2023xue,Hang:2024xas,He:2024olr}, and differential-equation calculations involving loop momentum integration are discussed in ref.~\cite{Benincasa:2024ptf}.

In this work, we show that the PDE calculations at loop level can be bypassed entirely: the physical wavefunction coefficients of interest can be obtained directly, without computing any auxiliary functions, for all two-site graphs of the theory \eqref{eq:FRW-action}. Concretely, since the twisted integral \eqref{eq:psi-FRW} is a linear functional of the flat-space wavefunction coefficient, it propagates the flat-space decomposition \eqref{eq:loop-decom} to its FRW counterpart, allowing us to read off all the FRW loops in terms of shifted FRW trees, as long as \eqrefe{eq:loop-decom} holds generically in flat space. The precise conditions, and the extension beyond power-law backgrounds, are discussed after \eqrefe{eq:wfc-loop-dS}. It is therefore enough in the rest of this work to focus on whether the flat-space wavefunction coefficient (integrand) $\psi^{(2;\ell_\tot)}$ satisfies \eqrefe{eq:loop-decom}. We start with the two-site bubble-like family, and generic topologies with possible tadpole loops will be studied in section~\ref{sec:5}.

\subsection{The Two-Site Tree}
\label{sec:2.1}

\begin{figure}[t]
\centering
\includegraphics[scale=1.1]{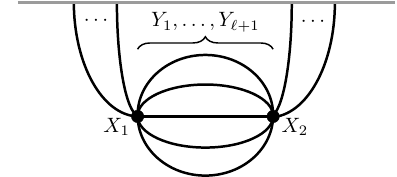}
\caption{The two-site $\ell$-loop diagram, in which the two vertices are connected by $\ell+1$ internal propagators with $\ell\geqq0$. 
Here $X_1,X_2$ denote the total energies flowing from the vertices to the late-time boundary, and $Y_j$ is the energy carried by the $j$-th internal line. 
The internal lines are numbered from bottom to top, with the lowest line labeled by $Y_1$ and the highest line labeled by $Y_{\ell+1}$.}
\label{fig:2siteloop}
\end{figure}

The simplest member of the two-site bubble-like family is the graph with a single internal line, corresponding to $\ell=0$ in the convention of figure~\ref{fig:2siteloop}.\ 
Its flat-space wavefunction coefficient is given as follows \cite{Hang:2024xas,Arkani-Hamed:2023kig}:
\begin{equation}
\label{eq:wfc-chain}
\psi^{(2;0)}(X_1,X_2;Y_1) \,=\, \frac{2Y_1}{(X_1+Y_1)(X_2+Y_1)(X_1+X_2)}\,,
\end{equation}
with $X_1,X_2$ denoting the total energies flowing from the vertices to the late-time boundary, and $Y_1$ the energy of the single internal line connecting the two vertices. 
By introducing the following divisors
\begin{equation}
\label{eq:TLD-chain}
L_1 = X_1 + Y_1 \,, \qquad~~~
L_2 = X_2 + Y_1 \,, \qquad~~~
D = X_1 +  X_2 \,,
\end{equation}
the wavefunction coefficient \eqref{eq:wfc-chain} can be expressed in terms of divisor variables
\begin{equation}
\label{eq:wfc-tree-divisor}
\psi^{(2;0)}(L_1,L_2;D) \,=\, \frac{(L_1+L_2)-D}{L_1\hs L_2\hs D}\,.
\end{equation}
The energy parameters $\{X_1,X_2,Y_1\}$ and the divisors $\{L_1,L_2,D\}$ are two sets of independent variables to express $\psi^{(2;0)}$.
Compared to the form calculated by the Feynman rules, we add an overall normalization factor of $2Y_1$ to flat-space coefficient \eqref{eq:wfc-chain} in the numerator.\
This choice gives a minus sign difference in the corresponding FRW coefficient obtained via twisted integral \eqref{eq:psi-FRW} compared with the convention in refs.~\cite{Hang:2024xas,Arkani-Hamed:2023kig}.\
The choice of normalization factor and overall sign is purely for later convenience when we display the all-loop shifted-tree formula.\ 
The two local divisors $L_1$ and $L_2$ are associated with the two bulk sites, while $D$ is the total-energy divisor measuring the energy entering the two-site subgraph from the late-time boundary.\ 
This distinction will be important below: in the shifted-tree decomposition, the local divisors are shifted in a common way by replacing the single tree energy $Y_1$ with the total internal energy, whereas the divisor $D$ carries the nontrivial subset-dependent shifts.

The flat-space tree coefficient \eqref{eq:wfc-chain} will serve as the seed from which all shifted-tree terms are generated.\ After establishing the flat-space identity \eqref{eq:loop-decom}, the FRW version follows simply by applying the twisted integral \eqref{eq:psi-FRW} on both sides, giving explicit forms of the loop-level FRW coefficients once the FRW tree is known. For example, in de Sitter (dS) space, a two-site tree coefficient is given in refs.~\cite{Arkani-Hamed:2015bza,Hillman:2019wgh} as:
\begin{equation}
\psi^{(2;0)}_{\ds}=\, \Li\!\[\frac{(X_1-Y_1)(X_2-Y_1)}{(X_1+Y_1)(X_2+Y_1)}\]-\Li\!\(\frac{X_1-Y_1}{X_1+Y_1}\)-\Li\!\(\frac{X_2-Y_1}{X_2+Y_1}\)+\zeta(2)\,,
\end{equation}
in the physical region $\{X_1,X_2\}>Y_1>0$. More complicated explicit expressions for the general FRW case are discussed in ref.~\cite{Arkani-Hamed:2023kig}.

\subsection{All-Loop FRW Bubbles from Shifted Flat-Space Trees}
\label{sec:2.2}

Now we review the key formula proposed in ref.~\cite{Hang:2024xas} that specifies the schematic decomposition formula \eqref{eq:loop-decom} to the bubble-like family.\ Consider a two-site graph with $\ell+1$ internal lines. Label the energy of the $j$-th internal line by $Y_j$ and collect these variables
\begin{equation}
\label{eq:two-site-energy-set}
\YY=\{Y_1,\ldots,Y_{\ell+1}\}\,,\qquad~~~
Y_\tot=\sum_{j=1}^{\ell+1}Y_j\,.
\end{equation}
We note that all the energy parameters throughout this paper are allowed to coincide numerically, but are always treated as distinct labels identifying different internal lines in all symbolic derivations, which makes $\YY$ a genuine set.\ 
For any finite set $V\subseteq\YY$, define the (doubled) total shift $\sigma[V]$ and the corresponding shifted total-energy divisor $D_V$ by
\begin{equation}
\label{eq:sigma-subset}
\sigma[V]\equiv 2\sum_{y\in V}y\,,\qquad~~~
D_V\equiv D+\sigma[V]\,.
\end{equation}
In particular, the total shift is additive under disjoint union,
\begin{equation}
\label{eq:additive-total-shift}
\sigma[V_1\sqcup V_2]=\sigma[V_1]+\sigma[V_2]\,, \qquad~~~ 
V_1\cap V_2=\emp\,,
\end{equation}
and
\begin{equation}
\sigma[\emp]=0\,,\qquad
\sigma[\YY]=2Y_\tot\,,\qquad
D_{\emp}=D\,,\qquad
D_{\YY}=D+2Y_\tot\,,
\end{equation}
with $Y_\tot$ given in \eqrefe{eq:two-site-energy-set}.

Specifying \eqrefe{eq:loop-decom} to the bubble-like family is the shifted-tree formula proposed in ref.~\cite{Hang:2024xas} which expresses the two-site $\ell$-loop wavefunction coefficient as the following alternating subset sum
\begin{shadowequation}
\begin{equation}
\label{eq:wfc-loop}
\psi^{(2;\ell)}(X_1,X_2;\YY) \,=\, \sum_{V\subseteq\YY}(-1)^{|V|}\,
\psi^{(2;0)}(X_1,X_2;Y_\tot)\big|_{D\to D_V}\,,\qquad
\ell\geqq0\,.
\end{equation}
\vspace*{-3mm}
\end{shadowequation}
\noindent
In the above formula, the lhs denotes the flat-space wavefunction coefficient for the two-site bubble-like graph at $\ell$ loops, which can be obtained (up to an overall normalization factor that can be determined easily) by applying the Feynman rules or summing over combinatorial tubings of the underlying two-site diagrams \cite{Arkani-Hamed:2017fdk}.\ 
On the right-hand side (rhs), each term is obtained from the standard tree expression in energy basis by replacing the internal energy with $Y_\tot$, then switching to the divisor basis and shifting $D\to D_V$ for a subset $V\subseteq\YY$ while keeping $L_1,L_2$ fixed. Equivalently, in the energy basis each term is the tree coefficient evaluated at shifted physical energies, $\psi^{(2;0)}(X_1,X_2;Y_\tot)\big|_{D\to D_V}=\psi^{(2;0)}\big(X_1+\tfrac12\sigma[V],\,X_2+\tfrac12\sigma[V];\,\tfrac12\sigma[\YY\setminus V]\big)$.
The divisor basis is advantageous in displaying the subset shifts. We comment here that the original formula proposed in ref.~\cite{Hang:2024xas} is in terms of a twisted integrand as in \eqrefe{eq:psi-FRW}, i.e., with the external energies shifted by $X_i\to X_i+x_i$ on both sides, and is only valid for the loop case $\ell\geqq1$ due to the choice of the normalization factor, and also does not include the $V=\YY$ term in the summation. We have made three cleanups here. (i)\,Since the formula is independent of the twist-variable shifts, we exhibit it in terms of a property of flat-space wavefunction coefficients, which clearly encapsulates the twisted version that can be applied inside the twisted integral \eqrefe{eq:psi-FRW}. (ii)\,We adopt the positive sign convention for the normalization factor in \eqrefe{eq:wfc-chain},  extending the formula to $\ell=0$ where the rhs reproduces the consistent tree coefficient $\psi^{(2;0)}$ on the left. (iii)\,We extend the summation to the full power set of $\YY$ including the trivial $V=\YY$ term, which will be clear when we reorganize the rhs at the end of this section.

The flat-space shifted-tree formula \eqref{eq:wfc-loop} is powerful in directly obtaining the explicit form of the loop-level FRW wavefunction coefficients. Since the twisted integral \eqref{eq:psi-FRW} is linear in terms of the functional dependence of the flat-space coefficient $\psi^{(2;\ell)}$ and \eqrefe{eq:wfc-loop} does not care about twist shifts, the twisted integral can be applied term by term on the rhs of \eqrefe{eq:wfc-loop}. This gives an FRW shifted-tree formula in the exact same form because the divisor shifts commute with the twisted integral. Indeed, in the divisor basis $\{L_1,L_2,D\}$ the twisted integral of the tree coefficient is
\begin{equation}
\psi^{(2;0)}_\frw(L_1,L_2;D)\,=\,\int_0^{\infty}\prod_{a=1}^2(\td x_a x_a^{\ga_a})\,\psi^{(2;0)}(L_1\!+\!x_1,L_2\!+\!x_2;D\!+\!x_1\!+\!x_2)\,.
\end{equation}
The replacement $Y_1\to Y_\tot$ changes the local divisors but commutes with the twisted integral.\
With $L_1,L_2$ held fixed, the divisor shift $D\to D_V$ in each flat-space term just propagates directly to the final FRW coefficient
\begin{equation}
\int_0^{\infty} \prod_{a=1}^2
(\td x_a x_a^{\ga_a})\,\psi^{(2;0)}(L_1\!+\!x_1,L_2\!+\!x_2;D\!+\!x_1\!+\!x_2)\big|_{D\to D_V}\!=\,\psi^{(2;0)}_\frw(L_1,L_2;D_V)\,,
\end{equation}
which gives
\begin{equation}
\label{eq:FRW-wfc-loop}
\psi^{(2;\ell)}_\frw \,=\, \sum_{V\subseteq\YY}(-1)^{|V|}\,\psi^{(2;0)}_\frw(X_1,X_2;Y_\tot)\big|_{D \hs\to D_V}\,,\qquad
\ell\geqq0\,.
\end{equation}
The same commutation is also manifest in the energy basis, since the subset shifts are independent of the twist variables.
Therefore, once the two-site FRW tree is known, one can directly obtain the loop-level FRW coefficients in an explicit closed form without solving any new differential-equation system at the loop level. It is enlightening to show the power of the formula by a simple example for the dS case. For convenience, we first express the dS tree in the divisor basis $\{L_1,L_2,D\}$, where the shift $D\to D_V$ is simplest:
\begin{equation}
\psi^{(2;0)}_\ds=\,\Li\!\[\frac{(D\!-\!L_1)(D\!-\!L_2)}{L_1\hs L_2}\]-\Li\!\(\frac{D\!-\!L_2}{L_1}\)-\Li\!\(\frac{D\!-\!L_1}{L_2}\) +\zeta(2)\,.
\end{equation}
Then we are ready to apply the FRW shifted-tree formula \eqrefe{eq:FRW-wfc-loop} inherited directly from the flat-space version, which gives an explicit form of the dS bubble loops without a single calculation:
\begin{align}
\label{eq:wfc-loop-dS}
\psi^{(2;\ell)}_{\ds}\,=\,&\sum_{V\subseteq\YY}(-1)^{|V|}\LB\Li\!\[\frac{(D_V\!-\!L_1)(D_V\!-\!L_2)}{L_1\hs L_2}\]-\Li\!\(\frac{D_V\!-\!L_2}{L_1}\)
\right.\nn\\
&\left.\hspace{2cm}
-\,\Li\!\(\frac{D_V\!-\!L_1}{L_2}\)\!+\zeta(2)\RB,
\end{align}
valid for all $\ell\geqq0$ with local divisors understood as $L_{1,2}=X_{1,2}+Y_\tot$.

Although we have presented the power-law case, the above derivation uses only two properties of the twisted integral \eqref{eq:psi-FRW}: it is linear, and it acts on the flat-space coefficient only by translating the external energies.\ 
It therefore extends to a generic FRW background whenever each time-dependent coupling in \eqrefe{eq:time-dep-flat-action} admits a spectral representation $\lam_n(\eta)=\int_0^\infty\td x\,\rho_n(x)\,e^{\ii x\eta}$ (cf.~eqs.~(2.9)--(2.10) of ref.~\cite{Arkani-Hamed:2017fdk}), possibly in the distributional sense, in which case \eqrefe{eq:psi-FRW} holds with $x_a^{\ga_a}\td x_a$ replaced by $\rho_{n_a}(x_a)\,\td x_a$, where $n_a$ is the valence of vertex $a$; constant couplings correspond to the Dirac measure $\rho_n(x)\,\td x=\lam_n\,\delta(x)\,\td x$ with its full weight at the endpoint $x=0$, recovering flat space. What the power-law case adds is closed-form tree solutions, the twisted-cohomology structure underlying the differential equations mentioned above, and a natural regularization through analytic continuation in $\ga_a$.

Two remarks apply when using these FRW formulas. First, applying the integral term by term requires each shifted-tree term to be integrable, or a common linear regularization of both sides, such as the analytic continuation in $\ga_a$. Second, since the measures are fixed by the vertices of the graph on the lhs, the FRW tree on the rhs is understood as integrated with the same measures: it is the physical tree of the same theory with the same vertices, obtained by opening all but one internal line into external legs, rather than the tree with the same external legs. This is the convention in the dS expression \eqref{eq:wfc-loop-dS} and in the tadpole-decorated formulas of section~\ref{sec:5}. Neither remark affects the flat-space identities \eqref{eq:wfc-loop} and, for tadpole-decorated graphs, \eqref{eq:shifted-tree-decorated} of section~\ref{sec:5}, which are what we prove in this work and which hold without any such assumption.

We will prove the formula \eqref{eq:wfc-loop} rigorously for all $\ell$ and give a novel geometric interpretation in the following sections.\ 
For future convenience, we first reorganize the rhs in an explicit form.\ 
After replacing $Y_1\to Y_\tot$, the divisors $L_1$ and $L_2$ in \eqrefe{eq:wfc-chain} become
\begin{equation}
\label{eq:L1-L2-def}
L_1=X_1+Y_\tot\,,\qquad~~~
L_2=X_2+Y_\tot\,,
\end{equation}
and they satisfy
\begin{equation}
\label{eq:L1+L2=DY}
L_1+L_2=D_{\YY}\,.
\end{equation}
Therefore, in the divisor basis each shifted-tree wavefunction coefficient takes the form
\begin{equation}
\label{eq:wfc-chain-shifted}
\psi^{(2;0)}\Big|_{D \hs\to D_V}\,=\,\frac{(L_1+L_2)-D_V}{L_1L_2D_V}
\,=\, \frac{D_{\YY}-D_V}{L_1L_2D_V}\,,
\end{equation}
which vanishes when $V=\YY$. Therefore, to exhibit a clear combinatorial structure, it is convenient to extend the original summation in ref.~\cite{Hang:2024xas} to include $\YY$ itself. We will see below that this seemingly immaterial observation plays a significant role in organizing the proof and providing geometric insight.\
Now, substituting \eqrefe{eq:wfc-chain-shifted} into \eqrefe{eq:wfc-loop} gives
\begin{align}
\psi^{(2;\ell)}&=\sum_{V\subseteq\YY}
(-1)^{|V|}\,\frac{D_{\YY}-D_V}{L_1L_2D_V}
\nn\\
&=\frac{D_{\YY}}{L_1L_2}\sum_{V\subseteq\YY}\frac{(-1)^{|V|}}{D_V}-\frac{1}{L_1L_2}\sum_{V\subseteq\YY}(-1)^{|V|}\,,
\end{align}
where the second sum vanishes by the binomial identity
\begin{equation}
\sum_{V\subseteq\YY}(-1)^{|V|}
=\sum_{k=0}^{\ell+1}\binom{\ell+1}{k}(-1)^k
=(1-1)^{\ell+1}=0\,.
\end{equation}
What remains is the common two-site prefactor multiplied by an alternating sum over the shifted total-energy divisors, so the shifted-tree formula \eqref{eq:wfc-loop} reduces to
\begin{equation}
\label{eq:wfc-loop-Ds}
\psi^{(2;\ell)}=\frac{D_{\YY}}{L_1L_2}\sum_{V\subseteq\YY} \frac{(-1)^{|V|}}{D_V}\,,
\qquad \ell\geqq0\,.
\end{equation}
In this form, all dependence on the shifted tree terms has collapsed into an alternating subset sum of inverse shifted divisors $D_V$.\ 
The sign factor $(-1)^{|V|}$ coincides with the M\"obius function $\mu(\emp,V)$ on the subset $V$.\
This form will be analyzed in the following sections.

\section{Boolean Lattices, Maximal Chains, and Permutohedra}
\label{sec:3}

The proof of \eqrefe{eq:wfc-loop} for arbitrary $\ell$ proceeds in two steps. Section~\ref{sec:4} proves that the alternating subset sum in \eqrefe{eq:wfc-loop-Ds} equals a maximal-chain expansion. Section~\ref{sec:6} identifies this expansion with the graph-tubing representation derived from the all-loop recursion relation, establishing that the shifted-tree sum reproduces the physical wavefunction coefficient. Here we state the identity and illustrate its organization on the Boolean lattice and the associated permutohedra through low-loop examples.

We first express the alternating sum as a product of commuting finite-difference operators.\
For any shift parameter $a$ and any function $f(D)$, define the translation and finite-difference operators by
\begin{equation}
\label{eq:def-finite-diff}
E_a f(D)\equiv f(D+a)\,,\qquad~~~
\Delta_a \equiv \id - E_a \,,
\end{equation}
where $\id$ denotes the identity operator. For a finite set $A$, we use the product notation
\begin{equation}
\label{eq:set-operator-products}
E_A \equiv \prod_{a\in A}E_a \,,\qquad~~~
\Delta_A \equiv \prod_{a\in A}\Delta_a \,,\qquad~~~
E_\emp \equiv \Delta_\emp \equiv \id \,.
\end{equation}
These products are well-defined since they do not depend on the ordering of $A$. Indeed, we have
\begin{equation}
E_{a_1}E_{a_2}f(D)\,=\,f(D + a_2 + a_1) \,=\,E_{a_2}E_{a_1}f(D)\,.
\end{equation}
Thus, the translation operators commute, which also implies that the finite-difference operators $\Delta_a$ commute.\
Since the divisor shifts appearing below are of the form $2a$, for any finite set $A$ of shift parameters we define $2A\equiv\{2a\mid a\in A\}$.\ 
The notation in eq.~\eqref{eq:set-operator-products} then gives $E_{2A}=\prod_{a\in A}E_{2a}$ and $\Delta_{2A}=\prod_{a\in A}\Delta_{2a}$.

Then consider the action of the operator $\Delta_{2\YY}$ on the seed $1/D$ according to the previous product notation. Expanding $\Delta_{2\YY}$ amounts to choosing either $\id$ or $-E_{2y}$ from each factor.\
Each term is therefore specified by a subset $V\subseteq\YY$ and has sign $(-1)^{|V|}$. Acting with $E_{2V}$ on $1/D$ shifts $D$ to $D_V$.\
Summing over all such choices gives
\begin{equation}
\label{eq:finite-diff-to-vertex-sum}
\Delta_{2\YY}\bigg(\frac1D\bigg)=\sum_{V\subseteq\YY}\frac{(-1)^{|V|}}{D_V}\, ,
\end{equation}
which allows us to rewrite \eqrefe{eq:wfc-loop-Ds} as:
\begin{equation}
\label{eq:wfc-loop-fin-diff}
\psi^{(2;\ell)} \,=\, \frac{L_1+L_2}{L_1L_2}
\Delta_{2\YY}\bigg(\frac1D\bigg)=\Delta_{2\YY}\bigg(\psi^{(2;0)}(L_1,L_2;D)\bigg)\,, \qquad 
\ell\geqq0 \,,
\end{equation}
where in the first equality we use $D_{\YY}=L_1+L_2$ to rewrite the expression in terms of the independent divisor variables $\{L_1,L_2,D\}$, while in the second equality we use \eqrefe{eq:wfc-tree-divisor} together with the fact that the finite-difference operator acts only on $D$ and therefore commutes through the factors $L_1$ and $L_2$. This is the finite-difference form of \eqrefe{eq:wfc-loop}: expanding $\Delta_{2\YY}$ reproduces the alternating sum over $V\subseteq\YY$ in \eqrefe{eq:wfc-loop-Ds}.\
We find that \eqrefe{eq:finite-diff-to-vertex-sum} has a natural combinatorial organization on the Boolean lattice $\PP(\YY)$: a partially ordered set whose elements are all the subsets $V\subseteq\YY$, with the order defined as the inclusion relation of these subsets.\ In its Hasse diagram, two vertices are connected when one subset is obtained from the other by adding a single element.\ The rhs of \eqrefe{eq:finite-diff-to-vertex-sum} is therefore a sum indexed by all elements of $\PP(\YY)$. Examples of Boolean lattices can be found in \eqrefe{eq:bool-bubble} and \eqrefe{eq:bool-sunset}.\
More mathematical exposition is summarized in appendix~\ref{sec:B}.

The maximal-chain decomposition states that the same finite-difference expression can be decomposed into a sum over the maximal chains of $\PP(\YY)$.
A maximal chain is a sequence of subsets running from $\emp$ to $\YY$ by adding one element of $\YY$ at each step. It is therefore determined by the order in which the elements are added.\ We denote this order by a permutation $\pi=(\pi_1,\ldots,\pi_{|\YY|})\in\SS_{|\YY|}$. The corresponding chain is given by
\begin{equation}
\label{eq:ctr-max-chain}
C[\pi]:~\emp=C[\pi]_0 \subsetneq 
C[\pi]_1 \subsetneq \cdots \subsetneq 
C[\pi]_{|\YY|-1} \subsetneq 
C[\pi]_{|\YY|} =\YY \,.
\end{equation}
The rank-$r$ element of this chain and its associated (doubled) cumulative shift are
\begin{equation}
\label{eq:ctr-max-chain-r-vertex-sigma}
C[\pi]_r \equiv \{Y_{\pi_1},\ldots,Y_{\pi_r}\}\,,\qquad~~~
\sigma[\pi]_r \equiv \sigma[C[\pi]_r] =2\sum_{j=1}^{r}Y_{\pi_j}\,,
\end{equation}
where $0\leqq r\leqq|\YY|$. In particular, $\sigma[\pi]_0=0$ and $\sigma[\pi]_{|\YY|}=2Y_{\tot}$.

With the above bookkeeping, the maximal-chain decomposition states:
\begin{shadowequation}
\vspace{-1mm}
\begin{equation}
\label{eq:fin-diff-sum-max-chain}
\Delta_{2\YY}\bigg(\frac1D\bigg)\,=\,\Bigg(\prod_{y\in\YY}2y\Bigg)\sum_{\pi\in\SS_{|\YY|}}\prod_{r=0}^{|\YY|}\frac1{D_{C[\pi]_r}}\,,
\end{equation}
\vspace{-3mm}
\end{shadowequation}
\noindent
where $D_{C[\pi]_r}$ denotes the double-shifted divisor induced by the subset $C[\pi]_r$ as in \eqrefe{eq:sigma-subset}.\
Just as the rhs of \eqrefe{eq:finite-diff-to-vertex-sum} is a sum over the vertices of the Boolean lattice $\PP(\YY)$, the rhs of \eqrefe{eq:fin-diff-sum-max-chain} has an interpretation on the same lattice in two related ways.
(i)\,For each contribution specified by a permutation $\pi$, the denominator is the product of the shifted divisors encountered along the corresponding maximal chain from $\emp$ to $\YY$, with $|\YY|=\ell+1$ steps. Each chain is specified by an ordering of the $\ell+1$ internal energies, so there are $(\ell+1)!$ maximal chains. This is reminiscent of a path integral: the total contribution sums over all such paths between the opposite vertices $\emp$ and $\YY$.
(ii)\,Identifying subsets with the vertices of the $(\ell+1)$-dimensional cube, the vertices of each maximal chain span an $(\ell+1)$-simplex. These simplices form the Freudenthal--Kuhn (FK) triangulation of the cube \cite{Freudenthal1942,Kuhn1960}. The maximal-chain sum in \eqrefe{eq:fin-diff-sum-max-chain} has a geometric realization as the sum of the corresponding simplex integrals.
We will prove \eqrefe{eq:fin-diff-sum-max-chain} algebraically in section~\ref{sec:4.1} and geometrically in section~\ref{sec:4.2}. The algebraic proof is an induction on the finite-difference product, while the geometric proof represents its action on $1/D$ as a weighted integral over the Boolean cube and the maximal-chain expansion as the simplex decomposition of the same integral. In addition, section~\ref{sec:5} extends \eqrefe{eq:fin-diff-sum-max-chain} to multiple divisors.

Substituting \eqrefe{eq:fin-diff-sum-max-chain} into \eqrefe{eq:wfc-loop-fin-diff}, the prefactor $L_1+L_2=D_{\YY}$ cancels the final divisor $D_{C[\pi]_{|\YY|}}=D_{\YY}$ in each chain contribution, giving
\begin{shadowequation}
\begin{equation}
\label{eq:wfc-loop-max-chain}
\psi^{(2;\ell)} \,=\, \Bigg(\prod_{y\in\YY}2y\Bigg)\frac{1}{L_1L_2}\,\sum_{\pi\in\SS_{|\YY|}}\prod_{r=0}^{|\YY|}\frac1{D_{C[\pi]_r}}\,, \qquad \ell\geqq0\,.
\end{equation}
\end{shadowequation}
\noindent
The finite-difference product thus bridges the vertex sum in \eqrefe{eq:finite-diff-to-vertex-sum} and the chain sum in \eqrefe{eq:fin-diff-sum-max-chain}, relating the shifted-tree sum \eqref{eq:wfc-loop} and the maximal-chain decomposition \eqref{eq:wfc-loop-max-chain} of the physical wavefunction. The agreement of \eqrefe{eq:wfc-loop-max-chain} with the all-loop perturbative recursion \cite{Arkani-Hamed:2017fdk} is established in section~\ref{sec:6}, completing the proof of \eqrefe{eq:wfc-loop}. This agreement is illustrated by the examples in section~\ref{sec:7}.

Returning to the main algebraic thread and before turning to the general proof of \eqrefe{eq:fin-diff-sum-max-chain}, it is useful to collect the three equivalent representations in one formula:
\begin{equation}
\label{eq:subset-findiff-maxchain}
\underbrace{\sum_{V\subseteq\YY}\frac{(-1)^{|V|}}{D_V}}_{\text{subset sum}}\quad=
\underbrace{\Delta_{2\YY}\bigg(\frac1D\bigg)}_{\text{finite-difference form}}=\quad
\underbrace{\bigg(\prod_{y\in\YY}2y\bigg)
\sum_{\pi\in\SS_{|\YY|}}\prod_{r=0}^{|\YY|}
\frac{1}{D_{C[\pi]_r}}}_{\text{maximal-chain sum}}\,.
\end{equation}
Beyond the Boolean lattice, the above identity admits a natural organization on the associated permutohedron, whose vertices correspond to the maximal chains of $\PP(\YY)$.
For $|\YY|\geqq2$, let $P_{|\YY|}$ be the $(|\YY|-1)$-dimensional permutohedron, defined as the convex hull of all coordinate permutations of $(1,\ldots,|\YY|)$ \cite{postnikov2009permutohedra}, with coordinates labeled by the elements of $\YY$. The vertex $v_\pi$, whose coordinate labeled by $Y_{\pi_r}$ equals $r$, corresponds to the maximal chain $C[\pi]$.

For each nonempty proper subset $V\subsetneq\YY$, the convex hull of the vertices whose corresponding chains contain $V$ is a facet $F_V$ of dimension $|\YY|-2$.\ Hence the facets meeting at $v_\pi$ are indexed by the $|\YY|-1$ intermediate subsets of $C[\pi]$, and since these nested subsets determine $\pi$ uniquely, their intersection is the single vertex $v_\pi$:
\begin{equation}
\label{eq:permutohedron-chain-facets}
v_\pi=\bigcap_{r=1}^{|\YY|-1}F_{C[\pi]_r}\,.
\end{equation}
We label each facet $F_V$ by $D_V$ and each vertex $v_\pi$ by the product of the labels of its incident facets.\ 
By \eqrefe{eq:permutohedron-chain-facets}, the reciprocal of the vertex label is then the product of inverse intermediate divisors in \eqrefe{eq:wfc-loop-max-chain}.\ 
The endpoint subsets $\emp$ and $\YY$ do not label facets: after multiplication by the wavefunction prefactor, $D_{\YY}$ cancels and $D_\emp$ remains in the common denominator.\ 
This mirrors the shifted-tree expansion, where the empty-subset term is retained while the full-subset term vanishes by \eqrefe{eq:wfc-chain-shifted}. The M\"obius signs $(-1)^{|V|}$ from the subset sum are carried by the facet labels but not by the vertex labels.

The following examples work out \eqrefe{eq:subset-findiff-maxchain} explicitly for $|\YY|=2$ and $|\YY|=3$, showing how the subset expansion is reorganized into maximal chains on the Boolean lattice and how these chains are further arranged on the permutohedra $P_2$ and $P_3$. Their graph-tubing interpretation is given in section~\ref{sec:6}.

\subsection{Example: One-Loop Bubble}
\label{sec:3.1}

\begin{figure}[tp!]
\centering
\begin{subfigure}{0.48\textwidth}
\centering
\hspace*{-1mm}
\includegraphics[scale=1.]{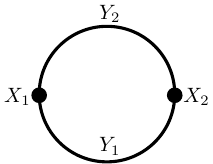}
\vs
\caption{}
\label{fig:1loop-bubble}
\end{subfigure}
\hspace*{-1.cm}
\begin{subfigure}{0.48\textwidth}
\centering
\hspace*{-1mm}
\includegraphics[scale=1.]{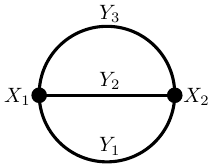}
\vs
\caption{}
\label{fig:2loop-sunset}
\end{subfigure}
\vspace{-2mm}
\caption{(a)\,Two-site one-loop bubble diagram.
(b)\,Two-site two-loop sunset diagram.
The late-time boundary and external legs are omitted.}
\end{figure}

We first consider the one-loop bubble shown in figure~\ref{fig:1loop-bubble}. 
In that case, we have $\ell=1$, and the set of internal energies is $\YY=\{Y_1,Y_2\}$ with $|\YY|=2$.
The associated Boolean lattice of $\PP(\YY)$ contains two rank-one nodes and gives $2!=2$ maximal chains. The corresponding Hasse diagram is given by
\begin{equation}
\label{eq:bool-bubble}
\begin{tikzpicture}[
scale=1.2,
baseline=(current bounding box.center),
every node/.style={inner sep=2pt},
edge/.style={line width=1.1pt}
]
\node (empty) at (0,0) {$\emp$};
\node (x) at (-1,1) {$\{Y_1\}$};
\node (y) at (1,1) {$\{Y_2\}$};
\node (z) at (0,2) {$\YY$};
\draw[edge] (empty) -- (x);
\draw[edge] (empty) -- (y);
\draw[edge] (x) -- (z);
\draw[edge] (y) -- (z);
\end{tikzpicture}
\end{equation}
and the corresponding maximal chains are
\begin{equation}
\label{eq:max-chain-bubble}
C[(1,2)]:~ \emp\subsetneq\{Y_1\}\subsetneq\YY\,,
\qquad~~~
C[(2,1)]:~ \emp\subsetneq\{Y_2\}\subsetneq \YY\,.
\end{equation}
Using these two chains and the energy set $\YY$, the general identity \eqref{eq:subset-findiff-maxchain} specializes to
\begin{align}
\label{eq:fin-diff-bubble}
\Delta_{2Y_2}\Delta_{2Y_1}\bigg(\frac1D\bigg)&
=\frac1{D_\emp}-\frac1{D_{\{Y_1\}}}-\frac1{D_{\{Y_2\}}}+\frac1{D_{\YY}}
\nn\\
&= 4Y_1Y_2\(\frac1{D_\emp D_{\{Y_1\}}D_{\YY}}+\frac1{D_\emp D_{\{Y_2\}}D_{\YY}}\).
\end{align}
To convert \eqrefe{eq:fin-diff-bubble} back into the wavefunction coefficient, it is convenient to label the proper subsets of $\YY$ by
\begin{equation}
\(V_0,V_1,V_2\)=\(\emp,\,\{Y_1\},\,\{Y_2\}\).
\end{equation}
The corresponding shifted divisors then take the explicit form:
\begin{align}
\label{eq:DV-bub}
D_0 &\equiv D_{V_0} = D = X_1+X_2\,,
\nn\\
D_1 &\equiv D_{V_1}=D + 2Y_1\,,\qquad
D_2 \equiv D_{V_2}=D + 2Y_2\,,
\nn\\
D_{\YY} &= D+2Y_\tot = D + 2(Y_1+Y_2)\,.
\end{align}
The full-set divisor $D_{\YY}$ is not assigned an additional label since it cancels when the common prefactor in \eqrefe{eq:wfc-loop-fin-diff} is restored.\
The shifted divisors $L_1$ and $L_2$ in this prefactor are obtained by replacing the single energy $Y_1$ by the total internal energy, $Y_1\to Y_\tot=Y_1+Y_2$
\begin{equation}
L_1 = X_1+Y_1+Y_2\,, \qquad~~~
L_2= X_2+Y_1+Y_2 \,.
\end{equation}
Then, multiplying \eqrefe{eq:fin-diff-bubble} by $D_{\YY}/(L_1L_2)$, we can obtain the maximal-chain form of the one-loop bubble wavefunction coefficient as
\begin{equation}
\label{eq:max-chain-l1}
\psi^{(2;1)}\,=\,\frac{4Y_1Y_2}{L_1L_2D_0}\(\frac{1}{D_1}+\frac{1}{D_2}\).
\end{equation}
With the same divisor labels, the shifted-tree expression \eqref{eq:wfc-loop} for the one-loop bubble case can be obtained by using the M\"obius signs determined by $(-1)^{|V_j|}$ with $j=0,1,2$\,:
\begin{equation}
\label{eq:wfc-1loop}
\psi^{(2;1)}
\,=\,\frac{(L_1+L_2)-D_0}{L_1L_2D_0}
-\frac{(L_1+L_2)-D_1}{L_1L_2D_1}
-\frac{(L_1+L_2)-D_2}{L_1L_2D_2} \,,
\end{equation}
which is equal to \eqrefe{eq:max-chain-l1}.

For $|\YY|=2$, the permutohedron $P_2$ is a line segment:
\begin{equation}
\label{eq:P2}
\begin{tikzpicture}
[scale=0.9,
baseline=(current bounding box.center),
dot/.style={circle, fill=red, draw=red, inner sep=1.6pt},
edge/.style={line width=1.2pt},
sym/.style={red!90!black}]
\coordinate (v1) at (0,0);
\coordinate (v2) at (2.4,0);
\draw[edge] (v1)--(v2);
\node[dot,sym,label={[text=red!90!black]above:{$D_1^-$}}] at (v1) {};
\node[dot,sym,label={[text=red!90!black]above:{$D_2^-$}}] at (v2) {};
\node[text=gray] at (1.2,0.6) {$D_0^+$};
\end{tikzpicture}
\end{equation}
where the two endpoints represent the maximal chains $C[(1,2)]$ and $C[(2,1)]$ in \eqrefe{eq:max-chain-bubble}. Their unique intermediate subsets, $\{Y_1\}$ and $\{Y_2\}$, index the facets $F_{\{Y_1\}}$ and $F_{\{Y_2\}}$. Since the facets of a line segment are its endpoints, the divisor labels $D_{\{Y_1\}}=D_1$ and $D_{\{Y_2\}}=D_2$ label both facets and maximal-chain vertices.
These correspondences are summarized below, with $\longmapsto$ indicating the divisor labels:
\begin{equation}
C[(1,2)] \,\longleftrightarrow\, v_{(1,2)}= F_{\{Y_1\}}\,\longmapsto\, D_1\,,\qquad~~
C[(2,1)] \,\longleftrightarrow\, v_{(2,1)}=F_{\{Y_2\}} \,\longmapsto\, D_2\,.
\end{equation}
Multiplying the reciprocal labels $1/D_1$ and $1/D_2$ by the common factor $4Y_1Y_2/(L_1L_2D_0)$ gives the contributions of $C[(1,2)]$ and $C[(2,1)]$, respectively, in \eqrefe{eq:max-chain-l1}.\ 
Only the intermediate divisors distinguish the two vertices: the bottom and top elements $\emp$ and $\YY$ are shared by both chains and do not index facets. The top divisor $D_{\YY}$ has canceled, while the bottom divisor $D_\emp=D_0$ remains in the common factor.\ 
The central label $D_0$ records this bottom divisor and the empty-subset term.\ 
In addition, the superscripts $\pm$ on the divisors record the overall signs appearing in each term of \eqrefe{eq:wfc-1loop}.

\subsection{Example: Two-Loop Sunset}
\label{sec:3.2}

For the two-site two-loop sunset shown in figure~\ref{fig:2loop-sunset}, we have $\ell=2$, and the set of internal energies is $\YY=\{Y_1,Y_2,Y_3\}$ with $|\YY|=3$.\
The Boolean lattice $\PP(\YY)$ now contains three rank-one nodes and three rank-two nodes, so it has $3!=6$ maximal chains.\
The corresponding Hasse diagram is given by
\begin{equation}
\label{eq:bool-sunset}
\begin{tikzpicture}[
scale=1.2,
baseline=(current bounding box.center),
every node/.style={inner sep=2pt},
edge/.style={line width=1.1pt}]
\node (empty) at (0,0) {$\emp$};
\node (x) at (-2,1.2) {$\{Y_1\}$};
\node (y) at (0,1.2) {$\{Y_2\}$};
\node (z) at (2,1.2) {$\{Y_3\}$};
\node (xy) at (-2,2.6) {$\{Y_1,Y_2\}$};
\node (xz) at (0,2.6) {$\{Y_1,Y_3\}$};
\node (yz) at (2,2.6) {$\{Y_2,Y_3\}$};
\node (xyz) at (0,3.8) {$\YY$};
\draw[edge]  (empty) -- (x);
\draw[edge]  (empty) -- (y);
\draw[edge]  (empty) -- (z);
\draw[edge]  (x) -- (xy);
\draw[edge] (x) -- (xz);
\draw[edge]  (y) -- (xy);
\draw[edge]  (y) -- (yz);
\draw[edge]  (z) -- (xz);
\draw[edge]  (z) -- (yz);
\draw[edge]  (xyz) -- (xy);
\draw[edge]  (xyz) -- (yz);
\draw[edge]  (xyz) -- (xz);
\end{tikzpicture}
\end{equation}
and the corresponding maximal chains are
\begin{align}
\label{eq:max-chain-sunset}
C[(1,2,3)]:&~\emp\subsetneq\{Y_1\}\subsetneq\{Y_1,Y_2\}\subsetneq\YY,\quad~~
C[(1,3,2)]:\emp\subsetneq\{Y_1\}\subsetneq\{Y_1,Y_3\}\subsetneq\YY,
\nn\\[1mm]
C[(2,1,3)]:&~\emp\subsetneq\{Y_2\}\subsetneq\{Y_1,Y_2\}\subsetneq\YY,\quad~~
C[(2,3,1)]:\emp\subsetneq\{Y_2\}\subsetneq\{Y_2,Y_3\}\subsetneq\YY,
\nn\\[1mm]
C[(3,1,2)]:&~\emp\subsetneq\{Y_3\}\subsetneq\{Y_1,Y_3\}\subsetneq \YY,\quad~~
C[(3,2,1)]:\emp\subsetneq\{Y_3\}\subsetneq\{Y_2,Y_3\}\subsetneq\YY.
\end{align}
Applying \eqrefe{eq:subset-findiff-maxchain}, we can obtain
\begin{align}
\label{eq:finite-diff-sum-sun}
&\Delta_{2\YY}\bigg(\frac1D\bigg)=\frac1{D_\emp}
-\frac1{D_{\{Y_1\}}} -\frac{1}{D_{\{Y_2\}}}
-\frac1{D_{\{Y_3\}}}
+\frac1{D_{\{Y_1,Y_2\}}}
+\frac1{D_{\{Y_1,Y_3\}}}
+\frac1{D_{\{Y_2,Y_3\}}}
-\frac1{D_{\YY}}
\nn\\
&=8Y_1Y_2Y_3\Bigg(
\frac1{D_\emp D_{\{Y_1\}}D_{\{Y_1,Y_2\}}D_{\YY}}
+\frac1{D_\emp D_{\{Y_1\}} D_{\{Y_1,Y_3\}}D_{\YY}} 
+\frac{1}{D_\emp D_{\{Y_2\}} D_{\{Y_1,Y_2\}}D_{\YY}}
\nn\\
&\quad~+\hs\frac{1}{D_\emp D_{\{Y_2\}} D_{\{Y_2,Y_3\}}D_{\YY}}
+\frac{1}{D_\emp D_{\{Y_3\}} D_{\{Y_1,Y_3\}}D_{\YY}} 
+\frac{1}{D_\emp D_{\{Y_3\}} D_{\{Y_2,Y_3\}}D_{\YY}} \Bigg)\,.
\end{align}
Similarly, by choosing an ordering for the proper subsets and labeling them as
\begin{equation}
\(V_0,\ldots,V_6\)=\(\emp,\,\{Y_1\},\,\{Y_2\},\,\{Y_3\},\,\{Y_1,Y_2\},\,\{Y_1,Y_3\},\,\{Y_2,Y_3\} \),
\end{equation}
we can write down the shifted divisors
\begin{align}
D_0 &\equiv D_{\emp} = D = X_1+X_2\,,
\nn\\
D_1 &\equiv D_{\{Y_1\}} = D+2Y_1 \,, 
\qquad
D_2 \equiv D_{\{Y_2\}} = D+2Y_2 \,, 
\nn\\
D_3 &\equiv D_{\{Y_3\}} = D+2Y_3 \,, 
\nn\\
D_4 &\equiv D_{\{Y_1,Y_2\}} = D+2(Y_1+Y_2) \,, 
\qquad
D_5 \equiv D_{\{Y_1,Y_3\}}= D+2(Y_1+Y_3) \,, 
\nn\\
D_6 &\equiv D_{\{Y_2,Y_3\}}= D+2(Y_2+Y_3)\,,
\nn\\
D_{\YY} &= D+2Y_\tot = D + 2(Y_1+Y_2+Y_3)\,.
\end{align}
Multiplying \eqrefe{eq:finite-diff-sum-sun} by $D_{\YY}/(L_1L_2)$, we obtain
\begin{equation}
\label{eq:max-chain-l2}
\psi^{(2;2)}\,=\,\frac{8Y_1Y_2Y_3}{L_1L_2D_0}\(\frac{1}{D_1 D_4}+\frac{1}{D_1 D_5} 
+\frac{1}{D_2 D_4} +\frac{1}{D_2 D_6} +\frac{1}{D_3 D_5} +\frac{1}{D_3 D_6} \),
\end{equation}
where the shifted divisors $L_1$ and $L_2$ are given by
\begin{equation}
L_1 = X_1+Y_1+Y_2+Y_3 \,, \qquad~~~
L_2= X_2+Y_1+Y_2+Y_3 \,.
\end{equation}
With the same labels, the shifted-tree expression \eqref{eq:wfc-loop} for $\ell=2$ becomes
\begin{equation}
\label{eq:wfc-2loop}
\psi^{(2;2)}\,=\, \frac{(L_1+L_2)-D_0}{L_1L_2D_0}
-\sum_{j=1}^{3} \frac{(L_1+L_2)-D_j}{L_1L_2D_j} +\sum_{j=4}^{6}\frac{(L_1+L_2)-D_j}{L_1L_2D_j}\,,
\end{equation}
which is equal to \eqrefe{eq:max-chain-l2}.

For $|\YY|=3$, the permutohedron $P_3$ is a hexagon:
\begin{equation}
\label{eq:P3}
\centering
\begin{tikzpicture}
[scale=0.9,
baseline=(current bounding box.center),
dot/.style={circle, fill=black, inner sep=1.6pt},
edge/.style={very thick},
sym/.style={red!90!black, line width=2pt}]
\coordinate (O) at (0,0);
\coordinate (v1) at (-1,{sqrt(3)});
\coordinate (v2) at ( 1,{sqrt(3)});
\coordinate (v3) at ( 2,0);
\coordinate (v4) at ( 1,{-sqrt(3)});
\coordinate (v5) at (-1,{-sqrt(3)});
\coordinate (v6) at (-2,0);

\draw[sym,edge] (v1)--(v2)--(v3)--(v4)--(v5)--(v6)--(v1);

\node[dot, label=above:{$D_{1,4}$}] at (v1) {};
\node[dot, label=above:{$D_{1,5}$}] at (v2) {};
\node[dot, label=right:{$D_{3,5}$}] at (v3) {};
\node[dot, label=below:{$D_{3,6}$}] at (v4) {};
\node[dot, label=below:{$D_{2,6}$}] at (v5) {};
\node[dot, label=left:{$D_{2,4}$}] at (v6) {};
\node[sym] at (0,2.28)        {$D_1^-$};
\node[sym] at (1.95,1.10)     {$D_5^+$};
\node[sym] at (1.95,-1.10)    {$D_3^-$};
\node[sym] at (0,-2.15)       {$D_6^+$};
\node[sym] at (-1.95,-1.10)   {$D_2^-$};
\node[sym] at (-1.95,1.10)    {$D_4^+$};
\node[text=gray] at (0,0) {$D_0^+$};
\end{tikzpicture}
\end{equation}
where $D_{i,j}\equiv D_iD_j$.\ 
In \eqrefe{eq:P3}, the six vertices represent the maximal chains in \eqrefe{eq:max-chain-sunset}. Each chain passes through two intermediate subsets, which index its two incident facets (edges). 
By \eqrefe{eq:permutohedron-chain-facets}, their intersection is the corresponding vertex, whose divisor label is the product of their labels.\ 
These correspondences are summarized below
\begin{align}
C[(1,2,3)] &\,\longleftrightarrow\, v_{(1,2,3)}=F_{\{Y_1\}}\cap F_{\{Y_1,Y_2\}}
\,\longmapsto\, D_{1,4}\,,
\nn\\[1mm]
C[(1,3,2)] &\,\longleftrightarrow\, v_{(1,3,2)}=F_{\{Y_1\}}\cap F_{\{Y_1,Y_3\}}
\,\longmapsto\, D_{1,5}\,,
\nn\\[1mm]
C[(2,1,3)] &\,\longleftrightarrow\, v_{(2,1,3)}=F_{\{Y_2\}}\cap F_{\{Y_1,Y_2\}}
\,\longmapsto\, D_{2,4}\,,
\nn\\[1mm]
C[(2,3,1)] &\,\longleftrightarrow\, v_{(2,3,1)}=F_{\{Y_2\}}\cap F_{\{Y_2,Y_3\}}
\,\longmapsto\, D_{2,6}\,,
\nn\\[1mm]
C[(3,1,2)] &\,\longleftrightarrow \,v_{(3,1,2)}=F_{\{Y_3\}}\cap F_{\{Y_1,Y_3\}}
\,\longmapsto\, D_{3,5}\,,
\nn\\[1mm]
C[(3,2,1)] &\,\longleftrightarrow\, v_{(3,2,1)}=F_{\{Y_3\}}\cap F_{\{Y_2,Y_3\}}
\,\longmapsto\, D_{3,6}\,.
\end{align}
Multiplying the six reciprocal vertex labels by the common factor $8Y_1Y_2Y_3/(L_1L_2D_0)$ gives the six chain contributions in \eqrefe{eq:max-chain-l2}. The bottom and top elements $\emp$ and $\YY$ are shared by all chains and do not index facets: $D_{\YY}$ has canceled, while $D_\emp=D_0$ remains in the common factor and is displayed at the center as $D_0$.

Conversely, each facet $F_V$ contains precisely the vertices whose chains pass through $V$, as stated before \eqrefe{eq:permutohedron-chain-facets}. In $P_3$, these are the two endpoints of its edge. The following pairs list the two vertices of each facet together with its divisor label:
\begin{alignat}{4}
v_{(1,2,3)}\,,v_{(1,3,2)}\in{}&F_{\{Y_1\}}
&&\longmapsto D_1\,, \qquad~~~
&v_{(2,1,3)}\,,v_{(2,3,1)}\in{}&F_{\{Y_2\}}
&&\longmapsto D_2\,,
\nn\\[1mm]
v_{(3,1,2)}\,,v_{(3,2,1)}\in{}&F_{\{Y_3\}}
&&\longmapsto D_3\,, \qquad~~~
&v_{(1,2,3)}\,,v_{(2,1,3)}\in{}&F_{\{Y_1,Y_2\}}
&&\longmapsto D_4\,,
\nn\\[1mm]
v_{(1,3,2)}\,,v_{(3,1,2)}\in{}&F_{\{Y_1,Y_3\}}
&&\longmapsto D_5\,, \qquad~~~
&v_{(2,3,1)}\,,v_{(3,2,1)}\in{}&F_{\{Y_2,Y_3\}}
&&\longmapsto D_6\,.
\end{alignat}
For example, the top edge $F_{\{Y_1\}}$ carries $D_1$, the divisor shared by its endpoint labels $D_{1,4}=D_1D_4$ and $D_{1,5}=D_1D_5$. Thus the vertices encode the six maximal-chain terms, while the six facet labels together with the central empty-subset label $D_0$ organize the seven shifted-tree terms.\ 
Again, the superscripts $\pm$ on the divisors of \eqrefe{eq:P3} record the overall signs appearing in each term of \eqrefe{eq:wfc-2loop}.

The three-loop banana example is given in appendix~\ref{app:A1} and we can continue these examples for higher loops.\
By using {\sc Mathematica}, we have checked \eqrefe{eq:fin-diff-sum-max-chain} on large random primes up to $\ell=9$.\
In the next section, we will provide the rigorous proof of \eqrefe{eq:fin-diff-sum-max-chain} for arbitrary $\ell$.

\section{Proofs of the Maximal-Chain Formula}
\label{sec:4}

Having identified the maximal-chain formula through the low-loop examples above, we now prove the identity in full generality.\
We provide two complementary arguments.\
The first is algebraic: it treats \eqrefe{eq:fin-diff-sum-max-chain} as an identity of rational functions and proves it by induction, with the induction step implemented by the Leibniz rule that inserts one new shift into every possible position of a chain.\
The second is geometric: it rewrites both sides as the same cubical integral, where the maximal chains supply the simplex decomposition of the Boolean cube.\
Together, these two proofs show that the maximal-chain formula is first a combinatorial identity on the Boolean lattice, and then a geometric simplex decomposition of the associated Boolean cube.

\subsection{Algebraic Proof via Induction}
\label{sec:4.1}

We first prove the maximal-chain identity~\eqref{eq:fin-diff-sum-max-chain} algebraically, by induction on the number of internal energies, $|\YY|=\ell+1$. Using the cumulative shifts $\sigma[\pi]_r$ defined in \eqrefe{eq:ctr-max-chain-r-vertex-sigma}, the identity can be expressed as
\begin{equation}
\label{eq:finite-diff-sum-max-chain-b}
\Delta_{2\YY}\bigg(\frac1D\bigg)=\Bigg(\prod_{i=1}^{\ell+1}2Y_i\Bigg)
\sum_{\pi\in\SS_{\ell+1}}
\prod_{r=0}^{\ell+1}\frac{1}{D+\sigma[\pi]_r} \,.
\end{equation}

\paragraph{Proof.}
For $\ell=0$, we have the internal energy set $\YY=\{Y_1\}$ and a unique permutation $\pi\in\SS_1$, so the base case holds trivially:
\begin{equation}
\Delta_{2Y_1}\bigg(\frac1{D}\bigg)
\,=\, \frac1{D}-\frac1{D+2Y_1}
\,=\, \frac{2Y_1}{D(D+\sigma[\pi]_1)} \,.
\end{equation}

Assume the formula holds for $\YY=\{Y_1,\ldots,Y_{\ell+1}\}$.\
To show it holds after adjoining $Y_{\ell+2}$, applying $\Delta_{2Y_{\ell+2}}$ to both sides of \eqrefe{eq:finite-diff-sum-max-chain-b} yields
\begin{equation}
\label{eq:finite-diff-sum-max-chain-c}
\Delta_{2(\YY\sqcup\hs\{Y_{\ell+2}\})}\bigg(\frac1{D}\bigg)=\Bigg(\prod_{i=1}^{\ell+1}2Y_i\Bigg)
\sum_{\pi\in\SS_{\ell+1}}\Delta_{2Y_{\ell+2}}\!
\(~\prod_{r=0}^{\ell+1}\frac1{D+\sigma[\pi]_r}\) .
\end{equation}
The only nontrivial step is the following elementary identity.

\paragraph{Lemma.}
For any fixed values $\sigma_0,\ldots,\sigma_M$ independent of $D$, one has
\begin{equation}
\label{eq:finite-difference-product-identity}
\Delta_a\!\(\,\prod_{r=0}^{M}\frac{1}{D+\sigma_r}\)=
\,a\sum_{j=0}^{M}
\(\,\prod_{r=0}^{j}\frac{1}{D+\sigma_r}\)
\(\,\prod_{r=j}^{M}\frac{1}{D+\sigma_r+a}\).
\end{equation}

Assuming the lemma for the moment, apply it with $\sigma_r=\sigma[\pi]_r$, $a=2Y_{\ell+2}$, and $M=\ell+1$ to every chain contribution $\pi\in\SS_{\ell+1}$ in the induction hypothesis. We then obtain, on the rhs of \eqrefe{eq:finite-diff-sum-max-chain-c},
\begin{align}
&\Bigg(\prod_{i=1}^{\ell+1}2Y_i\Bigg)
\sum_{\pi\in\SS_{\ell+1}}\Delta_{2Y_{\ell+2}}\!
\(~\prod_{r=0}^{\ell+1}\frac1{D+\sigma[\pi]_r}\)
\nn\\[1mm]
=&\(\,\prod_{i=1}^{\ell+2}2Y_i\)
\sum_{\pi\in\SS_{\ell+1}} \sum_{j=0}^{\ell+1}
\Bigg(\prod_{r=0}^{j}\frac1{D+\sigma[\pi]_r}\Bigg)\Bigg(\prod_{r=j}^{\ell+1}\frac1{D+\sigma[\pi]_r+2Y_{\ell+2}}\Bigg)\,.
\end{align}
The lemma states that applying one more finite difference to an existing chain is equivalent to inserting the new shift at every possible position of that chain. These insertions generate all maximal chains of the enlarged Boolean lattice exactly once. Explicitly, for a fixed permutation
$\pi$, the index $j$ specifies the position after which the new element $Y_{\ell+2}$ is inserted into the chain, so that the corresponding term can be written in terms of the augmented permutation as
\begin{equation}
\Bigg(\prod_{r=0}^{j}\frac1{D+\sigma[\pi]_r}\Bigg)\Bigg(\prod_{r=j}^{\ell+1}\frac1{D+\sigma[\pi]_r+2Y_{\ell+2}}\Bigg)=\,\prod_{r=0}^{\ell+2}\frac1{D+\sigma[	\pi^{\langle j\rangle}]_r}\,,
\end{equation}
where $\pi^{\langle j\rangle}$ denotes the augmented permutation obtained by inserting the new label $\ell+2$ after the $j$-th entry of $\pi$:
\begin{equation}
\pi^{\langle j\rangle} \equiv\left(\pi_1, \ldots, \pi_j, \ell+2, \pi_{j+1}, \ldots, \pi_{\ell+1}\right)\in \SS_{\ell+2}\,, \quad 0 \leqq j \leqq \ell+1\,.
\end{equation}
As $\pi$ runs over $\SS_{\ell+1}$ and $j$ runs from 0 to $\ell+1$, these insertions produce every element of $\SS_{\ell+2}$ exactly once. Thus the induction step follows, provided the lemma holds.

\paragraph{Proof of the lemma.}
The lemma can be proved using the Leibniz rule for finite-difference operators. For an arbitrary shift on generic functions, it reads
\begin{equation}
\label{eq:general-leibniz}
\Delta(AB)=(\Delta A)(EB)+A(\Delta B)\,.
\end{equation}
Indeed, with $\Delta\equiv\id-E$ we have
\begin{align}
\Delta(AB)&=AB-E(AB)=AB-(EA)(EB)
\nn\\
&=(A-EA)(EB)+A(B-EB)=(\Delta A)(EB)+A(\Delta B)\,.
\end{align}
The finite-difference operator distributes asymmetrically among the products.\ 
We choose the above convention: when $\Delta$ acts on a given factor, all factors to its right are acted on by $E$. For a product of $Q$ functions $f_1,\ldots,f_Q$, iterating the Leibniz rule gives
\begin{equation}
\label{eq:general-product-leibniz}
\Delta\Bigg(\prod_{q=1}^{Q}f_q\Bigg)\,=\,\sum_{q=1}^{Q}\Bigg(\prod_{i<q} f_i\Bigg)
(\Delta f_q)\Bigg(\prod_{i>q}E \hs f_i\Bigg)\,,
\end{equation}
which amounts to choosing the factor $f_q$ hit by the finite difference and summing over all such choices, while shifting all factors to its right by $E$.
Applying to the functions in the lemma yields
\begin{align}
&\Delta_a\!\(\,\prod_{r=0}^{M}\frac{1}{D+\sigma_r}\)=
\sum_{j=0}^{M}
\(\,\prod_{r=0}^{j-1}\frac{1}{D+\sigma_r}\) \(\Delta_a \frac1{D+\sigma_j}\)
\(\,\prod_{r=j+1}^{M}E_a\frac{1}{D+\sigma_r}\)
\nn\\
&\qquad =\,\sum_{j=0}^{M}
\(\,\prod_{r=0}^{j-1}\frac{1}{D+\sigma_r}\) \( \frac{a}{(D+\sigma_j)(D+\sigma_j+a)}\)
\(\,\prod_{r=j+1}^{M}\frac{1}{D+\sigma_r+a}\),
\end{align}
which is the desired equality.

Together with the induction argument above, this completes the algebraic proof of \eqrefe{eq:finite-diff-sum-max-chain-b}.\
As a consistency check, we also formalized \eqrefe{eq:fin-diff-sum-max-chain} in {\sc Lean 4} \cite{moura2021lean}, where the equality is verified by induction on the list of shifts. The corresponding code is provided in the GitHub repository \href{https://github.com/yf-hang/formal-physics-lean}{\textcolor{black}{\faGithub}}.

\paragraph{Comment on the Leibniz convention}
Another equivalent convention places the translation operator on the first factor, so that, upon iteration, all factors to the left of $\Delta$ action are shifted. Since the factors commute in our application, interchanging them converts one convention into the other. However, we choose the convention that is naturally aligned with the direction of filtrations: the shifts propagate in the same direction as the chains viewed as ascending filtrations. More explicitly, if the factors fit into an ascending filtration $\cdots f_q\subseteq f_{q+1}\cdots$, the product will be more naturally written as $\cdots f_qf_{q+1}\cdots$, so that every term in the Leibniz expansion is an ordered sequence $\left(f_1, \ldots, f_{q-1}, \Delta f_q, E f_{q+1}, \ldots, E f_Q\right)$. This has the interpretation that the ``past'' of the action is untouched, while the ``future'' is shifted. Throughout the paper, when such a filtration structure appears, we will always view it in ascending direction from left to right.

\subsection{Geometric Proof via Simplex Decomposition}
\label{sec:4.2}

As the induction proof may not fully capture the intrinsic geometric feature of the formula, here we provide a geometric proof that makes explicit the simplex triangulation mentioned in section~\ref{sec:3}.

We use the same notation as in section~\ref{sec:4.1}, and set $M=|\YY|=\ell+1$ throughout this section. On the rhs, the sum over all orderings suggests a simplicial interpretation. For a fixed permutation $\pi\in\SS_M$, the corresponding maximal-chain contribution has the Laplace representation
\begin{equation}
\label{eq:geo-proof-laplace-rep}
\prod_{r=0}^{M}\frac1{D+\sigma[\pi]_r}
\,=\,\int_{t_r\geqq0}\,\prod_{r=0}^{M}\td t_r\exp\!\[-\sum_{r=0}^{M}\(D+\sigma[\pi]_r\)t_r\],
\end{equation}
where we temporarily restrict ourselves to the convergent region $\re(D+\sigma[\pi]_r)>0$.\
The exponent satisfies
\begin{equation}
\sum_{r=0}^{M}\sigma[\pi]_r t_r
=\sum_{k=1}^{M}\sum_{r=1}^{k}2Y_{\pi_r}t_k
=\sum_{r=1}^{M}2Y_{\pi_r}\sum_{k=r}^{M}t_k\,,\qquad~ \sigma[\pi]_0=0\,,
\end{equation}
which suggests switching to cumulative variables
\begin{equation}
u_r\equiv \sum_{k=r}^{M} t_k\quad(1\leqq r\leqq M)\,, \qquad~~~
u_0\equiv \sum_{k=0}^{M}t_k\,, \qquad~~~
u_{M+1}\equiv 0\,.
\end{equation}
Equivalently, the inverse transformation is
\begin{equation}
\label{eq:t-as-u}
t_0=u_0-u_1\,,\qquad~~
t_r=u_r-u_{r+1}\,,\qquad~~
1\leqq r\leqq M\,.
\end{equation}
The linear change of variables has unit Jacobian, so the integration measure is unchanged.\
Since $t_r\geqq0$, the new variables obey
\begin{equation}
0\leqq u_M\leqq u_{M-1}\leqq\cdots\leqq u_1\leqq u_0\,.
\end{equation}
Thus, for fixed $u_0$, the variables $\(u_1,\ldots,u_M\)$ lie in the $M$-dimensional ordered simplex \cite{nakahara2018geometry},
\begin{equation}
S(u_0) = \LB\(u_1, \ldots, u_M\) \in \[0, u_0\]^M \mid 0 \leqq u_M \leqq \cdots \leqq u_1 \leqq u_0\RB.
\end{equation}

Thus each maximal-chain contribution can be written as
\begin{align}
\prod_{r=0}^{M} \frac{1}{D+\sigma[\pi]_r}
\,=\, \int_0^\infty \td u_0
\int_{S(u_0)} \prod_{r=1}^{M}\td u_r
\exp\!\(-D\hs u_0-\sum_{r=1}^{M}2Y_{\pi_r}u_r\).
\end{align}
Now perform the coordinate permutation $u_r\to u_{\pi_r}$.\
This change of variables has unit Jacobian.\ It removes the permutation from the exponent, but transfers it to the integration region:
\begin{shadowequation}
\begin{equation}
\label{eq:simplex-contribution}
\prod_{r=0}^{M}\frac1{D+\sigma[\pi]_r}
\,=\, \int_0^\infty \td u_0
\int_{S[\pi](u_0)} \prod_{r=1}^{M}\td u_r
\exp\!\(-D\hs u_0-\sum_{r=1}^{M}2Y_ru_r\),
\end{equation}
\end{shadowequation}
\noindent
where
\begin{equation}
\label{eq:permuted-simplex}
S[\pi](u_0) = \LB\(u_1, \ldots, u_M\) \in \[0, u_0\]^M \mid 0\leqq u_{\pi_M}\leqq\cdots\leqq u_{\pi_1}\leqq u_0\RB.
\end{equation}
For fixed $u_0>0$, the simplices $S[\pi](u_0)$ with $\pi\in\SS_M$ form the FK triangulation of the cube $[0,u_0]^M$: each permutation selects one coordinate-ordering sector, and distinct simplices meet only along common boundary faces. We denote the decomposition of the cube into these simplices, regarded as separate copies indexed by $\pi$, and the inverse gluing by
\beqs
\label{eq:FK-cut-glue}
\begin{align}
\mathcal{T}^{\rm{FK}}\big([0,u_0]^M\big)&=\coprod_{\pi\in\SS_M}S[\pi](u_0)\,,
\\
\mathcal{G}\Big(\coprod_{\pi\in\SS_M}\!S[\pi](u_0)\Big)&\equiv\Big(\coprod_{\pi\in\SS_M}\!S[\pi](u_0)\Big)\Big/\!\sim\;\cong\;[0,u_0]^M\,,
\end{align}
\eeqs
where $\sim$ identifies the common boundary faces of adjacent simplices, so that $\mathcal{G}\circ\mathcal{T}^{\rm{FK}}$ returns the cube. Since the identified faces have measure zero, gluing does not affect integrals, and summing the simplex integrals gives the cubical integral:
\begin{equation}
\label{eq:integral-triangulation}
\sum_{\pi \in \SS_M} \int_{S[\pi](u_0)}=\int_{\coprod\limits_{\pi\in\SS_M}\!S[\pi](u_0)}=\int_{\mathcal{G}\Big(\coprod\limits_{\pi\in\SS_M}\!S[\pi](u_0)\Big)}=\,\int_{[0, u_0]^M} \,.
\end{equation}
Applying this identity to the simplex representation of each maximal-chain contribution, the sum over chains is converted into a single cubical integral:
\begin{equation}
\sum_{\pi\in\SS_M}\prod_{r=0}^{M}\frac1{D+\sigma[\pi]_r} \,=
\int_0^\infty \td u_0 \int_{[0,u_0]^M}\prod_{r=1}^{M}\td u_r\, \exp\!\(-D\hs u_0-\sum_{r=1}^{M}2Y_r u_r\).
\end{equation}
This cubical integral region is an analytic realization of the Boolean cube associated with the combinatorial lattice $\PP(\YY)$: the direction $u_i$ is labeled by the doubled shift $2Y_i$ in the exponent, and the decomposition \eqref{eq:integral-triangulation} is labeled by permutations.\
Under the identification above, the one-to-one correspondence between permutations, maximal chains, and the simplicial regions spanned by these chains is
\begin{equation}
\label{eq:perm-chain-simplex-corr}
\pi ~\Longleftrightarrow~C[\pi]~\Longleftrightarrow~S[\pi]\,.
\end{equation}
This correspondence also gives a sharper form of the ``path integral'' analogy mentioned previously. Since both the total-energy divisor $D$ and the internal energies $Y_i$ have dimensions of energy, their Laplace-conjugate variables $u_0$ and $u_i$ have dimensions of time. More precisely, each $Y_i$ is conjugate to an emergent Euclidean time $u_i$, whereas $D$ is conjugate to the total duration $u_0$ that bounds any particular ordering of these auxiliary times, as encoded by \eqrefe{eq:permuted-simplex}. The exponential weight in \eqrefe{eq:simplex-contribution} therefore resembles a Euclidean Boltzmann factor.
In this sense, each simplicial contribution \eqref{eq:simplex-contribution} can be interpreted as the integration of the Boltzmann factor on a specific time ordering path $S[\pi]\cong C[\pi]$, and the total contribution is a sum over all such paths on the emergent multi-dimensional Euclidean time cube $[0,u_0]^M$ identified with the combinatorial object $\PP(\YY)$, integrating over the total duration $u_0$\footnote{%
This emergent Euclidean-time interpretation is different from the original Lorentzian bulk-time integral that generates the wavefunction. In the two-site case there are only two bulk time variables for any number of loops.}.

From this geometric perspective, the formula holds naturally since the lhs can also be represented by the same cubical integral. Indeed, for $\re(D)>0$ we have
\begin{equation}
\Delta_{2\YY}\bigg(\frac{1}{D}\bigg)
=\,\Delta_{2\YY}\!\(\int_0^{\infty}\td u_0\, e^{-D\hs u_0}\)
=\,\int_0^{\infty}\td u_0\, e^{-D\hs u_0}\prod_{i=1}^{M}\(1-e^{-2Y_iu_0}\).
\end{equation}
\begin{figure}[t]
\centering
\captionsetup[subfigure]{labelfont=normalfont}
\newcommand{\cubeUnionPanel}[4]{%
\begin{subfigure}{\textwidth}
\centering
\begin{tabular}{
m{0.35\textwidth}
m{0.18\textwidth}
m{0.35\textwidth}
}
\centering
\includegraphics[scale=0.40]{#2}
&\centering
\cubeHarpoons{\mathcal{T}^{\rm{FK}}}{\mathcal{G}}
&\centering
\includegraphics[scale=0.40]{#3}
\end{tabular}
\vspace*{-8mm}
\caption{}
\label{#4}
\end{subfigure}%
}
\cubeUnionPanel{2}{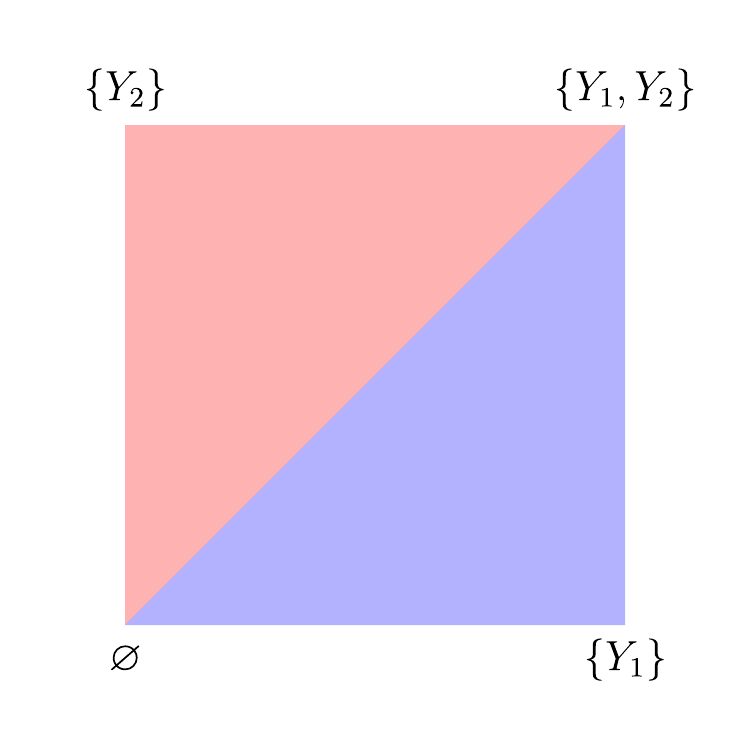}{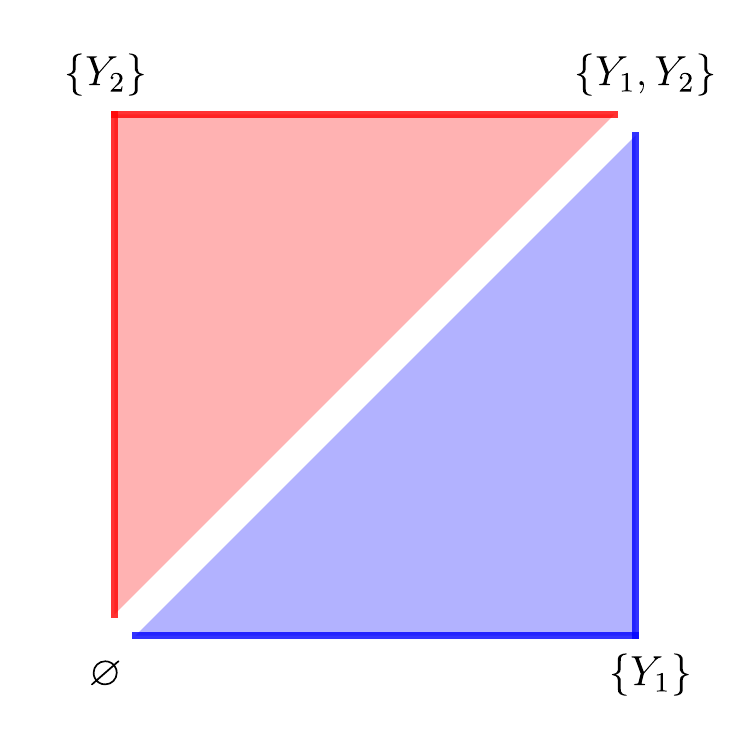}{fig:2-cube-union}
\cubeUnionPanel{3}{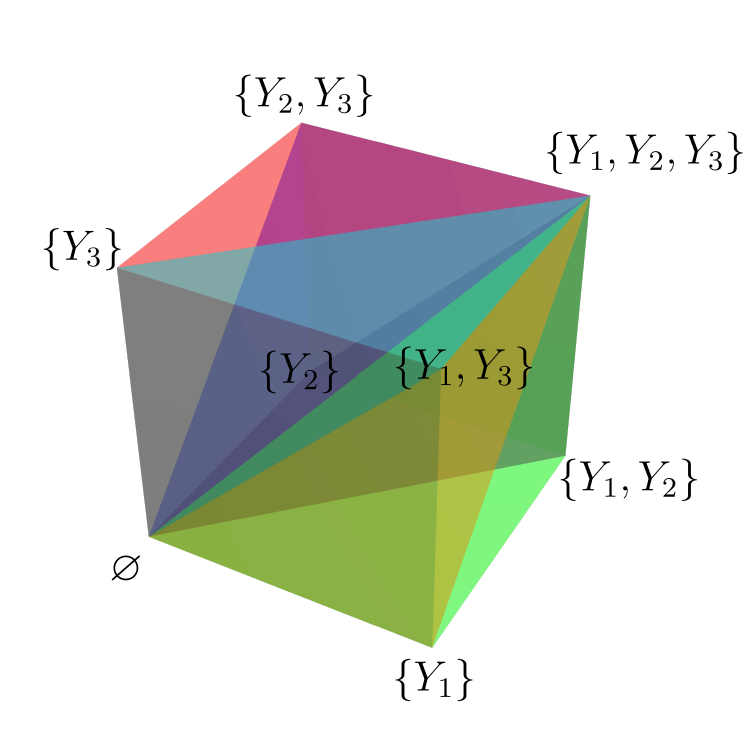}{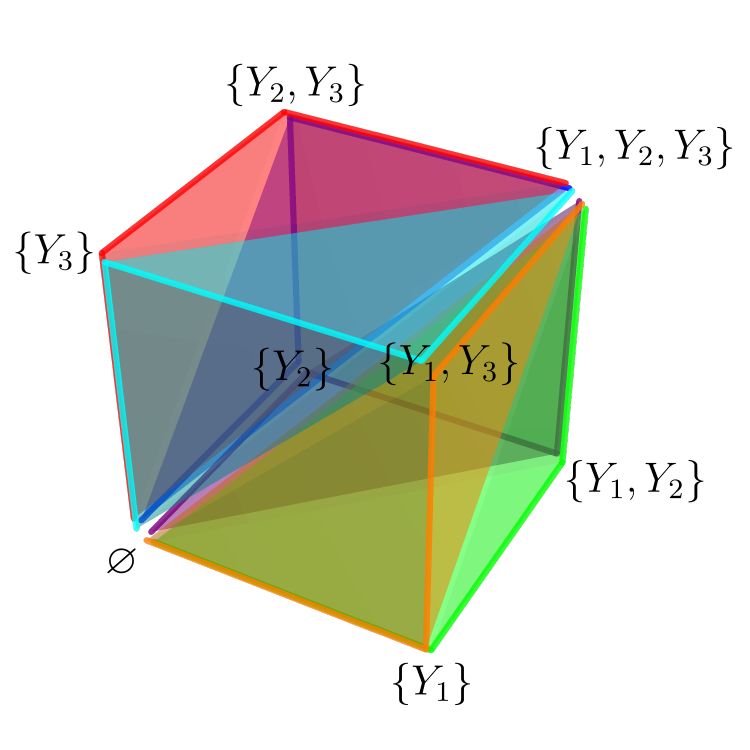}{fig:3-cube-union}
\caption{Examples of the FK triangulation of the Boolean cube $[0,u_0]^{\ell+1}$. The upper arrow $\mathcal{T}^{\rm{FK}}$ cuts the cube into the disjoint union of simplex domains $\coprod_{\pi\in\SS_{\ell+1}}S[\pi](u_0)$, and the lower arrow $\mathcal{G}$ glues them back by identifying common boundary faces, as defined in \eqrefe{eq:FK-cut-glue}. (a)\,For $\ell=1$, the square is divided into two triangles. (b)\,For $\ell=2$, the cube is divided into six tetrahedra. In both panels, the cubical integral on the left corresponds to the shifted-tree decomposition on the lhs of \eqrefe{eq:fin-diff-sum-max-chain}, while the simplex contributions on the right correspond to the maximal-chain sum on the rhs. Bold lines denote the maximal chains spanning the simplices.}
\label{fig:3}
\end{figure}
Each factor in the product can be opened into an integral over one auxiliary variable by the elementary identity
\begin{equation}
1-e^{-2Y_iu_0}\,=\,2Y_i\int_0^{\hs u_0}\td u_i\, e^{-2Y_iu_i} \,.
\end{equation}
Applying it independently for all $i$ turns the product into an integral over the full cube:
\begin{shadowequation}
\begin{equation}
\Delta_{2\YY}\!\bigg(\frac{1}{D}\bigg) =\Bigg(\prod_{i=1}^{\ell+1}2Y_i\Bigg)
\int_0^{\infty}\td u_0
\int_{[0,u_0]^{\ell+1}}\prod_{r=1}^{\ell+1}\td u_r\,
\exp\!\(-D\hs u_0-\sum_{r=1}^{\ell+1}2Y_ru_r\).
\end{equation}
\end{shadowequation}
\noindent
This is the same cubical integral obtained from the simplex decomposition of the rhs, proving the formula \eqrefe{eq:fin-diff-sum-max-chain} in the convergent region of the Laplace transformations. Finally, since the two sides of \eqrefe{eq:fin-diff-sum-max-chain} are rational functions of $D$ and $Y_i$, the formula holds generically by analytic continuation, as long as the singularities are not encountered.

This proof provides a sort of ``dual'' geometric representation of the decomposition of a loop wavefunction coefficient.\ 
In ref.~\cite{Hang:2024xas}, a loop wavefunction coefficient represents a weighted area carved out by its codimension-2 residues (after corresponding twist-variable shifts) on the plane of twist variables $(x_1,x_2)\cong\mathbb{R}^2$, and the shifted-tree decomposition is geometrized as the triangulation of this area into an alternating sum of partially overlapping triangles representing canonical forms of 2-simplices. In comparison, the present proof shows that the loop wavefunction coefficient can be viewed as a sum over simplicial pieces in the triangulation of the $(\ell+1)$-dimensional Boolean-lattice cube, while the decomposition into trees is represented by gluing the disjoint union of these simplex domains back into the whole cubical integral, identifying shared boundary faces. We illustrate the current ``dual'' picture on the Boolean lattice in figure~\ref{fig:3}, and the previous picture in the plane of twist variables can be found in ref.~\cite{Hang:2024xas}.\ 
The two representations are equivalent and complementary: the picture in the plane of twist variables remains two-dimensional, enabling visualization on paper, though it renders the validity of the underlying formula geometrically implicit; the Boolean-lattice picture quickly becomes higher-dimensional, but it keeps the triangulation manifest, as its constituent simplices have disjoint interiors.

\section{All-Loop All-Topology at Two-Site from Shifted Trees}
\label{sec:5}

\begin{figure}[t]
\centering
\includegraphics[scale=1.1]{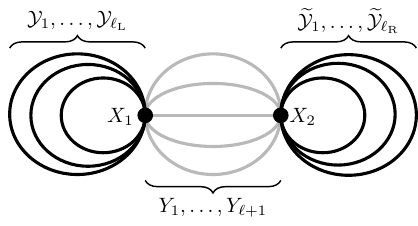}
\caption{A tadpole-decorated two-site graph with the late-time boundary and external legs omitted.\ 
The gray propagators form the central bubble-like block and carry the internal energies $Y_i$.\ 
The black tadpole loops attached to the left and right sites carry the energies $\mY_i$ and $\tmY_i$, respectively.}
\label{fig:tadpole-two-site}
\end{figure}

A generic two-site graph topology may also include tadpole loops, as illustrated in figure~\ref{fig:tadpole-two-site}.\ 
In this section, we concretize the schematic tree decomposition \eqref{eq:loop-decom} in this complete generality.\ 
We similarly collect the tadpole energies into the corresponding sets of different labels (where equal numerical values are again allowed),
\begin{equation}
\label{eq:tadpole-sets}
\bmY_\L = \big\{\mY_1,\ldots,\mY_{\ell_\L}\big\}\,,\qquad~~~~
\bmY_\R = \big\{\tmY_1,\ldots,\tmY_{\ell_\R}\big\}\,,\qquad~~~
\bmY_\L\sqcup\bmY_\R\equiv\bT\,.
\end{equation}
Here $\ell_\L$ and $\ell_\R$ denote the numbers of left and right tadpole loops, respectively; they should not be confused with the number of central bubble loops, $\ell$.

As was proposed in ref.~\cite{Hang:2024xas} and checked up to two loops, the $\psi^{(2;\ell,\ell_\L,\ell_\R)}$ for such a generic graph admits a decomposition into an alternating sum of the wavefunction coefficients $\psi^{(2;\ell)}$ of the bubble-like family shifted by the tadpole energies,
\begin{shadowequation}
\begin{equation}
\label{eq:shifted-tree-decorated}
\psi^{(2;\ell,\ell_\L,\ell_\R)}\,=\,\sum_{U_\L\subseteq\bmY_\L}\sum_{U_\R\subseteq\bmY_\R} (-1)^{|U_\L|+|U_\R|}\, \psi^{(2;\ell)}\,\Big|{\substack{\\[0mm]
X_1\to X_{1}+\sigma[U_\L]
\\[1mm]
X_2\to X_{2}+\sigma[U_\R]}}~,
\end{equation}
\end{shadowequation}
\noindent
where $\sigma[U_\L]$ and $\sigma[U_\R]$ are total shifts induced by tadpole energy sets as defined in \eqrefe{eq:sigma-subset}. Similar to \eqrefe{eq:wfc-loop}, two cleanups have been made compared with the original form in ref.~\cite{Hang:2024xas}. (i)\,Irrelevant twist-variable shifts are removed, exhibiting the formula as a property satisfied by the flat-space wavefunction coefficients. (ii)\,A different sign convention is chosen for $\psi^{(2;\ell,\ell_\L,\ell_\R)}$ in order to get back consistently to the bubble and tree cases for $\ell_\L=\ell_\R=0$. In the above formula, $\psi^{(2;\ell)}$ is understood as a function of independent energy variables $\{X_1,X_2\}\sqcup\YY$ to perform the shift, as the tadpole shifts take simpler forms in this basis. This is the same basis as in the twisted integral \eqref{eq:psi-FRW}, which makes it obvious that integration commutes with the variable shifts. With the same argument of linearity, applying the twisted integral term by term on both sides makes \eqrefe{eq:shifted-tree-decorated} valid for FRW wavefunction coefficients as well.

Combining the above tadpole-decorated decomposition \eqref{eq:shifted-tree-decorated} with the shifted-tree formula \eqref{eq:wfc-loop}, or combining the corresponding FRW version, we are able to obtain the complete shifted-tree decomposition from any generic two-site wavefunction coefficient all the way down to a sum of trees, valid for both flat space and FRW background: 
\begin{shadowequation}
\vspace*{-2mm}
\begin{equation}
\label{eq:shifted-tree-decorated-completely}
\psi^{(2;\ell,\ell_\L,\ell_\R)}_{\rm{Flat/FRW}}=\!
\sum_{U_\L\subseteq\bmY_\L}\sum_{U_\R\subseteq\bmY_\R} \sum_{ V\subseteq\YY}(-1)^{|U_\L|+|U_\R|+|V|}\hs
\psi^{(2;0)}_{\rm{Flat/FRW}}\big(L_{1,U_\L},L_{2,U_\R};D_{V,U_\L,U_\R}\big)\,.
\end{equation}
\end{shadowequation}
\noindent
Here we recast all the tadpole shifts into the divisor basis with shifted divisors defined as
\begin{align}
\label{eq:shifted-divisors-tadpole}
L_{1,U_\L} &\equiv L_1+\sigma[U_\L]\,, \qquad~~~ L_{2,U_\R}\equiv L_2+\sigma[U_\R]\,, 
\nn\\[1mm]
D_{V,U_\L,U_\R} &\equiv D+\sigma[V \sqcup U_\L\sqcup U_\R ]\,,
\end{align}
with $L_{1,2}=X_{1,2}+Y_\tot$.\ 
As can be seen in the full decomposition formula, whereas the bubble-like shifted-tree sum and its maximal-chain expansion \eqref{eq:fin-diff-sum-max-chain} are controlled by the single lattice $\PP(\YY)$, the tadpole-decorated version is organized schematically by subset combinatorics of a factorized product of Boolean lattices
\begin{equation}
\label{eq:PY-PT}
\PP(\YY) \times \PP(\bmY_\L) \times \PP(\bmY_\R) ~\cong~ \PP(\YY) \times \PP(\bT) \,,
\end{equation}
where $\PP(\YY)$ describes the central bubble-like block, whereas $\PP(\bmY_\L)$ and $\PP(\bmY_\R)$ separately encode the left and right tadpole attachments. To rigorously describe the direct product of Boolean lattices and the chain structure within, we need the proper language of basic lattice theory which is summarized in appendix~\ref{sec:B}. For now, it suffices to note that each summand in the above decomposition can be labeled by a vertex on this product lattice; a detailed exposition is given at the end of section~\ref{sec:5.1}.

By combining the complete decomposition formula \eqref{eq:shifted-tree-decorated-completely} with known FRW tree wavefunction coefficients, one can get explicit expressions for all-loop all-topology two-site FRW wavefunction coefficients directly. As a simple example in dS space with shifted divisors defined previously, we have 
\begin{align}
\label{eq:wfc-generic-loop-dS}
\psi^{(2;\ell,\ell_\L,\ell_\R)}_{\ds}=&\,\sum_{U_\L\subseteq\bmY_\L}\sum_{U_\R\subseteq\bmY_\R}  \sum_{V\subseteq\YY}(-1)^{|U_\L|+|U_\R|+|V|}
\nn\\
&\times \Bigg\{\zeta(2)+\Li\!\[\frac{(D_{V,U_\L,U_\R}\!-\! L_{1,U_\L})(D_{V,U_\L,U_\R}\!-\!L_{2,U_\R})}{L_{1,U_\L}\hs  L_{2,U_\R}}\]
\nn\\
&\quad -\Li\!\(\frac{D_{V,U_\L,U_\R}\!-\!L_{2,U_\R}}{ L_{1,U_\L}}\)-\Li\!\(\frac{D_{V,U_\L,U_\R}\!-\! L_{1,U_\L}}{L_{2,U_\R}}\)\!\Bigg\}\,,
\end{align}
which is an explicit form ready for analytic and numerical evaluation.

We will prove the formula \eqref{eq:shifted-tree-decorated} for arbitrary tadpole-decorated two-site graphs at arbitrary loop order and identify the detailed underlying combinatorial structure.\ 
We will find that it is organized explicitly by local maximal chains within the product Boolean lattice.\ 
The main observation is that, for the pure bubble-like family studied in sections~\ref{sec:3}--\ref{sec:4}, all finite-difference shifts act on the same divisor $D$ of a tree summand in the divisor basis, whereas each tadpole operator shifts one external energy and changes the corresponding local divisor and all total-energy divisors, while leaving the opposite local divisor unchanged.\ 
The strategy is to iterate \eqrefe{eq:fin-diff-sum-max-chain} on these divisors using the Leibniz rule, thereby converting the alternating sum on the rhs of \eqrefe{eq:shifted-tree-decorated} into a sum over chains. For a suitable divisor ordering, the resulting chains are then placed in one-to-one correspondence with graph tubings, proving that the formula reproduces the correct wavefunction on the lhs.\
In addition, we will also see that different divisor orderings give different but equivalent graph-tubing rules that produce the same physical wavefunction coefficient, suggesting that the decomposition on the rhs of \eqrefe{eq:shifted-tree-decorated} is more fundamental than any of its equivalent chain expansions. For reference, we collect the crucial logical steps of the proof below.

\subsection*{Summary of proof idea}
\begin{enumerate}[label=(\roman*.),wide=0pt]
\item Rewrite the alternating subset sum on the rhs of \eqrefe{eq:shifted-tree-decorated} using tadpole finite-difference operators, and insert the central-chain expression \eqrefe{eq:wfc-loop-max-chain}. This gives the rhs of \eqrefe{eq:tadpole-wrap-up}, with the divisor forms left generic.

\item Apply the Leibniz rule to each product of inverse divisors in \eqrefe{eq:tadpole-wrap-up}, distributing the tadpole operators among the factors. The resulting operator assignments are organized by ordered partitions of $\bT$, as in \eqrefe{eq:tadpole-W-leibniz}.

\item Apply the single-divisor identity \eqrefe{eq:tad-max-chain-single-divisor}, generalized from \eqrefe{eq:fin-diff-sum-max-chain}, to each assigned partition block and include the cumulative translations from preceding blocks to obtain \eqrefe{eq:operator-action-on-single-divisor}.

\item Substitute \eqrefe{eq:operator-action-on-single-divisor} into \eqrefe{eq:tadpole-W-leibniz} and sum over the central chains to obtain the generic chain expression on the rhs of \eqrefe{eq:tadpole-chain-expansion}.

\item Specialize this expression to the divisors and ordering in \eqrefe{eq:natural-divisor-ordering}, retain the nonzero ordered partitions, and collect the common factor \eqrefe{eq:Vol} to obtain \eqrefe{eq:all-loop-wfc-all}.

\item In section~\ref{sec:6}, match the weights in \eqrefe{eq:all-loop-wfc-all} term by term with the graph-tubing contributions supplied by the all-loop recursion, including their common normalization. This identifies the chain sum with the physical wavefunction coefficient and proves \eqrefe{eq:shifted-tree-decorated}.
\end{enumerate}

\subsection{Tadpole Shifts and Generic Maximal-Chain Formula}
\label{sec:5.1}

To begin with, we derive a generic version of the maximal-chain decomposition applicable for tadpole shifts.\ 
The first step is still to rewrite the alternating subset sum on the rhs of \eqrefe{eq:shifted-tree-decorated} in terms of the action of finite-difference operators. The translation operator introduced in \eqrefe{eq:def-finite-diff} now acts on the external energies $X_1$ and $X_2$ associated with the endpoints to which the tadpole is attached. Thus, we can define
\begin{equation}
\label{eq:tadpole-shift-operators}
E_{2t} f(X_1,X_2)\equiv
\begin{cases}
f(X_1+2t,X_2)\,,&t\in\bmY_\L\,,\\
f(X_1,X_2+2t)\,,&t\in\bmY_\R\,,
\end{cases}
\qquad~~~
\Delta_{2t}\equiv \id-E_{2t}\,.
\end{equation}
The translation and finite-difference operators commute, as before.\ 
Using the above definitions and \eqrefe{eq:wfc-loop-fin-diff}, \eqrefe{eq:shifted-tree-decorated} can be recast compactly as:
\begin{equation}
\label{eq:tadpole-compact}
\psi^{(2;\ell,\ell_\L,\ell_\R)}
=\Delta_{2\bmY_\L}\Delta_{2\bmY_\R}\!\(\psi^{(2;\ell)}\)
=\Delta_{2\bT}\!\(\psi^{(2;\ell)}\)
=\Delta_{2\bT}\hs\Delta_{2\YY}\!\(\psi^{(2;0)}\),
\end{equation}
which is another incarnation of a factorized product representation of the Boolean lattice \eqref{eq:PY-PT} that organizes the full mixed bubble-tadpole configuration.\
The central bubble and tadpole shifts add energy increments that do not depend on the divisor values. Their order of application therefore does not affect the result, and the corresponding finite-difference operators commute, $\[\Delta_{2\YY},\,\Delta_{2\bT}\] = 0$.

We aim to derive a chain expansion for the successive action of the finite-difference operators of tadpoles, expanding $\Delta_{2\bT}\!\(\psi^{(2;\ell)}\)$ in \eqrefe{eq:tadpole-compact} in analogy to \eqrefe{eq:wfc-loop-max-chain}.\
We take the chain sum \eqref{eq:wfc-loop-max-chain} and study the action of these operators on each summand. As each tadpole finite-difference operator shifts multiple divisors, the expansion is more involved than the pure bubble-like case in \eqrefe{eq:fin-diff-sum-max-chain} where only a single divisor is shifted.\ 
To clarify the question, the part of the maximal-chain expansion relevant to the tadpole shifts can be generically collected as follows.\ 
The central bubble wavefunction coefficient is a sum over products of inverse divisors,
\begin{equation}
\label{eq:tadpole-central-block}
\psi^{(2;\ell)}=\sum_{\pi \in\SS_{|\YY|}}W[\pi](X)\,,\qquad~~~ 
W[\pi](X) = N[\pi] \prod_{q=1}^{Q_\pi} \frac{1}{H[\pi]_q(X)} \,,
\end{equation}
where $X=(X_1,X_2)$ and $\pi$ labels the corresponding central-chain contribution $W[\pi](X)$\,\footnote{%
The dependence on $\pi$ is indicated in square brackets, as in section~\ref{sec:4}, to avoid cluttering the subscripts.}.\
For a fixed $\pi$, the symbol $N[\pi]$ is the $X$-independent prefactor, $H[\pi]_q$ denotes the $q$-th divisor, and $Q_\pi$ is the total number of divisors.\ 
Each divisor is affine-linear in $X$\,\footnote{%
$H[\pi]_q(X)=a_qX_1+b_qX_2+c_q$, where $c_q$ is independent of $X$ but may depend on the internal energies.}, while its dependence on the central internal energies $Y_i$ is left implicit. 
The displayed argument $X$ therefore indicates only the variables on which the tadpole translation operators act.

Hence, to apply the maximal-chain formula to the divisors affected by tadpole shifts, it is convenient to write a generic form valid for each $H[\pi]_q$. 
Since each divisor $H[\pi]_q$ is affine-linear in $X$, shifting $X$ changes it by an amount independent of $X$. For a shift $2t$, define this shift increment by
\begin{equation}
\label{eq:delta-increment}
\delta_{2t}H \equiv - \Delta_{2t}H = E_{2t}H-H \,.
\end{equation}
For any subset $U\subseteq\bT$, let $\sigma[U]$ denote the total shift introduced in \eqrefe{eq:sigma-subset}.\ We then define the corresponding divisor-dependent shift by
\begin{equation}
\label{eq:U-shift-sigma}
\sigma[U](H)\equiv\sum_{t\in U}\delta_{2t}H \,.
\end{equation}
When $\delta_{2t}H=2t$ for every $t\in U$, this reduces to $\sigma[U](H)=\sigma[U]$.\
Further, to keep track of chains restricted to a certain subset, we generalize the definition \eqref{eq:ctr-max-chain} to be dependent on the underlying set, defining the maximal chain of $\PP(U)$ by
\begin{equation}
\label{eq:local-max-chain}
C[\lam|U]:~\emp=C[\lam|U]_0
\,\subsetneq\, C[\lam|U]_1
\,\subsetneq\, \cdots
\,\subsetneq\, C[\lam|U]_{|U|-1}
\,\subsetneq\, C[\lam|U]_{|U|} = U \,,
\end{equation}
where $\lam=(\lam_1,\ldots,\lam_{|U|})\in\SS_{|U|}$ and the notation $|U$ records the underlying set $U$ explicitly, so that the chain specified by $\lam$ is understood as a maximal chain in $\PP(U)$. For the central chains of section~\ref{sec:3}, the underlying set is fixed by the set on which $\pi$ acts and is therefore suppressed; here it is displayed explicitly because the local chains may be associated with different subsets $U\subseteq\bT$.
The rank-$r$ element of this chain and its associated (doubled) cumulative shift are
\beqs
\begin{align}
\label{eq:local-max-chain-r}
C[\lam|U]_r &\equiv \big\{t_{\lam_1},\ldots,t_{\lam_r}\big\}\,,
\\
\label{eq:U-sigma-r}
\sigma[\lam|U]_r &\equiv \sigma[C[\lam|U]_r]
=2\sum_{j=1}^{r}t_{\lam_j}\,,
\\
\label{eq:U-sigma-r-H}
\sigma[\lam|U]_r(H) &\equiv \sigma[C[\lam|U]_r](H)
=\sum_{j=1}^{r}\delta_{2t_{\lam_j}}H \,,
\end{align}
\eeqs
where $0\leqq r\leqq |U|$.\
At $r=0$, both \eqrefe{eq:U-sigma-r} and \eqrefe{eq:U-sigma-r-H} vanish, while at $r=|U|$ they reduce to $\sigma[U]$ and $\sigma[U](H)$ in \eqrefe{eq:U-shift-sigma}, respectively.\
These terminal values are independent of the ordering $\lam$.

With the notation introduced above and the definition
\begin{equation}
H_U \equiv H+\sigma[U](H)\,,\qquad 
U\subseteq\bT \,,
\end{equation}
the action of the finite
differences indexed by $U$ on a single inverse divisor takes the same maximal-chain form as \eqrefe{eq:fin-diff-sum-max-chain}:
\begin{equation}
\label{eq:tad-max-chain-single-divisor}
\Delta_{2U}\bigg(\frac1H\bigg)
=\prod_{t\in U}\big(\delta_{2t}H\big)
\sum_{\lam\in\SS_{|U|}}\prod_{r=0}^{|U|}\frac1{H_{C[\lam|U]_r}}\,.
\end{equation}
This identity generalizes the central maximal-chain formula \eqref{eq:fin-diff-sum-max-chain} to a generic divisor $H$, and follows from either proof in section~\ref{sec:4}, since both rely only on the affine linearity of $H$. Algebraically, affine linearity gives $\Delta_{2t^\pp}(\delta_{2t}H)=0,\,\forall t^\pp,t\in U$. Geometrically, affine linearity makes each $\delta_{2t}H$ a constant that replaces $2Y_r$ in the cubical integral of section~\ref{sec:4.2}, and the rest of the proof is unchanged\,\footnote{
The integrals converge when all $H_V$, $V\subseteq U$, have positive real part; since both sides are rational, the identity holds everywhere by analytic continuation.}.
Eq.~\eqref{eq:fin-diff-sum-max-chain} is recovered by setting $H=D\,,\,U=\YY$, and $\lam=\pi$, for which $\delta_{2y}D=2y$ and $D_{C[\pi]_r}=D+\sigma[\pi]_r$. 
From this point on, we also suppress the argument $X$ when no confusion can arise.

Note that \eqrefe{eq:tad-max-chain-single-divisor} gives the contribution of a
fixed subset of shifts assigned to only one divisor. The full tadpole operator acts instead on a product of divisors,
\begin{equation}
\label{eq:tadpole-wrap-up}
\psi^{(2;\ell,\ell_\L,\ell_\R)}
=\Delta_{2\bT}\!\(\psi^{(2;\ell)}\)
=\sum_{\pi\in\SS_{|\YY|}}\!N[\pi]\,\Delta_{2\bT}\!\(~\prod_{q=1}^{Q_\pi}\frac1{H[\pi]_q}\).
\end{equation}
Its expansion requires the finite-difference Leibniz rule \eqref{eq:general-leibniz}.\
Each resulting term assigns a subset $T_{(q)}\subseteq\bT$ to every divisor $H[\pi]_q$, whose contribution is evaluated by \eqrefe{eq:tad-max-chain-single-divisor}. To elaborate, the consecutive action of multiple finite-difference operators on the product can be expanded by iterating \eqrefe{eq:general-product-leibniz}, and the resulting summation can be organized by ordered partitions of the index set of these operators.\ In detail, a straightforward application of \eqrefe{eq:general-product-leibniz} for a single $\Delta_{2t}$ gives
\begin{equation}
\label{eq:tadpole-product-leibniz}
\Delta_{2t}\Bigg(\prod_{q=1}^{Q}f_q\Bigg)\,=\,\sum_{q=1}^{Q}\Bigg(\prod_{i<q} f_i\Bigg)
(\Delta_{2t} f_q)\Bigg(\prod_{i>q}E_{2t}\hs f_i\Bigg)\,.
\end{equation}
As the Leibniz rule convention is asymmetric, we give an ordering to the functions $f_1,\ldots,f_Q$ as bookkeeping, so that the consecutive action of $|\bT|$ operators can be expanded by distributing $|\bT|$ distinct items into $Q$ ordered containers, allowing empty containers.\
Explicitly, let $\op_Q(\bT)$ denote the set of ordered partitions $\TT=(T_{(1)},\ldots,T_{(Q)})$ with $T_{(1)}\sqcup\cdots\sqcup T_{(Q)}=\bT$, allowing empty blocks.\
Then the consecutive action can be expanded as follows:
\begin{equation}
\label{eq:tadpole-iterated-leibniz}
\Delta_{2\bT}
\Bigg(\prod_{q=1}^{Q}f_q\Bigg)=\sum_{\TT\in\op_{\hsm Q}(\bT)}\prod_{q=1}^{Q}\Big[E_{2T_{(<q)}}\Delta_{2T_{(q)}}(f_q)\Big]\,,
\end{equation}
where each function $f_q$ is hit by all finite-difference operators indexed by elements in $T_{(q)}$, and by all translation operators indexed by elements to the left of $T_{(q)}$ in $\TT$, collected in the subset $T_{(<q)}\equiv T_{(1)}\sqcup\cdots\sqcup T_{(q-1)}$ for $2\leqq q\leqq Q$ and $T_{(<1)}=\emp$. The product $E_{2T_{(<q)}}=\prod_{t\in T_{(<q)}}E_{2t}$ is the cumulative effect of the translation operators in \eqrefe{eq:tadpole-product-leibniz}.

Specifically, applying \eqrefe{eq:tadpole-iterated-leibniz} to a single central-chain contribution in \eqrefe{eq:tadpole-wrap-up} gives
\begin{equation}
\label{eq:tadpole-W-leibniz}
\Delta_{2\bT}\!\(~\prod_{q=1}^{Q_\pi}\frac1{H[\pi]_q}\)=\sum_{\TT\in\op_{ Q_\pi}\!(\bT)}\hs\prod_{q=1}^{Q_\pi}\!\[E_{2T_{(<q)}}\Delta_{2T_{(q)}}\(\frac1{H[\pi]_q}\)\],
\end{equation}
which is the form that allows us to apply \eqrefe{eq:tad-max-chain-single-divisor} to the consecutive action of finite-difference operators on each divisor and obtain chain expansions. 
This expansion requires additional bookkeeping for the chains associated with the subsets $T_{(q)}\subseteq\bT$.
To evaluate each factor on the rhs of \eqrefe{eq:tadpole-W-leibniz}, we apply \eqrefe{eq:tad-max-chain-single-divisor} and incorporate the cumulative shifts associated with the blocks to the left of $T_{(q)}$. This gives
\begin{align}
\label{eq:operator-action-on-single-divisor}
&E_{2T_{(<q)}}\Delta_{2T_{(q)}}\!\(\frac1{H[\pi]_q}\)
\nn\\
&=\prod_{t\in T_{(q)}}\!\big(\delta_{2t}H[\pi]_q\big)
\sum_{\lam_{(q)}\in\SS_{|T_{(q)}|}}\prod_{r=0}^{|T_{(q)}|}
\frac1{H[\pi]_q+{\sigma[\lam_{(q)}|T_{(q)}]_r}(H[\pi]_q)+ \sigma[T_{(<q)}](H[\pi]_q)}\,,
\end{align}
where $E_{2T_{(<q)}}$ leaves the prefactor $\prod_{t\in T_{(q)}}\!\big(\delta_{2t}H[\pi]_q\big)$ unchanged by affine linearity, and $\lam_{(q)}\in\SS_{|T_{(q)}|}$ labels a local maximal chain in $\PP(T_{(q)})$.

Substituting these local-chain expansions into \eqrefe{eq:tadpole-wrap-up} and summing over the maximal chains of $\PP(\YY)$, we obtain the following chain expansion for the generic two-site wavefunction coefficient $\psi^{(2;\ell,\ell_\L,\ell_\R)}$:
\begin{shadowequation}
\vspace*{-3mm}
\beqs
\label{eq:tadpole-chain-expansion}
\begin{align}
\psi^{(2;\ell,\ell_\L,\ell_\R)}
&=\sum_{\pi \in\SS_{|\YY|}}
\Delta_{2\bT}\Big(W[\pi]\Big)
=\sum_{\pi\in\SS_{|\YY|}}\hs \sum_{\TT\in\op_{\hsm Q_\pi}\!(\bT)}\hs \sum_{\blam\in\Pi(\TT)} \W_{\pi;\TT\!,\hs\blam} \,,
\\[0.5mm]
\label{eq:tadpole-chain-weight}
\W_{\pi;\TT\!,\hs\blam} &= N[\pi]
\prod_{q=1}^{Q_\pi}\[\hs\prod_{t\in T_{(q)}}\!\big(\delta_{2t}H[\pi]_q\big)
\right.\nn\\
&\left.\hspace{.6cm}
\times\prod_{r=0}^{|T_{(q)}|}\frac1{H[\pi]_q+\sigma[\blam|T_{(q)}]_r(H[\pi]_q)+\sigma[T_{(<q)}](H[\pi]_q)}\].
\end{align}
\eeqs
\end{shadowequation}
\noindent
Here we introduce the subset-restriction notation $\blam|T_{(q)}$ in the following sense. For a fixed ordered partition $\TT=\big(T_{(1)},\ldots,T_{(Q_\pi)}\big)$, we collect the blockwise selected local permutations into the direct product
\begin{equation}
\label{eq:young-subgroup-partition}
\blam \equiv \big(\lam_{(1)},\ldots,\lam_{(Q_\pi)}\big) \in
\Pi(\TT) \equiv \SS_{|T_{(1)}|}\times\cdots\times\SS_{|T_{(Q_\pi)}|}
\subseteq \SS_{|\bT|}\,.
\end{equation}
The fixed block enumerations, concatenated in the order of $\TT$, identify $\Pi(\TT)$ with the Young subgroup of $\SS_{|\bT|}$ that preserves each block setwise. Applying each $\lam_{(q)}$ to its block and concatenating the resulting label lists gives a global ordering of $\bT$, also denoted by $\blam$. Its restriction to each block, expressed in that block's local numbering, is
\begin{equation}
\label{eq:permutation-restriction}
\blam|T_{(q)} = \lam_{(q)} \in
\SS_{|T_{(q)}|}\,.
\end{equation}
Accordingly, $\sigma[\blam|T_{(q)}]_r(H[\pi]_q)$ is the local cumulative shift in \eqrefe{eq:operator-action-on-single-divisor}, evaluated with $\lam_{(q)}=\blam|T_{(q)}$ on the underlying set $T_{(q)}$.

In the geometric picture of chains and lattices, in light of the correspondence \eqref{eq:perm-chain-simplex-corr}, each local permutation $\lam_{(q)}$ labels a maximal chain in the Boolean lattice $\PP(T_{(q)})$,
\begin{equation}
\label{eq:local-chain-main-text}
C[\lam_{(q)}]:~\emp = C[\lam_{(q)}]_0
\subsetneq \cdots \subsetneq
C[\lam_{(q)}]_{|T_{(q)}|} = T_{(q)}\,.
\end{equation}
Here and below, the annotation of the underlying set $|T_{(q)}$ is omitted for simplicity.\ 
By corollary~\ref{product-of-local-maximal-chains}, the direct product of the local maximal chains,
\begin{equation}
\label{eq:product-of-local-chains}
\LL_{\blam} \equiv C[\lam_{(1)}]
\times\cdots\times C[\lam_{(Q_\pi)}] \subseteq
\PP(T_{(1)}) \times\cdots\times \PP(T_{(Q_\pi)}) \,,
\end{equation}
is a distributive sublattice of the direct-product Boolean lattice. Under the disjoint-union isomorphism \eqref{Aeq:powerset-product-isomorphism}, it is mapped isomorphically onto the distributive sublattice
\begin{equation}
\Phi(\LL_{\blam}) \subseteq \PP(\bT) \,.
\end{equation}
In the following, we identify $\LL_{\blam}$ with its image $\Phi(\LL_{\blam})$ whenever no confusion can arise, and regard $\LL_{\blam}$ as the distributive sublattice of the total Boolean lattice $\PP(\bT)$.

\begin{figure}[t]
\centering
\captionsetup[subfigure]{labelfont=normalfont}
\begin{subfigure}{0.49\textwidth}
\centering
\hspace*{-3.5mm}
\includegraphics[width=\linewidth]{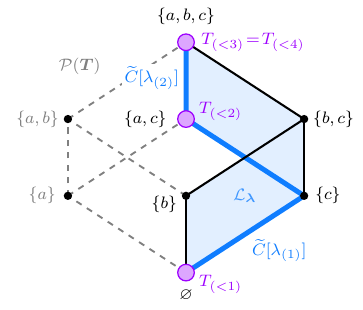}
\caption{}
\label{fig:lifted-chain-a}
\end{subfigure}%
\hfill
\begin{subfigure}{0.49\textwidth}
\centering
\hspace*{-5.5mm}
\includegraphics[width=\linewidth]{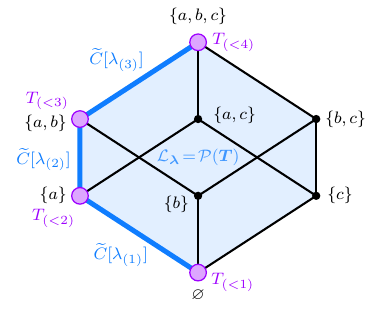}
\caption{}
\label{fig:lifted-chain-b}
\end{subfigure}
\caption{The three layers of the chain picture for $\bT=\{a,b,c\}$ and $Q_\pi=3$ divisors, so that $\TT\in\op_3(\bT)$. In both panels, the Hasse diagram of the full Boolean lattice $\PP(\bT)$ is the 1-skeleton of a $3$-cube. The shaded region is the sublattice $\LL_{\blam}$, the bold line is the distinguished maximal chain $C[\blam]$ glued from the lifted chains $\tC[\lam_{(q)}]$, and the shaded dots mark the cuts $T_{(<q)}$, through which $C[\blam]$ passes. (a)\,$\TT=(\{a,c\},\{b\},\emp)$ with local ordering $\lam_{(1)}=(2,1)$. The sublattice $\LL_{\blam}$ is the image under $\Phi$ of the product of local maximal chains $C[\lam_{(1)}]=\{\emp,\{c\},\{a,c\}\}$, $C[\lam_{(2)}]=\{\emp,\{b\}\}$, and $C[\lam_{(3)}]=\{\emp\}$, which is a rectangular $3\times2$ grid as an abstract lattice and appears folded in this layout; edges outside it are dashed. The empty last block gives $\tC[\lam_{(3)}]=\{\bT\}$, a single point, so that $T_{(<3)}=T_{(<4)}$. (b)\,$\TT=(\{a\},\{b\},\{c\})$. All blocks are singletons, so every local ordering is trivial and $\Pi(\TT)$ has a single element. The product of the two-element local chains is the whole lattice, $\LL_{\blam}=\PP(\bT)$. Each lifted chain is a single edge, so every element of $C[\blam]$ is a cut, and the partition alone fixes the chain.}
\label{fig:lifted-chain}
\end{figure}

Strictly speaking, $\LL_{\blam}$ is not itself a chain when more than one block is nonempty.\ 
Nevertheless, the blockwise concatenation prescribed by $\TT$ selects a distinguished maximal chain in $\LL_{\blam}$. 
Lift each local chain by adjoining the union of the preceding blocks. The resulting lifted chain $\tC[\lam_{(q)}]$ has rank-$r$ element defined by
\begin{equation}
\tC[\lam_{(q)}]_r \equiv T_{(<q)} \sqcup C[\lam_{(q)}]_r\,, \qquad
0\leqq r\leqq |T_{(q)}|\,,
\end{equation}
with $T_{(<q)}$ defined previously. Equivalently, it has the form
\begin{equation}
\tC[\lam_{(q)}]:~
T_{(<q)} \subsetneq \cdots \subsetneq (T_{(<q)} \sqcup C[\lam_{(q)}]_r ) \subsetneq\cdots \subsetneq T_{(<q+1)} \quad \cong \quad C[\lam_{(q)}]\,,
\end{equation}
with the chain isomorphism $\tC[\lam_{(q)}]\cong C[\lam_{(q)}]$ defined by the map
\begin{equation}
T_{(<q)}\sqcup C[\lam_{(q)}]_r \longmapsto C[\lam_{(q)}]_r \,.
\end{equation} 
Each lifted chain is a saturated chain in $\PP(\bT)$ (definition~\ref{def-saturated-chain}), since successive elements differ by exactly one label and hence form covering pairs by \eqrefe{Aeq:powerset-cover-relation}.
Under the inverse isomorphism \eqref{Aeq:powerset-product-isomorphism-inverse}, it becomes a chain in the product sublattice $ \LL_{\blam}$:
\begin{equation}
\label{eq:components-under-inverse-iso}
\Phi^{-1} \big( \tC[\lam_{(q)}]_r \big) =\big(T_{(1)},\ldots,T_{(q-1)},C[\lam_{(q)}]_r,
\emp,\ldots,\emp \big) \in \LL_{\blam} \,.
\end{equation}
The lifted chains have matching endpoints:
\begin{equation}
\tC[\lam_{(q)}]_{|T_{(q)}|} =
T_{(<q+1)} = \tC[\lam_{(q+1)}]_0 \,, \qquad 
1\leqq q<Q_\pi.
\end{equation}
Furthermore, every element of $\tC[\lam_{(q)}]$ is contained in every element of $\tC[\lam_{(p)}]$ whenever $q<p$. Their union therefore forms
the global saturated chain
\begin{equation}
\label{eq:global-chain-from-local-chains}
C[\blam] \equiv \bigcup_{q=1}^{Q_\pi}
\tC[\lam_{(q)}] \subseteq \Phi( \LL_{\blam}) \subseteq \PP(\bT) \,.
\end{equation}
This chain begins at $\emp$ and ends at $\bT$. Hence, by theorem~\ref{Maximal-chains-in-a-finite-bounded-poset}, it is a maximal chain in $\PP(\bT)$, and therefore also in the sublattice $\LL_{\blam}$, since any chain in $\LL_{\blam}$ is a chain in $\PP(\bT)$. By construction, it is precisely the chain associated with the blockwise-concatenated global ordering $\blam$. This construction is illustrated in figure~\ref{fig:lifted-chain} for $\bT=\{a,b,c\}$, where the sublattice $\LL_{\blam}$ is shaded and the global chain $C[\blam]$ is drawn in bold.

Naturally, restriction of the global maximal chain to a subset is defined by intersection,
\begin{equation}
\label{eq:global-chain-restriction}
C[\blam]\big|{T_{(q)}} \equiv
\left\{ A\cap T_{(q)} \,\middle|\, A\in C[\blam] \right\}\,.
\end{equation}
It recovers the corresponding local maximal chain,
\begin{equation}
\label{eq:global-to-local-chain-restriction}
C[\blam]\big|{T_{(q)}} = C[\lam_{(q)}] = C\big[\blam|{T_{(q)}}\big] \subseteq \PP(T_{(q)}),
\end{equation}
in direct parallel with the permutation restriction
\eqref{eq:permutation-restriction}.

The lifted chains are more than a bookkeeping device. Since $\sigma[U](H)$ is additive over disjoint unions, the denominators in \eqrefe{eq:tadpole-chain-weight} are precisely the divisors shifted by the elements of the lifted chains:
\begin{equation}
\label{eq:lifted-chain-shift}
H[\pi]_q+\sigma[\blam|T_{(q)}]_r(H[\pi]_q)+\sigma[T_{(<q)}](H[\pi]_q) =H[\pi]_q+\sigma\big[\tC[\lam_{(q)}]_r\big](H[\pi]_q)\,.
\end{equation}
That is, the cumulative shift from the preceding blocks and the local shift within $T_{(q)}$ combine into a single shift by an element of the global chain $C[\blam]$. For a fixed ordered partition $\TT$, the weight $\W_{\pi;\TT\!,\hs\blam}$ is therefore read off directly from $C[\blam]$: the partition cuts the chain into the $Q_\pi$ consecutive segments $\tC[\lam_{(q)}]$, and the elements of the $q$-th segment shift the $q$-th divisor $H[\pi]_q$. In this way, the lifted chains glue the local chains and their permutation labels into a single geometric object in $\PP(\bT)$ that governs the combinatorics of \eqrefe{eq:tadpole-chain-expansion}.

Therefore, the formula \eqref{eq:tadpole-chain-expansion} can be interpreted as a generalized ``path integral'' over central maximal chains, ordered partitions, and the associated local-chain tuples, where for a fixed ordered partition each tuple can be identified with its blockwise-concatenated global maximal chain in $\PP(\bT)$. In this language, the combinatorial sum can be recast as a single sum over paths:
\begin{equation}
\label{eq:chain-path-sum}
\sum_{\pi\in\SS_{|\YY|}}\hs \sum_{\TT\in\op_{\hsm Q_\pi}\!(\bT)}\hs \sum_{\blam\in\Pi(\TT)} \W_{\pi;\TT\!,\hs\blam}=\sum_{(\Gamma_{\YY},\TT,\Gamma_{\bT})}\W[\Gamma_{\YY};\TT,\Gamma_{\bT}]\,,
\end{equation}
where the sum on the right-hand side runs over all triples $(\Gamma_{\YY},\TT,\Gamma_{\bT})$ with $\Gamma_{\YY}\in\ch_{\rm m}(\PP(\YY))$, $\TT\in\op_{Q_\pi}(\bT)$ for the ordering $\pi$ labeling $\Gamma_{\YY}=C[\pi]$, and $\Gamma_{\bT}\in\ch_{\rm m}(\PP(\bT))$ passing through every \emph{cut} $T_{(<q)}$ (shaded dots in figure~\ref{fig:lifted-chain}), i.e., $T_{(<q)}\in\Gamma_{\bT}$ for all $1\leqq q\leqq Q_\pi+1$.\ 
Here $\ch_{\rm m}(P)$ denotes the set of maximal chains of $P$, and $\W[\Gamma_{\YY};\TT,\Gamma_{\bT}]\equiv\W_{\pi;\TT\!,\hs\blam}$ for $\Gamma_{\YY}=C[\pi]$ and $\Gamma_{\bT}=C[\blam]$.\ 
This extends the single-lattice picture introduced in section~\ref{sec:3} and sharpened in section~\ref{sec:4.2}.\ 
Although \eqrefe{eq:chain-path-sum} is the natural setting for this geometric picture, since it treats the cumulative and local shifts on the same footing, we continue to work with \eqrefe{eq:tadpole-chain-expansion}, which keeps them separate. As we now show, this separation makes the role of the divisor ordering transparent.

\subsection{Divisor Orderings and All-Loop Chain Expansion}
\label{sec:5.2}

The advantage of keeping \eqrefe{eq:tadpole-chain-expansion} generic, with all functional dependence left unspecified, is that it leaves us the freedom to choose the order in which the divisor product of each maximal-chain contribution is written. This is the only genuine freedom, since the ordering of the blocks in $\TT$ is summed over. It matters because, by the asymmetry of the Leibniz rule \eqref{eq:general-leibniz}, the divisor ordering decides which blocks enter each cumulative shift $\sigma[T_{(<q)}](\bullet)$, so that different orderings give the same total but different summands. In the geometric picture of section~\ref{sec:5.1}, this admits two equivalent readings. First, one can hold each divisor fixed together with its block and local chain, which fixes the local factor $H[\pi]_q+\sigma[\blam|T_{(q)}]_r(H[\pi]_q)$; a change of ordering then glues these local pieces in a different order, so that each is lifted onto a different base $T_{(<q)}$, and the cuts and the resulting path change. Second, one can keep the paths in \eqrefe{eq:chain-path-sum} fixed: by \eqrefe{eq:lifted-chain-shift}, each segment of a path can be labeled by the divisor it shifts, and a change of ordering only permutes these divisor labels, which changes the weights and possibly which paths contribute. We adopt the latter view, fixing the partition data $\TT$ in each summand while leaving the divisor ordering, which may depend on $\pi$, undetermined.\
Eq.~\eqref{eq:tadpole-chain-expansion} then serves as a combinatorial skeleton of paths into which the divisors can be inserted in different orders, with a fixed labeling that is convenient for the matching proof in section~\ref{sec:6}. We will show that certain orderings lead to simpler summands and, more importantly, that one particular ordering yields a representation that matches old-fashioned perturbation theory term by term, thereby proving \eqrefe{eq:shifted-tree-decorated} at all loops. We proceed by exploring three consequences of specializing \eqrefe{eq:tadpole-chain-expansion} to the shifted-tree expansion \eqref{eq:wfc-loop-max-chain}.

To start with, the cumulative shifts simplify with a suitable choice of divisor ordering. Let us first examine the divisors produced by \eqrefe{eq:wfc-loop-max-chain} and their behavior under tadpole shifts. For each contribution labeled by $\pi$, there are three classes of divisors. The first two classes are given by
\beqs
\begin{align}
&L_\L\equiv L_1=X_1+Y_\tot\,, \qquad\, \delta_{2\tmY_i}L_\L=0 \,,\qquad
\sigma[U](L_\L)=0\,,\quad
\forall\, U\subseteq\bmY_\R
\\
&L_\R^{}\equiv L_2=X_2+Y_\tot\,,\qquad \delta_{2\mY_i}L_\R=0 \,, \qquad
\sigma[U](L_\R)=0\,,\quad
\forall\, U\subseteq\bmY_\L\,,
\end{align}
\eeqs
which shows that the left local divisor does not shift under the right tadpole operators and vice versa. The third class consists of the total-energy divisors which do shift under all tadpole operators,
\begin{align}
&D_{C[\pi]_j}=D+\sigma[\pi]_j=X_1+X_2+\sigma[\pi]_j\,,
\nn\\[1mm]
&\delta_{2\mY_i}D_{C[\pi]_j}=\delta_{2\mY_i}D\neq 0\,,\qquad
\delta_{2\tmY_i}D_{C[\pi]_j}=\delta_{2\tmY_i}D\neq 0\,,
\nn\\[1mm]
&\sigma[U]\big(D_{C[\pi]_j}\big)=\sigma[U](D)\neq0\,,\quad
\forall\,U\subseteq\bT\,, ~~ U\neq \emp\,.
\end{align}
Since the ordering affects only the assignment of $\sigma[T_{(<q)}](\bullet)$, while the total-energy divisors acquire nontrivial cumulative shifts from every set to the left of $T_{(q)}$, we may place the total-energy divisors at the leftmost positions to minimize the proliferation of $\sigma[T_{(<q)}](\bullet)$ and take advantage of the trivial actions on the local divisors. This leads naturally to the following divisor ordering\,\footnote{%
Recall that the common factor $D+ \sigma[\pi]_{|\YY|}$ is canceled by the numerator in \eqrefe{eq:wfc-loop-fin-diff}. The absence of this factor is not essential to the combinatorial structure discussed here.}
\begin{equation}
\label{eq:natural-divisor-ordering}
\(H[\pi]_1,\ldots, H[\pi]_{Q_\pi}\)=\(D+\sigma[\pi]_{0}, \ldots, D+\sigma[\pi]_{|\YY|-1},L_\L,L_\R\),
\end{equation}
with length $Q_\pi=|\YY|+2$ for each $\pi$.\
At this stage, there is no particular reason to prefer one ordering of the divisors $ D+\sigma[\pi]_j$ over another. The above ordering is ``natural'' only in the sense of the comment at the end of section~\ref{sec:4.1}, namely it aligns with the direction of the chain these divisors fit into. Other orderings work just as well since the assignment of $\sigma[T_{(<q)}](\bullet)$ will not be simplified or complicated, and algebraically all orderings give the same total sum. However, as explained previously, after investigating all such orderings simultaneously, we find this particular one gives a term-by-term correspondence with the representation of old-fashioned perturbation theory. Let us fix this ordering first and defer this technical proof to section~\ref{sec:6}.

The invariance of $L_\L$ and $L_\R$ under opposite-side tadpole shifts greatly constrains the nontrivial ordered partitions that can contribute to the generic chain expansion \eqref{eq:tadpole-chain-expansion}.\ 
Since the prefactor $N[\pi]$ is always nonzero, a weight $\W_{\pi;\TT,\blam}$ in \eqrefe{eq:tadpole-chain-weight} is nonzero precisely when every corresponding multiplier is nonzero.\ 
As the denominator product in \eqrefe{eq:tadpole-chain-weight} is nonzero away from singular loci, we have the following necessary and sufficient condition:
\begin{equation}
\label{eq:nonvanishing-weight-condition}
\W_{\pi;\TT,\blam} \neq0 \,\iff\,
\prod_{t\in T_{(q)}}(\delta_{2t}H[\pi]_q) \neq 0\,,\quad 1\leqq q \leqq Q_\pi\,.
\end{equation}
Algebraically, this condition identifies the non-vanishing Leibniz assignments. For each individual weight, it can be interpreted as the condition for non-vanishing residues on the associated combined-energy poles, thereby determining which combinatorial data contribute to the physical process.\
With the natural ordering \eqref{eq:natural-divisor-ordering}, the nontrivial ordered partitions in $\hop_{Q_\pi}(\bT)$ that satisfy \eqrefe{eq:nonvanishing-weight-condition} must take the form:
\begin{equation}
\label{eq:nontrivial-partition}
\hTT=\big(T_{(0)},\ldots,T_{(|\YY|-1)},T_{(\L)},T_{(\R)}\big) \in \hop_{Q_\pi}(\bT)\,,
\end{equation}
organized by first choosing two subsets $T_{(\L)}\subseteq \bmY_\L$ and $T_{(\R)}\subseteq\bmY_\R$, and then choosing an ordered partition of the remainder in $\op_{|\YY|}(\bT\setminus(T_{(\L)}\sqcup T_{(\R)}))$. 
The assignment of blocks to divisors, i.e., the partition with each block labeled by the divisor it is assigned to rather than by its position $q$, is independent of the choice of ordering in \eqrefe{eq:natural-divisor-ordering}, because changing that ordering affects the condition \eqref{eq:nonvanishing-weight-condition} only through a relabeling of $q$.\
This is also physically natural, since the divisor ordering is an auxiliary choice that should not affect the contributing physical residues. From now on, we focus on the nontrivial ordered partitions and adopt the notation in \eqrefe{eq:nontrivial-partition} in place of that used in the generic formula.\
Henceforth, in \eqrefe{eq:tadpole-chain-expansion}, we can replace the original $q$ indexing with the new $\al$ indexing as follows:
\vs
\begin{equation}
\renewcommand{\arraystretch}{1.25}
\setlength{\arraycolsep}{8pt}
\begin{array}{|c|c|c|c|c|}
\hline
\al&q(\al)&H[\pi]_{q(\al)}&T_{(q(\al))}&\lam_{(q(\al))}\\
\hline
j&j\!+\!1&D\!+\!\sigma[\pi]_{j}&T_{(j)}&\lam_{(j)}\\
\L&|\YY|+1&L_\L&T_{(\L)}&\lam_{(\L)}\\
\R& |\YY|+2&L_\R&T_{(\R)}&\lam_{(\R)}\\
\hline
\end{array}
\vs
\end{equation}
with $0\leqq j\leqq|\YY|-1$ and $1\leqq q\leqq|\YY|+2$. Accordingly, $T_{(<j)}=T_{(0)}\sqcup\cdots\sqcup T_{(j-1)}$ and $T_{(<0)}=\emp$.

The affine-linearity in $X$ of all divisors also simplifies the form of (doubled) shifts in the weight \eqref{eq:tadpole-chain-weight}. Explicitly, all the increments, if nontrivial, will take the same form:
\begin{equation}
\label{eq:delta-H-t}
\delta_{2t}H[\pi]_q=2\hs t \,,\qquad
t\in T_{(q)}\,.
\end{equation}
Combining \eqrefe{eq:delta-H-t} with eqs.~\eqref{eq:U-shift-sigma} and \eqref{eq:U-sigma-r-H}, all the divisor-dependent increments reduce to their divisor-independent forms:
\begin{equation}
\sigma[\cdots]_r(\bullet) = \sigma[\cdots]_r\,,\qquad~~~
\sigma[\cdots](\bullet) = \sigma[\cdots]\,.
\end{equation}
With this simplification, the nontrivial partition \eqref{eq:nontrivial-partition} and the natural divisor ordering \eqref{eq:natural-divisor-ordering} give the following cumulative shifts for the last two divisors:
\beqs
\begin{align}
\sigma[T_{(<|\YY|+1)}](H[\pi]_{|\YY|+1}) &\,=\, \sum_{t\in T_{(\leqq j)}}\delta_{2t}L_\L 
\,=\,\sum_{t\in(\bT \setminus(T_{(\L)}\sqcup T_{(\R)}))\cap\bmY_\L}2t
\,=\,\sigma[\bmY_\L\setminus T_{(\L)}]\,,
\\[1mm]
\sigma[T_{(<|\YY|+2)}](H[\pi]_{|\YY|+2})&\,=\,\sum_{t\in T_{(\leqq j)}\sqcup T_{(\L)}}\delta_{2t}L_\R
\,=\,\sum_{t\in(\bT\setminus T_{(\R)})\cap\bmY_\R}2t
\,=\,\sigma[\bmY_\R\setminus T_{(\R)}]\,,
\end{align}
\eeqs
where the leftmost expressions use the original $q$ indexing and the others use the new $\al$ indexing, with $T_{(\leqq j)}\equiv T_{(< j)}\sqcup T_{(j)}$. Here $j=|\YY|-1$ in both identities. The cumulative shifts for the remaining divisors in \eqrefe{eq:natural-divisor-ordering} are simplified to
\begin{equation}
\label{eq:total-energy-full-scalar-shift}
\sigma[T_{(<j)}]\big(D+\sigma[\pi]_{j}\big)
=\sigma[T_{(<j)}]\,.
\end{equation}

After these specializations,
the generic chain expansion \eqref{eq:tadpole-chain-expansion} takes the form:
\begin{shadowequation}
\vspace*{-2mm}
\beqs
\label{eq:all-loop-wfc-all}
\begin{align}
\label{eq:all-loop-wfc}
\psi^{(2;\ell,\ell_\L,\ell_\R)}
&\,=\,\sum_{\pi\in\SS_{|\YY|}}\,
\sum_{\hTT\in\hop_{\hsm|\YY|+2}(\bT)}\,
\sum_{\blam\in\Pi(\hTT)}
\W_{\pi;\hTT\!,\hs\blam} \,,
\\[1mm]
\label{eq:all-loop-wfc-weight}
\W_{\pi;\hTT\!,\hs\blam}  &\,=\,\Vol[\YY,\bT] \prod_{j=0}^{|\YY|-1}\prod_{k=0}^{|T_{(j)}|}\frac1{D+\sigma[\pi]_{j}+\sigma[\blam|T_{(j)}]_k+ \sigma[T_{(<j)}]}
\nn\\
&\quad~~~\times\prod_{\al\in\{\L,\R\}}\prod_{r=0}^{|T_{(\al)}|}\frac1{L_{\al}+\sigma[\blam|T_{(\al)}]_r+\sigma[\bmY_\al\setminus T_{(\al)}]}\,,
\end{align}
\eeqs
\end{shadowequation}
\noindent
where the second sum in \eqrefe{eq:all-loop-wfc} is
\begin{equation}
\sum_{\hTT\in\hop_{|\YY|+2}(\bT)}=\sum_{T_{(\L)}\subseteq \bmY_\L}~
\sum_{T_{(\R)}\subseteq\bmY_\R}~
\sum_{\TT\in\op_{|\YY|}(\bT\setminus(T_{(\L)} \sqcup\hs T_{(\R)}))}\,.
\end{equation}
The chain-independent volume factor is defined by
\begin{equation}
\label{eq:Vol}
\Vol[\YY,\bT] \equiv N[\pi]\Bigg(\prod_{q=1}^{|\YY|+2}\prod_{t\in T_{(q)}}(\delta_{2t}H[\pi]_q)\Bigg)=\Bigg(\prod_{y\in\YY}2y\Bigg)\Bigg(\prod_{t\in\bT}2t\Bigg)\,,
\end{equation}
where the equality follows from \eqrefe{eq:delta-H-t} and $\bT=T_{(1)}\sqcup\cdots\sqcup T_{(|\YY|+2)}$ 
in the original $q$ indexing, with $N[\pi]=\prod_{y\in\YY}2y$ read from \eqrefe{eq:wfc-loop-max-chain}.
This factor is independent of both the divisor ordering chosen in \eqrefe{eq:natural-divisor-ordering} and the partition data in \eqrefe{eq:nontrivial-partition}. It is the geometric volume of a scaled Boolean cube realizing the product lattice \eqref{eq:PY-PT}, with side lengths given by elements in $2\YY\sqcup 2\bT$.

\section{Maximal Chains as Descending Graphic Filtrations}
\label{sec:6}

The remaining technical ingredient required to complete the proofs of \eqrefe{eq:wfc-loop} and \eqrefe{eq:shifted-tree-decorated} is to show that their maximal-chain decompositions, \eqref{eq:wfc-loop-max-chain} and \eqref{eq:all-loop-wfc-all}, reproduce the correct physical flat-space wavefunction coefficients. We close this circle by showing that these chain expansions coincide with the representations obtained from old-fashioned perturbation theory, as encoded in the graph-tubing rule of ref.~\cite{Arkani-Hamed:2017fdk}. This rule is a graphical bookkeeping of the all-loop recursion derived there from the time-integral Feynman rules, in which internal edges are cut one at a time, shifting the energies of their endpoints\,\footnote{%
Although tadpoles are not discussed explicitly in ref.~\cite{Arkani-Hamed:2017fdk}, the derivation of the recursion in section~2.3 there applies verbatim to them: for an edge with both endpoints on the same vertex $v$, the non-time-ordered part of the propagator gives $e^{2\ii y\eta_v}$, so that cutting a tadpole of energy $y$ shifts the energy of $v$ by $2y$, while the time-ordered part, constant at coincident times, is annihilated by the time-translation operator.}, until only isolated vertices remain. Each complete tubing records one way of successively cutting all internal edges. More precisely, we establish a one-to-one correspondence between individual chain weights and graph-tubing contributions under a particular labeling choice. Since the graph-tubing rule follows directly from the Feynman rules, this term-wise matching completes the all-loop proof.

It is worth emphasizing that this term-wise matching is stronger than merely showing that the maximal-chain decompositions reproduce the correct total sum, since it identifies each individual summand. The existence of this stronger matching, despite some technical organization and labeling on the graph-tubing side, implies that the tubing rules and maximal chains encode essentially the same combinatorial structure. Mathematically, this identification can be abstracted as a correspondence between maximal ascending filtrations on the Boolean-lattice side and maximal descending filtrations on the graph topology side. More broadly, the two-site wavefunction admits complementary representations in terms of subset combinatorics and certain maximal-filtration decompositions, with maximal chains and graph-tubing rules as two oppositely oriented realizations of the latter.

In the presence of tadpoles, the correspondence is more involved than in the purely bubble-like case, since the tadpole shifts must additionally be assigned among the divisors associated with each central-chain contribution. We therefore give a generic proof for \eqrefe{eq:all-loop-wfc-all} that includes tadpoles, from which the bubble-like case \eqref{eq:wfc-loop-max-chain} follows straightforwardly. The idea of the proof is that,  since \eqrefe{eq:all-loop-wfc-all} is already an explicit all-loop analytic formula indexed by $\{\pi,\big(T_{(0)},\ldots,T_{(|\YY|-1)},T_{(\L)},T_{(\R)}\big),\blam\}$, we derive a corresponding all-loop expression directly from the graph-tubing rules using the same labeling, and show that it agrees with \eqrefe{eq:all-loop-wfc-all} via order reversal. The expression is established by conveniently viewing each graph-tubing configuration as an ascending filtration organized according to the smallest 2-site tube.

We first refer to tubes encircling only the left or right site as 1-site tubes and denote them by $\A^\L$ and $\A^\R$, respectively. Their diagrams $\onetubeL$ and $\onetubeR$ correspond to the divisors $L_\L+\cdots$ and $L_\R+\cdots$ according to the graph-tubing rules of ref.~\cite{Arkani-Hamed:2017fdk}.\ 
By simple graph topology, every 1-site tube must be encircled by a tube encircling both sites. We refer to the latter as a 2-site tube, denote it by $\B$, and associate its diagram $\twositetube$ with the divisor $D+\cdots$.\
These 2-site tubes are completely ordered (nested) by the encircling relation.\ 
Therefore, every graph-tubing contribution contains a smallest 2-site tube $\B_{\min}$ that encircles all 1-site tubes and is encircled by all the other 2-site tubes if there are any.

Next, we fix a graph-tubing configuration. The subset $T_{(\al)}\subseteq\bmY_\al$ can be identified with the tadpole lines on side $\al$ enclosed by the smallest 2-site tube $\B_{\min}$. Thus, $\B_{\min}$ encloses the tadpole line set $T_{(\L)}\sqcup\hs T_{(\R)}$, whereas the lines in $\bmY_\al\setminus T_{(\al)}$ lie outside $\B_{\min}$ and cannot be enclosed by any 1-site tube. Equivalently, the set $T_{(\al)}$ is the envelope of the tadpole edge sets enclosed in 1-site tubes on side $\al$.\
According to the graph-tubing rules, the 1-site tubes on each side must be nested as a graphical (ascending) maximal filtration. Their nesting ordering can be defined by the inclusion relation of their enclosed tadpole energy sets, therefore by a maximal chain $C[\lam_{(\al)}|T_{(\al)}]$ in $\PP(T_{(\al)})$:
\begin{equation}
\label{eq:1-site-tube-ordering}
\A^{\al}_{\emp}\subsetneq\cdots\subsetneq\A^{\al}_{C[\lam_{(\al)}|T_{(\al)}]_r}\subsetneq\cdots\subsetneq\A^{\al}_{T_{(\al)}} \subsetneq \B_{\min} \,, \qquad 
\al\in\{\L,\R\}\,,
\end{equation}
where the subscripts of $\A^\al_U$ denote the 1-site tube enclosing the tadpole energy set $U$, and $C[\lam_{(\al)}|T_{(\al )}]_r$ is defined in \eqrefe{eq:local-max-chain-r}.\ 
For each tube, the graph-tubing rules assign a divisor according to internal edges intersecting the tube, with the central-line contributions already included in $L_\al$ and the additional shifts supplied by tadpole lines on the same side not enclosed by the tube, each counted twice.\
For the 1-site tube $\A^\al_{C[\lam_{(\al)}|T_{(\al)}]_r}$, the tadpole lines on side $\al$ not enclosed by this tube are those in the envelope $T_{(\al)}$ but not in this tube and those outside of $T_{(\al)}$, altogether forming the disjoint union $(T_{(\al)}\setminus C[\lam_{(\al)}|T_{(\al )}]_r)\sqcup(\bmY_\al\setminus T_{(\al)})$. The graph-tubing rule therefore associates this tube with the inverse divisor shifted by intersecting tadpole lines:
\begin{align}
\label{eq:1-site-tube-corr}
\A^{\al}_{C[\lam_{(\al)}]_r} ~\longleftrightarrow~~&
\frac1{L_\al+\sigma[T_{(\al)}\setminus C[\lam_{(\al)}|T_{(\al )}]_r]+\sigma[\bmY_\al\setminus T_{(\al)}]}
\nn\\[1mm]
&~=\frac1{L_\al+\sigma[\lam_{(\al)}^c|T_{(\al)}]_{|T_{(\al)}|-r}+\sigma[\bmY_\al\setminus T_{(\al)}]}\,,
\end{align}
where on the first line we use the additive property \eqref{eq:additive-total-shift} of total shift on the disjoint union $(T_{(\al)}\setminus C[\lam_{(\al)}|T_{(\al)}]_r)\sqcup(\bmY_\al\setminus T_{(\al)})$, and on the second line we recast the $T_{(\al)}\setminus C[\lam_{(\al)}|T_{(\al)}]_r$ part into the equivalent shift induced by the complement chain as follows. Given a chain $C[\lam|U]$ with underlying set $U$, as defined in \eqrefe{eq:local-max-chain}, define its complement chain by its rank-$r$ element as
\begin{equation}
\label{eq:complement-chain-definition}
C[\lam^c|U]_r \equiv U\setminus C[\lam|U]_{|U|-r}\,, \qquad
0\leqq r\leqq |U| \,,
\end{equation}
where the reverse ordering is defined by $\lam^c\equiv(\lam_{|U|},\ldots,\lam_1)$.
Equivalently, a complement chain is a chain indexed by the reverse ordering in the following sense. Given a chain $C[\lam|U]$, consider the descending chain (i.e., descending filtration)
obtained by taking the complement in $U$ of each of its elements:
\begin{equation}
\label{eq:local-max-co-chain}
(U\setminus\emp) \supsetneq
(U\setminus C[\lam|U]_1) \supsetneq
\cdots \supsetneq
(U\setminus C[\lam|U]_{|U|-1}) \supsetneq
(U\setminus U)\,.
\end{equation}
Reading this descending chain in reverse order gives an ascending chain under $\subsetneq$, indexed by the reverse ordering $\lam^c$, with
\begin{equation}
C[\lam^c|U]_0=\emp, \qquad C[\lam^c|U]_r =
\{t_{\lam_{|U|}},t_{\lam_{|U|-1}},\ldots, t_{\lam_{|U|-r+1}}\}\,, \qquad
1\leqq r\leqq |U|\,.
\end{equation}
More mathematical exposition of complement chains can be found in corollary~\ref{coro-complementation-of-chains}.
Taking all such 1-site tubes on both sides into account gives the graph-tubing contribution
\begin{equation}
\prod_{\al\in\{\L,\R\}}\prod_{r=0}^{|T_{(\al)}|}\frac1{L_\al+\sigma[\lam_{(\al)}^c|T_{(\al)}]_{|T_{(\al)}|-r}+\sigma[\bmY_\al\setminus T_{(\al)}]}\,.
\end{equation}
Since reversal of the ordering is an involution,
\begin{equation}
(\lam^c)^c=\lam\,,
\end{equation}
the map $\lam\mapsto\lam^c$ defines a bijection between the chains indexed by $\lam$ and those indexed by $\lam^c$. Therefore, after making the change of index $r\mapsto |T_{(\al)}|-r$ in the product $\prod_{r=0}^{|T_{(\al)}|}$ and recovering local permutations via \eqrefe{eq:permutation-restriction} in \eqrefe{eq:all-loop-wfc-all}, the bijection $\lam\leftrightarrow\lam^c$ yields the desired one-to-one correspondence between the $1$-site tubes and the corresponding $L_\al+\cdots$ divisors in the weight.

We now match the nested 2-site tubes in this graph-tubing configuration with the $D+\cdots$ divisors in \eqrefe{eq:all-loop-wfc-weight}. Label a 2-site tube by $\B_{V,U}$, where $V\subseteq\YY$ and $U\subseteq\bT$ are the sets of central internal lines and attached tadpole lines enclosed by the tube.\ 
Since any two 2-site tubes contain both sites, compatibility of tubing rules requires them to be nested, with their ordering determined by the inclusion relation of both $V$ and $U$: $\B_{V,U}\subsetneq\B_{V^\pp,U^\pp} \iff (V\sqcup U)\subsetneq (V^\pp\sqcup U^\pp)$. The entire 2-site graphical maximal filtration is
\begin{equation}
\label{eq:2-site-tube-ordering}
\B_{\min}=\B_{\{Y_k\},T_{(\L)}\sqcup\hs T_{(\R)}}
\subsetneq\cdots\subsetneq\B_{V,U}\subsetneq\cdots\subsetneq\B_{\YY,\bT}=\B_{\max}\,.
\end{equation}
In a maximally nested tubing configuration, $\B_{\min}$ encloses exactly one central line $\{Y_k\}$ depending on the choice, and $\B_{\max}$ encloses all lines. As in section~\ref{sec:3}, after adjoining an auxiliary bottom element that does not represent a physical tube, the nesting ordering in \eqrefe{eq:2-site-tube-ordering} can be represented by the following isomorphic maximal chains:
\begin{equation}
\emp\sqcup\big(T_{(\L)}\sqcup T_{(\R)}\big)\subsetneq\cdots \subsetneq\YY\sqcup\bT
\quad \cong \quad
\emp\sqcup\emp\subsetneq\cdots \subsetneq\YY\sqcup\big(\bT\setminus(T_{(\L)}\sqcup T_{(\R)})\big)\,.
\end{equation}
We focus on the equivalent chain on the right obtained by removing the fixed set $T_{(\L)}\sqcup T_{(\R)}$ from every term on the left. The first step of the chain adds the chosen central line $\{Y_k\}$ and produces $\B_{\min}$. The remaining chain elements can be determined by the following choices on how to add the rest of internal lines. (i) A choice of $\pi\in\SS_{|\YY|}$ representing the ordering of adding central lines into the chain. (ii) A choice of ordered partition $(T_{(0)},\ldots,T_{(|\YY|-1)})$, with empty blocks allowed, that divides the tadpole lines in $\bT\setminus(T_{(\L)}\sqcup T_{(\R)})$ into $|\YY|$ consecutive blocks to be interleaved with the steps of $\pi$. (iii) Within each block, a choice of $\lam_{(j)}\in\SS_{|T_{(j)}|}$ representing the ordering of adding tadpole lines into the chain. These choices echo the combinatorial data labeling the weight of \eqrefe{eq:all-loop-wfc-all}.

There is relabeling freedom in the choice (ii). Namely, for a fixed underlying unordered partition, each ordering gives an equivalent interleaving choice. In order to term-wise match the weight from maximal-chain decomposition, we interleave these blocks in a reversed order. Namely, we add tadpole lines of block $T_{(|\YY|-j-1)}$ into the chain after the $(j+1)$-th step of $\pi$. Therefore, we may locate any 2-site tube by a group of indices $(j,k)$ with $0\leqq j\leqq|\YY|-1$ and $0\leqq k\leqq|T_{(|\YY|-j-1)}|$ representing the $(j+1)$-th step of $\pi$ and the $k$-th step of $\lam_{(|\YY|-j-1)}$. For example, $\B_{\min}$ is located by $(0,0)$ as it is after the first step of $\pi$ and at the $0$-th step of $\lam_{(|\YY|-1)}$. In general, a tube of location $(j,k)$ encloses the first $j+1$ central lines of the chain $\pi$, all tadpole lines in $T_{(|\YY|-1)},\ldots,T_{(|\YY|-j)}$, and the first $k$ lines of the chain $\lam_{(|\YY|-j-1)}$. Therefore, we have the following correspondence between location $(j,k)$ and enclosed central-line and tadpole sets $(V,U)$ of a 2-site tube $\B_{V,U}$:
\begin{equation}
\label{eq:V-U}
V=C[\pi]_{j+1}\,,\qquad
U=(T_{(\R)}\sqcup T_{(\L)})\sqcup T_{(|\YY|-1)}\sqcup\cdots\sqcup T_{(|\YY|-j)}
\sqcup C[\lam_{(|\YY|-j-1)}]_k\,,
\end{equation}
where the union $T_{(|\YY|-1)}\sqcup\cdots\sqcup T_{(|\YY|-j)}$ is empty for $j=0$ and the underlying set is omitted in the notation of the local tadpole chain. The form of \eqrefe{eq:V-U} echoes \eqrefe{eq:nontrivial-partition} under order reversal\,\footnote{The relative ordering between $T_{(\L)}$ and $T_{(\R)}$ is not important by definition, and is so displayed only to extend the order reversal to the whole partition. What matters is that we fix $T_{(\L)}$ and $T_{(\R)}$ before interleaving the blocks with $\pi$.}, and we note that the choice of (ii) with reversed interleaving order is equivalent to an ordered partition $(T_{(|\YY|-1)},\ldots,T_{(0)})$ with interleaving order as displayed. Similar to 1-site tubes, the graph-tubing rules assign a divisor according to internal edges intersecting the tube, and since 2-site tubes encircle both sites, these edges are just those not enclosed by the tube. Let $\pi^c$ and $\lam^c_{(|\YY|-j-1)}$ denote the reverse ordering of the corresponding chains. The lines outside this tube are
\begin{equation}
\YY\setminus V=C[\pi^c]_{|\YY|-1-j}\,,\qquad
\bT\setminus U=C[\lam^c_{(|\YY|-j-1)}]_{|T_{(|\YY|-j-1)}|-k}\sqcup T_{(<|\YY|-1-j)}\,.
\end{equation}
The graph-tubing rule therefore associates $\B_{V,U}$ with
\begin{align}
\B_{V,U}~\longleftrightarrow&~
\frac{1}{D+\sigma[\YY\setminus V]+\sigma[\bT\setminus U]}
\nn\\
&~=\frac{1}{D+\sigma[\pi^c]_{|\YY|-1-j}+\sigma[\lam^c_{(|\YY|-j-1)}]_{|T_{(|\YY|-j-1)}|-k}+\sigma[T_{(<|\YY|-1-j)}]}\,.
\end{align}
Taking all such 2-site tubings in this configuration into account gives
\begin{equation}
\prod_{j=0}^{|\YY|-1}\prod_{k=0}^{|T_{(|\YY|-j-1)}|}\frac{1}{D+\sigma[\pi^c]_{|\YY|-1-j}+\sigma[\lam^c_{(|\YY|-j-1)}]_{|T_{(|\YY|-j-1)}|-k}+\sigma[T_{(<|\YY|-1-j)}]}\,.
\end{equation}
Relabeling in the product index via $k\to |T_{(|\YY|-j-1)}|-k$ and $j\to |\YY|-1-j$ gives 
\begin{equation}
\prod_{j=0}^{|\YY|-1}\prod_{ k=0}^{|T_{(j)}|}\frac{1}{D+\sigma[\pi^c]_{j}+\sigma[\lam^c_{(j)}]_{k}+\sigma[T_{(<j)}]}\,,
\end{equation}
which up to bijections between chains and their complements yields the desired one-to-one correspondence between the $2$-site tubes and the corresponding $D+\cdots$ divisors in the weight.

Collectively, the previous procedure amounts to labeling the graph-tubing configuration via the choice $\{\pi,\big(T_{(\R)},T_{(\L)}, T_{(|\YY|-1)},\ldots,T_{(0)}\big),\blam\}$, which corresponds to the maximal-chain weight indexed by $\{\pi^c,\big(T_{(0)},\ldots,T_{(|\YY|-1)},T_{(\L)},T_{(\R)}\big),\blam^c\}$, with $\blam^c$ understood as global order reversal to match the reversed partition labels. Within each configuration, every tube one-to-one matches a factor inside the corresponding weight. By summing over the combinatorial index, one equivalently sums over all possible graph-tubing configurations.

Finally, having matched all divisor factors, it remains to check that the prefactor $\Vol[\YY,\bT]$ obtained from the chain expansion agrees with the graph-tubing normalization. In the graph-tubing approach, this prefactor is fixed by requiring unit residues, namely $\pm1$, on all codimension-2 boundaries in the $(x_1,x_2)$-plane after the shifts by the twist variables in \eqrefe{eq:psi-FRW}, for generic energy parameters such that the relevant poles do not coincide; examples through two loops can be found in ref.~\cite{Hang:2024xas}. The tree coefficient \eqrefe{eq:wfc-chain} has unit residues at the codimension-2 boundaries $\{L_1,L_2\}$, $\{L_1,D\}$, and $\{L_2,D\}$. On the rhs of \eqrefe{eq:wfc-loop}, the residues at $\{L_1,D_V\}$ and $\{L_2,D_V\}$ are $\pm1$ for every $V\subsetneq\YY$; at the common intersection $\{L_1,L_2\}$, summing the subset signs gives the residue $(-1)^{|\YY|}$, with residue coordinates ordered as $(L_1,L_2)$. The tadpole shifts in \eqrefe{eq:shifted-tree-decorated} act on $\psi^{(2;\ell)}$ as a whole and translate these boundaries, with residue signs determined by the tadpole subset signs. Thus $\psi^{(2;\ell,\ell_\L,\ell_\R)}$ has unit residues on the corresponding shifted boundaries. Since the divisor structure has already been matched, the tubing sum without its common prefactor has residue $(-1)^{|\YY|}/\Vol[\YY,\bT]$ at $\{L_1,L_2\}$. Matching this residue to $(-1)^{|\YY|}$ in the positive tree normalization of \eqrefe{eq:wfc-chain} fixes the prefactor to be $\Vol[\YY,\bT]$, while the remaining residues follow from the tree decomposition, or equivalently from the chain expansion.

With the above arguments, we complete the proof of the general validity of \eqrefe{eq:wfc-loop} and \eqrefe{eq:shifted-tree-decorated} at arbitrary loop order. In other words, the shifted-tree decomposition \eqref{eq:shifted-tree-decorated-completely} produces the exact physical wavefunction coefficient at arbitrary loop order and topology.

\paragraph{Comments on the matching proof} 
Since the matching proof above is technically involved and depends explicitly on several bookkeeping conventions adopted throughout this work, we summarize these choices here and isolate the conceptual essence from the technical details.
\begin{enumerate}[label=(\roman*.),wide=0pt]
\item We choose the filtration convention to run from left to right in the ascending direction. This choice is purely conventional. 

\item Given this choice, we adopt the Leibniz-rule convention \eqrefe{eq:general-leibniz}, which naturally aligns with the direction of filtration. As commented at the end of section~\ref{sec:4.1}, however, this alignment is just a matter of how commuting products are written, and is therefore also conventional. Importantly, this convention is what makes the tadpole cumulative shift appear as $\sigma[T_{(<q)}]$, rather than $\sigma[T_{(>q)}]$, in \eqrefe{eq:tadpole-chain-expansion} and its subsequent specifications. 

\item Once the preceding conventions are fixed, the choice of divisor ordering \eqref{eq:natural-divisor-ordering} genuinely matters. Other orderings can produce representations with different individual weights but the same total sum by construction, and correspond to different tubing rules on the graph-tubing side. The reason is that changing the divisor ordering reshuffles the assignment of tadpole cumulative shifts along the chain, thereby modifying the divisor factors appearing in each weight. For example, placing $\{L_{\L},L_{\R}\}$ at the leftmost positions adds a term $\sigma[T_{(\L)}\sqcup T_{(\R)}]$ to every $D+\cdots$ divisor and removes the term $\sigma[\bmY_{\al}\setminus T_{(\al)}]$ from every $L_\al+\cdots$ divisor. One may likewise permute the ordering of the divisors $D+\sigma[\pi]_{j}$. We emphasize the essential feature of the choice \eqref{eq:natural-divisor-ordering}.
\begin{center}
\emph{Eq.~\eqref{eq:natural-divisor-ordering} assigns a universal tadpole shift $\sigma[\bmY_{\al}\setminus T_{(\al)}]$ to the $L_\al+\cdots$ divisors at every step of the tadpole chains, while assigning progressively larger tadpole cumulative shifts via $\sigma[T_{(<q)}]$ to the $D+\cdots$ divisors along the ascending direction of the central chain.}
\end{center}
\vspace*{-1mm}

\item Graph-tubing configurations are naturally organized by the inclusion relation induced by the point-set topology of the tubes, whereas maximal chains are naturally ordered by inclusion of finite sets of internal-edge labels.\ From this viewpoint, as made explicit by the chain$\iff$complement-chain correspondence in the preceding proof, the graph-tubing filtration defined by encircling relation in \eqrefe{eq:1-site-tube-ordering} or \eqrefe{eq:2-site-tube-ordering} actually runs in the opposite direction to the cumulative-shift filtration defined by finite-set inclusion. An ascending filtration on one side is therefore mapped to a descending filtration on the other. This reversal is a consequence of the tubing rule assigning divisors according to internal edges intersecting the tube rather than those enclosed by it. 

\item The reversal of interleaving order in \eqrefe{eq:V-U} is only for an explicit term-wise matching of the combinatorial bookkeeping on the chain-weight side, as the graph-tubing output is invariant under relabeling of the ordered partition. The essential feature of the existing graph-tubing rule of ref.~\cite{Arkani-Hamed:2017fdk}, independent of labeling, is the one emphasized in (iii), which is precisely why the special choice \eqref{eq:natural-divisor-ordering} is required for the preceding proof.
\end{enumerate}
Having identified the essence of the matching proof as choosing the special divisor ordering that produces the feature emphasized above, it becomes clear that the tubing rule of ref.~\cite{Arkani-Hamed:2017fdk} is one of the equivalent representations of the physical wavefunction obtained from the $(|\YY|+2)!$ possible divisor orderings, some of which give identical individual weights. By contrast, the non-vanishing condition \eqref{eq:nonvanishing-weight-condition} is independent of this ordering up to simultaneous relabeling of the blocks and divisors, and determines which combinatorial data give nonzero weights. In this sense, its direct consequence---the assignment of blocks to divisors underlying \eqrefe{eq:nontrivial-partition}---may be regarded as the ``gauge-invariant'' organization of the physical wavefunction. In the next section, we present explicit examples that make manifest the term-wise correspondence between the maximal-chain weights and graph-tubing configurations, while also demonstrating how alternative divisor orderings give rise to distinct tubing rules and individual terms that nevertheless sum to the same physical wavefunction.

\section{Explicit Examples of the General Boolean Lattice Construction}
\label{sec:7}

We now illustrate the all-loop maximal-chain decomposition \eqref{eq:all-loop-wfc-all} via nontrivial two-site graphs. The weight sums are organized by the combinatorial data isolated in section~\ref{sec:5}: a central ordering $\pi$, a non-vanishing ordered partition $\hTT$, and a collection of local orderings $\blam$ associated with $\hTT$. We first revisit the one-loop bubble and two-loop sunset of sections~\ref{sec:3.1}--\ref{sec:3.2}. In the absence of tadpoles, the latter two types of data are trivial, so these examples reduce the general formula of section~\ref{sec:5} to the earlier maximal-chain expressions. We then turn to the simplest example in which both the ordered-partition and local-chain data are nontrivial: a central tree with one tadpole attached to each site. We compute the corresponding weights explicitly and verify that their sum agrees with the alternating shifted-tree sum.

These examples also provide an explicit illustration of the technical matching proof of section~\ref{sec:6}. For each algebraic weight, we identify its graphical counterpart through the underlying chain$\iff$complement-chain correspondence, thereby providing a nontrivial consistency check of the shifted-tree and maximal-chain formulas. In the final example, we further show that changing the divisor ordering generates a different graph-tubing rule, illustrating that the standard graph-tubing rule may be viewed as only one among equivalent ``gauge-dependent'' representations of the same physical wavefunction.

\subsection{Central Graphs without Tadpoles}
\label{sec:7.1}

For the bubble-like graphs, the tadpole set is empty, giving a unique ordered partition whose blocks are all empty with all local permutations trivial.\ 
Therefore, in \eqrefe{eq:all-loop-wfc-all}, we have $\bT=\emp\,,\,\Vol[\YY,\emp]=\prod_{y\in\YY}2y$, and all tadpole combinatorial data are discarded. 
The only remaining label is therefore the central ordering $\pi\in\SS_{|\YY|}$, and \eqrefe{eq:all-loop-wfc-weight} reduces to
\begin{equation}
\label{eq:central-weight-example}
\psi^{(2;\ell)}=\sum_{\pi\in\SS_{|\YY|}}\W_{\pi}\,,\qquad~~~
\W_{\pi}=\frac1{L_1L_2}\bigg(\prod_{y\in\YY}2y\bigg)\prod_{r=0}^{|\YY|-1}
\frac{1}{D+\sigma[\pi]_r}\,.
\end{equation}
This is exactly the central maximal-chain formula \eqref{eq:wfc-loop-max-chain}.  Thus, we continue the examples of sections~\ref{sec:3.1}--\ref{sec:3.2} by identifying their individual maximal-chain weights and corresponding graphical filtrations as the tadpole-free sector of the generic formula \eqref{eq:all-loop-wfc-all} and matching proof in section~\ref{sec:6}.

For the one-loop bubble, retain the divisor labels $D_0,D_1,D_2$ introduced in section~\ref{sec:3.1}. The two possible central orderings $\pi_1=(1,2)$ and $\pi_2=(2,1)$ give
\begin{equation}
\label{eq:bubble-central-weights}
\W_{(1,2)}=\frac{4Y_1Y_2}{L_1L_2D_0D_1}\,,
\qquad~~~
\W_{(2,1)}=\frac{4Y_1Y_2}{L_1L_2D_0D_2}\,.
\end{equation}
Their sum is precisely \eqrefe{eq:max-chain-l1}.\
Under the complementary-chain map described in section~\ref{sec:6}, these two weights correspond to graph-tubing configurations labeled by $\pi_1^c=(2,1)$ and $\pi_2^c=(1,2)$ respectively. Following the edge labels in figure~\ref{fig:1loop-bubble}, this correspondence can be displayed as
\beqs
\label{eq:bub-tubes}
\begin{align}
\W_{(1,2)} \,\iff~
\letterfig{}{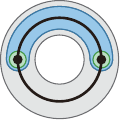}{0.75}{0cm}
\hspace*{1.5cm} &~=~
{\lgre\A_\emp^\L \, \A_\emp^\R}\,
{\lblu\B_{\{Y_2\}}}\,
{\lgray\B_{\YY}}\,,
\\
\W_{(2,1)}\,\iff~
\letterfig[180]{}{bub-tube}{0.75}{0cm} \hspace*{1.5cm}&~=~
{\lgre\A_\emp^\L \, \A_\emp^\R}\,
{\lblu\B_{\{Y_1\}}}\,
{\lgray\B_{\YY}}\,,
\end{align}
\eeqs
where the colors distinguish tube shapes only and carry no additional algebraic meaning. According to eqs.~\eqref{eq:1-site-tube-ordering}, \eqref{eq:2-site-tube-ordering} and \eqref{eq:V-U}, the two configurations arise from the following graphic filtrations
\begin{equation}
\label{eq:bubble-tube-inclusions}
\A^{\L/\R}_\emp\subsetneq\[
\begin{array}{c}
\B_{\{Y_2\}}\\
\B_{\{Y_1\}}
\end{array}\]
\subsetneq\B_{\YY}\,.
\end{equation}
We abbreviate $\B_{V,\emp}$ as $\B_V$ when no tadpoles are present; $\A^{\L/\R}_\emp$ means that the inclusion holds separately for the left and right 1-site tubes.
For the two central orderings $\pi_1^c$ and $\pi_2^c$, the middle tubes $\B_{\{Y_2\}}$ and $\B_{\{Y_1\}}$ are the smallest 2-site tubes $\B_{\min}$, respectively, while $\B_{\YY}$ is the common maximal tube $\B_{\max}$.
The associated factors for the tubes are given by
\begin{align}
\label{eq:bub-tube-factors}
\A_\emp^\L&=\frac1{L_1}\,,\quad~~
\A_\emp^\R=\frac1{L_2}\,,\quad~~
\B_{\{Y_1\}}=\frac1{D_2}\,,\quad~~
\B_{\{Y_2\}}=\frac1{D_1}\,,\quad~~
\B_{\YY}=\frac1{D_0}\,.
\end{align}
Multiplying the tube factors in each row of \eqrefe{eq:bub-tubes} by the volume factor $4Y_1Y_2$ reproduces the corresponding weight in \eqrefe{eq:bubble-central-weights}.

The same reduction applies to the two-loop sunset, with $\YY=\{Y_1,Y_2,Y_3\}$.  Using the divisor labels $D_0,\ldots,D_6$ of section~\ref{sec:3.2}, define a common factor
\begin{equation}
\label{eq:sunset-common-prefactor}
\mathcal{R}_{\rm{sun}}\equiv\frac{8Y_1Y_2Y_3}{L_1L_2D_0}\,.
\end{equation}
The six central orderings give the following weights
\begin{align}
\label{eq:sunset-central-weights}
\W_{(1,2,3)}&=\frac{\mathcal{R}_{\rm{sun}}}{D_1D_4}\,,&
\W_{(1,3,2)}&=\frac{\mathcal{R}_{\rm{sun}}}{D_1D_5}\,,&
\W_{(2,1,3)}&=\frac{\mathcal{R}_{\rm{sun}}}{D_2D_4}\,,
\nn\\
\W_{(2,3,1)}&=\frac{\mathcal{R}_{\rm{sun}}}{D_2D_6}\,,&
\W_{(3,1,2)}&=\frac{\mathcal{R}_{\rm{sun}}}{D_3D_5}\,,&
\W_{(3,2,1)}&=\frac{\mathcal{R}_{\rm{sun}}}{D_3D_6}\,.
\end{align}
Their sum recovers \eqrefe{eq:max-chain-l2}, term by term in the order of the six maximal chains \eqref{eq:max-chain-sunset}. Following the edge labels in figure~\ref{fig:2loop-sunset}, they correspond to the following graph-tubing configurations
\beqs
\begin{align}
\W_{(1,2,3)} \,\iff~
\letterfig{}{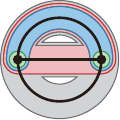}{0.75}{0cm}
\hspace*{1.5cm}&~=~
{\lgre\A_\emp^\L \, \A_\emp^\R}\,
{\lblu\B_{\{Y_3\}}}\,
{\red\B_{\{Y_2,Y_3\}}}\,
{\lgray\B_{\YY}}\,,
\\[1mm]
\W_{(1,3,2)} \,\iff~
\letterfig[0]{}{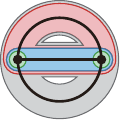}{0.75}{0cm} \hspace*{1.5cm}&~=~
{\lgre\A_\emp^\L \, \A_\emp^\R}\,
{\lblu\B_{\{Y_2\}}}\,
{\red\B_{\{Y_2,Y_3\}}}\,
{\lgray\B_{\YY}}\,,
\\[1mm]
\W_{(2,1,3)} \,\iff~
\letterfig{}{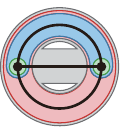}{0.75}{0cm} \hspace*{1.5cm}&~=~
{\lgre\A_\emp^\L \, \A_\emp^\R}\,
{\lblu\B_{\{Y_3\}}}\,
{\red\B_{\{Y_1,Y_3\}}}\,
{\lgray\B_{\YY}}\,, 
\\[1mm]
\W_{(2,3,1)} \,\iff~
\letterfig[180]{}{sun-a}{0.75}{0cm} \hspace*{1.5cm}&~=
{\lgre\A_\emp^\L \, \A_\emp^\R}\,
{\lblu\B_{\{Y_1\}}}\,
{\red\B_{\{Y_1,Y_3\}}}\,
{\lgray\B_{\YY}}\,,
\\[1mm]
\W_{(3,1,2)} \,\iff~
\letterfig[180]{}{sun-c}{0.75}{0cm} \hspace*{1.5cm}&~=~
{\lgre\A_\emp^\L \, \A_\emp^\R}\,
{\lblu\B_{\{Y_2\}}}\,
{\red\B_{\{Y_1,Y_2\}}}\,
{\lgray\B_{\YY}}\,,
\\[1mm]
\W_{(3,2,1)} \,\iff~
\letterfig[180]{}{sun-b}{0.75}{0cm} \hspace*{1.5cm} &~=~
{\lgre\A_\emp^\L \, \A_\emp^\R}\,
{\lblu\B_{\{Y_1\}}}\,
{\red\B_{\{Y_1,Y_2\}}}\,
{\lgray\B_{\YY}}\,.
\end{align}
\eeqs
These configurations come from the graphic filtrations defined by the six complement chains
\begin{equation}
\label{eq:sunset-tube-inclusions}
\A^{\L/\R}_\emp\subsetneq\[\begin{array}{c}
\B_{\{Y_3\}}\,\subsetneq\,\B_{\{Y_2,Y_3\}} \\
\B_{\{Y_2\}}\,\subsetneq\,\B_{\{Y_2,Y_3\}} \\
\B_{\{Y_3\}}\,\subsetneq\,\B_{\{Y_1,Y_3\}} \\
\B_{\{Y_1\}}\,\subsetneq\,\B_{\{Y_1,Y_3\}} \\
\B_{\{Y_2\}}\,\subsetneq\,\B_{\{Y_1,Y_2\}} \\
\B_{\{Y_1\}}\,\subsetneq\,\B_{\{Y_1,Y_2\}}
\end{array}\]\subsetneq\B_{\YY}\,,
\end{equation}
where the first 2-site tube $\B_{\{Y_j\}}$ in each row is the corresponding $\B_{\min}$, while $\B_{\YY}$ is the common outermost tube $\B_{\max}$. The associated factors for the tubes are given by
\begin{align}
\label{eq:sunset-tube-factors}
\A_\emp^\L&=\frac1{L_1}\,,\qquad~
\A_\emp^\R=\frac1{L_2}\,,\qquad~
\B_{\{Y_1\}}=\frac1{D_6}\,,\qquad~
\B_{\{Y_2\}}=\frac1{D_5}\,,\qquad~
\B_{\{Y_3\}}=\frac1{D_4}\,,
\nn\\
\B_{\{Y_1,Y_2\}}&=\frac1{D_3}\,,\qquad~
\B_{\{Y_1,Y_3\}}=\frac1{D_2}\,,\qquad~
\B_{\{Y_2,Y_3\}}=\frac1{D_1}\,,\qquad~
\B_{\YY}=\frac1{D_0}\,.
\end{align}
Multiplying the tube factors in each row by the volume factor $8Y_1Y_2Y_3$ reproduces the corresponding weight in \eqrefe{eq:sunset-central-weights}.

\subsection{Central Tree with Two Tadpoles}
\label{sec:7.2}

\begin{figure}[t]
\centering
\includegraphics[scale=1.]{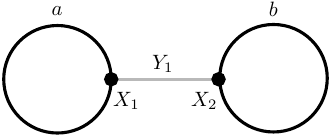}
\caption{A central tree decorated with one tadpole at each site.\ 
$Y_1$, $a$, and $b$ are the energies carried by the internal line, left tadpole and right tadpole, respectively.}
\label{fig:dumbell}
\end{figure}

We next illustrate the genuinely new data contained in \eqrefe{eq:all-loop-wfc-all}, relative to \eqrefe{eq:wfc-loop-max-chain}, through the simplest example shown in figure~\ref{fig:dumbell}: a central tree with a tadpole loop attached to each of its left and right sites. We label the central and tadpole internal energy sets by $\YY=\{Y_1\}\,,\,\bmY_\L=\{a\}\,,\,\bmY_\R=\{b\}$, and $\bT=\bmY_\L \sqcup \bmY_\R= \{a,b\}$. 
Unlike the examples in the previous subsection, the finite-difference Leibniz rule now distributes the two tadpole shifts among divisors from the central tree.

Organizing the central-tree contribution \eqref{eq:wfc-chain} via \eqrefe{eq:tadpole-central-block} gives
\begin{equation}
\label{eq:tadpole-tree-central}
\psi^{(2;0)} = W[\pi] =2Y_1\frac{1}{L_1L_2D}\,,
\end{equation}
where the $X$-dependence and the unique rank-one central ordering $\pi\in\SS_1$ will be suppressed in this section. 
The natural divisor ordering \eqref{eq:natural-divisor-ordering} becomes
\begin{equation}
\label{eq:tadpole-tree-divisor-list}
\(H_1,H_2,H_3\)=\(D,L_1,L_2\).
\end{equation}
The left-tadpole shift $\delta_{2a}$ can act nontrivially only on $L_1$ and $D$, whereas the right-tadpole shift $\delta_{2b}$ can act nontrivially only on $L_2$ and $D$:
\begin{equation}
\label{eq:tadpole-tree-shifts}
\delta_{2a}L_1=\delta_{2a}D=2a\,,\qquad~~
\delta_{2b}L_2=\delta_{2b}D=2b\,.
\end{equation}
The non-vanishing condition \eqref{eq:nonvanishing-weight-condition} therefore leaves exactly four nontrivial ordered partitions,
\begin{align}
\label{eq:dub-nv-partition}
\hTT_{(1)}&=\big(\, \emp\,, \,\{a\}\,,\,\{b\}\,\big)\,,
\qquad~~~
\hTT_{(2)}=\big(\,\{b\}\,, \,\{a\}\,,\,\emp\,\big)\,,
\nn\\[1mm]
\hTT_{(3)}&=\big(\,\{a\}\,,\,\emp\,,\,\{b\}\,\big)\,,
\qquad~~~
\hTT_{(4)}=\big(\,\{a,b\}\,,\,\emp\,,\,\emp\,\big)\,,
\end{align}
where each $\hTT_{(j)}=(T_{(0)},T_{(\L)},T_{(\R)})\in\hop_3(\bT)$.\
For the first three partitions, every nonempty block is a singleton, so the possible local ordering is rank-one and unique, which will be suppressed.\ 
The fourth partition assigns both tadpole shifts to $D$; its nonempty block $T_{(0)}=\{a,b\}$ has two possible local orderings $(1,2)$ and $(2,1)$ which correspond to the two local maximal chains:
\begin{equation}
C[(1,2)]:\emp\subsetneq\{a\}\subsetneq\{a,b\}\,,\qquad~~~
C[(2,1)]: \emp\subsetneq\{b\}\subsetneq\{a,b\} \,.
\end{equation}
In other words, $\PP(T_{(0)})$ is a Boolean lattice isomorphic to \eqref{eq:bool-bubble}. Therefore, \eqrefe{eq:all-loop-wfc-weight} gives five non-vanishing weights
\beqs
\label{eq:dub-weights}
\begin{align}
\label{eq:tad-eg1-weights-1}
\W_{\hTT_{(1)}}
&\,=\,\Vol\frac1{L_1(L_1+2a)L_2(L_2+2b)D}\,,
\\[1mm]
\label{eq:tad-eg1-weights-2}
\W_{\hTT_{(2)}}
&\,=\,\Vol\frac1{L_1(L_1+2a)(L_2+2b)D(D+2b)}\,,
\\[1mm]
\label{eq:tad-eg1-weights-3}
\W_{\hTT_{(3)}}
&\,=\,\Vol\frac1{(L_1+2a)L_2(L_2+2b)D(D+2a)}\,,
\\[1mm]
\label{eq:tad-eg1-weights-4-ab}
\W_{\hTT_{(4)},(1,2)}
&\,=\,\Vol\frac1{(L_1+2a)(L_2+2b)D(D+2a)(D+2a+2b)}\,,
\\[1mm]
\label{eq:tad-eg1-weights-4-ba}
\W_{\hTT_{(4)},(2,1)}
&\,=\,\Vol\frac1{(L_1+2a)(L_2+2b)D(D+2b)(D+2a+2b)}\,,
\end{align}
\eeqs
where $\Vol\equiv\Vol[\{Y_1\},\{a,b\}]=8Y_1ab$.\
The weight sum yields the full wavefunction coefficient, which can be checked explicitly to coincide with the alternating shifted-tree sum:
\begin{align}
\psi^{(2;0,1,1)}&=\W_{\hTT_{(1)}}+\W_{\hTT_{(2)}}+\W_{\hTT_{(3)}}+\W_{\hTT_{(4)},(1,2)}+\W_{\hTT_{(4)},(2,1)}
\nn\\[1mm]
&=\Delta_{2b}\Delta_{2a}\!\(\psi^{(2;0)}\)
\nn\\[1mm]
&=\frac{2Y_1}{L_1 L_2 D}- \frac{2Y_1}{(L_1+2a) L_2 (D+2a)} - \frac{2Y_1}{L_1 (L_2+2b) (D+2b)} 
\nn\\
&\quad+ \frac{2Y_1}{(L_1+2a) (L_2+2b) (D+2a+2b)} \,.
\end{align}
This example makes the two layers of the construction transparent: the ordered partitions determine which tadpole shifts are assigned to each divisor, while the local chains resolve the shifts associated with each individual divisor.

Following the matching proof in section~\ref{sec:6}, the five weights correspond to the following tubing configurations:
\beqs
\label{eq:tad-eg1-tubing}
\begin{align}
\W_{\hTT_{(1)}}
&\,\iff~\letterfig{0mm}{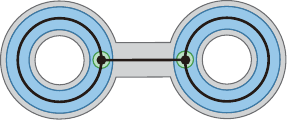}{0.75}{0cm}
\hspace*{3.7cm}
~=~{\lgre\A^\L_\emp}\,{\lblu\A^\L_{\{a\}}}\,
{\lgre\A^\R_\emp}\,{\lblu\A^\R_{\{b\}}}\,
{\lgray\B_{\YY,\{a,b\}}}\,,
\\
\W_{\hTT_{(2)}}
&\,\iff~ \letterfig{0mm}{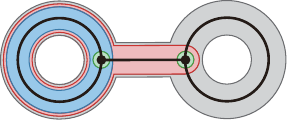}{0.75}{0cm}
\hspace*{3.7cm}
~=~{\lgre\A^\L_\emp}\,{\lblu\A^\L_{\{a\}}}\,
{\lgre\A^\R_\emp}\,
{\red\B_{\YY,\{a\}}}\,
{\lgray\B_{\YY,\{a,b\}}}\,,
\\
\W_{\hTT_{(3)}}
&\,\iff~ \letterfig[180]{0mm}{dub-2}{0.75}{0cm}
\hspace*{3.7cm}
~=~{\lgre\A^\L_\emp}\,{\lgre\A^\R_\emp}\,
{\lblu\A^\R_{\{b\}}}\,
{\red\B_{\YY,\{b\}}}\,
{\lgray\B_{\YY,\{a,b\}}}\,,
\\
\W_{\hTT_{(4)},(1,2)}
&\,\iff~ \letterfig[180]{1mm}{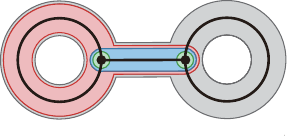}{0.75}{0cm}
\hspace*{3.7cm}
~=~{\lgre\A^\L_\emp}\,{\lgre\A^\R_\emp}\,
{\lblu\B_{\YY,\emp}}\,
{\red\B_{\YY,\{b\}}}\,
{\lgray\B_{\YY,\{a,b\}}}\,,
\\
\W_{\hTT_{(4)},(2,1)}
&\,\iff~ \letterfig{-1mm}{dub-3}{0.75}{0cm}
\hspace*{3.7cm}
~=~{\lgre\A^\L_\emp}\,{\lgre\A^\R_\emp}\,
{\lblu\B_{\YY,\emp}}\,
{\red\B_{\YY,\{a\}}}\,
{\lgray\B_{\YY,\{a,b\}}}\,.
\end{align}
\eeqs
These configurations arise from the graphic filtrations organized by the four possible types of smallest 2-site tubes. According to eqs.~\eqref{eq:1-site-tube-ordering}, \eqref{eq:2-site-tube-ordering} and \eqref{eq:V-U}, the corresponding graphic filtrations are defined in terms of complement chains as
\beqs
\begin{align}
\W_{\hTT_{(1)}}\iff\quad
&\left.
\begin{gathered}
\A^\L_{\emp}\subsetneq \A^\L_{\{a\}}\\
\A^\R_{\emp}\subsetneq \A^\R_{\{b\}}
\end{gathered}
\right\}
\subsetneq
\underbrace{\B_{\YY,\{a,b\}}}
_{\B_{\min}(=\B_{\max})},
\\[2mm]
\W_{\hTT_{(2)}}\iff\quad
&\left.
\begin{gathered}
\A^\L_{\emp}\subsetneq \A^\L_{\{a\}}\\
\A^\R_{\emp}
\end{gathered}
\right\}
\subsetneq
\underbrace{\B_{\YY,\{a\}}}_{\B_{\min}}
\subsetneq
\underbrace{\B_{\YY,\{a,b\}}}_{\B_{\max}},
\\[2mm]
\W_{\hTT_{(3)}}\iff\quad
&\left.
\begin{gathered}
\A^\L_{\emp}\\
\A^\R_{\emp}\subsetneq \A^\R_{\{b\}}
\end{gathered}
\right\}
\subsetneq
\underbrace{\B_{\YY,\{b\}}}_{\B_{\min}}
\subsetneq
\underbrace{\B_{\YY,\{a,b\}}}_{\B_{\max}},
\\[2mm]
\W_{\hTT_{(4)},(1,2)}\iff\quad
&\A^{\L/\R}_{\emp}
\subsetneq
\underbrace{\B_{\YY,\emp}}_{\B_{\min}}
\subsetneq
\B_{\YY,\{b\}}
\subsetneq
\underbrace{\B_{\YY,\{a,b\}}}_{\B_{\max}},
\\[2mm]
\W_{\hTT_{(4)},(2,1)}\iff\quad
&\A^{\L/\R}_{\emp}
\subsetneq
\underbrace{\B_{\YY,\emp}}_{\B_{\min}}
\subsetneq
\B_{\YY,\{a\}}
\subsetneq
\underbrace{\B_{\YY,\{a,b\}}}_{\B_{\max}}.
\end{align}
\eeqs
In the above, we mark the four possible types of smallest 2-site tubes by $\B_{\min}$ and the common largest 2-site tube by $\B_{\max}$. In $\W_{\hTT_{(1)}}$, $\B_{\min}$ and $\B_{\max}$ coincide.\ We now explain the chain$\iff$complement-chain correspondence underlying the above bijection. In the weight labels, since the central energy set is a singleton, its suppressed unique rank-one ordering determines a rank-one maximal chain that is self-complementary, and hence maps to the same chain on the graphic filtration side. The same is true for the nonempty singleton blocks appearing in the four nontrivial partitions of the tadpole energy set: their suppressed rank-one orderings determine self-complementary rank-one local maximal chains.
A non-identity complement map appears only for the partition $\hTT_{(4)}$, where $C[(1,2)]\iff C^c[(1,2)]=C[(2,1)]$ and $C[(2,1)]\iff C^c[(2,1)]=C[(1,2)]$. This exchange is reflected by the tubes $\B_{\YY,\{b\}}$ and $\B_{\YY,\{a\}}$ in the last two lines. 

The associated factors for the tubes are given by
\begin{align}
\label{eq:dumbbell-tube-factors}
\A_\emp^\L&=\frac1{L_1\!+\!2a}\,,\quad~~
\A_{\{a\}}^\L=\frac1{L_1}\,,\quad~~
\A_\emp^\R=\frac1{L_2\!+\!2b}\,,\quad~~
\A_{\{b\}}^\R=\frac1{L_2}\,,
\nn\\[1mm]
\B_{\YY,\emp}&=\frac1{D\!+\!2a\!+\!2b}\,,\quad~~
\B_{\YY,\{a\}}=\frac1{D\!+\!2b}\,,\quad~~
\B_{\YY,\{b\}}=\frac1{D\!+\!2a}\,,\quad~~
\B_{\YY,\{a,b\}}=\frac1{D}\,.
\end{align}
Multiplying the tube factors in each contribution of \eqrefe{eq:tad-eg1-tubing} by the volume factor $8Y_1ab$ reproduces the corresponding algebraic weight in \eqrefe{eq:dub-weights}, whose sum yields the dumbbell coefficient given in ref.~\cite{Hang:2024xas}.

Furthermore, to illustrate explicitly the sense in which the assignment of tadpoles to divisors should be regarded as the ``gauge-invariant'' content of \eqrefe{eq:nonvanishing-weight-condition}, let us reconsider the same example using a non-natural divisor ordering
\begin{equation}
\(H_1,H_2,H_3\)=\(L_2,D,L_1\).
\end{equation}
The underlying tadpole assignments remain unchanged: for a nontrivial contribution, $a$ can still be assigned only to $L_1$ or $D$, while $b$ can be assigned only to $L_2$ or $D$.\ 
The nontrivial  ordered partitions are simply the previous ones permuted according to the new divisor positions,
\begin{align}
\hTT_{(1)}^{\, \pp}&=\big(\,\{b\}\,,\, \emp\,, \,\{a\}\,\big)\,, \qquad~~~
\hTT_{(2)}^{\, \pp}=\big(\,\emp\,,\,\{b\}\,, \,\{a\}\,\big)\,,
\nn\\[1mm]
\hTT_{(3)}^{\, \pp}&=\big(\,\{b\}\,,\,\{a\}\,,\,\emp\,\big)\,,
\qquad~~~
\hTT_{(4)}^{\,\pp}=\big(\,\emp\,,\,\{a,b\}\,,\,\emp\,\big)\,,
\end{align}
with the same two local orderings $(1,2)$ and $(2,1)$ for $\hTT_{(4)}^{\,\pp}$. Since the divisors are no longer in natural ordering, we instead use the order-independent general chain expansion \eqref{eq:tadpole-chain-expansion} for the corresponding weights
\beqs
\label{eq:dub-weights-prime}
\begin{align}
\W_{\hTT_{(1)}^{\, \pp}}^{\, \pp} &=\Vol
\frac1{L_1(L_1+2a)(D+2b)L_2(L_2+2b)}\,,
\\
\W_{\hTT_{(2)}^{\, \pp}}^{\, \pp} &=\Vol
\frac1{L_1(L_1+2a)D(D+2b)L_2}\,,
\\
\W_{\hTT_{(3)}^{\, \pp}}^{\, \pp}&=\Vol
\frac1{(L_1+2a)(D+2b)(D+2a+2b)L_2(L_2+2b)}\,,
\\
\W_{\hTT_{(4)}^{\, \pp},(1,2)}^{\, \pp} &=\Vol
\frac1{(L_1+2a)D(D+2a)(D+2a+2b)L_2}\,,
\\
\W_{\hTT_{(4)}^{\, \pp},(2,1)}^{\, \pp} &=\Vol
\frac1{(L_1+2a)D(D+2b)(D+2a+2b)L_2}\,,
\end{align}
\eeqs
where the order-independent volume factor $\Vol\equiv\Vol[\{Y_1\},\{a,b\}]=8Y_1ab$ follows from the same analysis as before via \eqrefe{eq:Vol}. These terms are not a permutation of the weights \eqref{eq:dub-weights} obtained from the natural ordering. For instance, the first contribution now contains $1/(D+2b)$ rather than $1/D$, since the tadpole $b$, assigned to the leftmost divisor $L_2$, contributes to the cumulative tadpole shift of $D$.\ Nevertheless, it is straightforward to verify that the sum of the weights in \eqrefe{eq:dub-weights-prime}  is identical to that in \eqrefe{eq:dub-weights}.

Inspecting the functional forms of the two representations \eqref{eq:dub-weights} and \eqref{eq:dub-weights-prime}, the weights arising from the natural ordering are not, a priori, simpler than those obtained from the alternative ordering. The practical advantage of the natural ordering is instead that its individual weights admit the term-by-term graph-tubing correspondence displayed above. The alternative ordering, however, yields the same total sum directly from the finite-difference/maximal-chain construction, without relying on the existing tubing rule. Equivalently, each divisor ordering may be viewed as generating its own tubing rule, giving rise to equivalent representations of the same result. In this sense, the graph-tubing rule is simply one convenient representation of the underlying combinatorial structure of the physical wavefunction coefficient.

This provides a concrete realization of the ordering-independent, or ``gauge-invariant'', organization of the chain expansion.\ Changing the divisor ordering redistributes the cumulative shifts along the chains and thereby modifies the individual rational terms, while leaving the non-vanishing assignment data and the final wavefunction coefficient unchanged. For higher-loop cases, the agreement of the resulting total sums can involve nontrivial partial-fraction identities, but the order-independent formula \eqref{eq:tadpole-chain-expansion} makes these identities manifest. Importantly, the corresponding shifted-tree decomposition is insensitive to all these different ``gauge-dependent'' representations and remains canonically organized on the product Boolean lattice. This further reinforces the ``path-integral'' interpretation of the maximal-chain decomposition: changing the divisor ordering changes the weights assigned to a fixed set of paths, while the sum over paths computes the same wavefunction coefficient.

Appendix~\ref{app:A2} provides an additional example of a one-loop central bubble with a single right tadpole.\ In this case, the central sector admits two rank-two maximal chains, and for each central contribution the tadpole shift can be distributed among three eligible divisors.\ We compute the corresponding maximal-chain weights, verify the shifted-tree decomposition, and identify the graphical counterpart of each weight through the nontrivial complement maps acting on the central chains.

\section{Conclusion and Outlook}
\label{sec:8}

In this work, we have proved the shifted-tree decomposition for two-site wavefunction coefficients at arbitrary loop order, including arbitrary left and right tadpole decorations.\
For the conformally coupled scalar theories considered here, the complete formula \eqrefe{eq:shifted-tree-decorated-completely} expresses each graph coefficient as a finite signed sum of two-site tree coefficients evaluated at shifted energies.\
This establishes the all-loop validity of the decomposition proposed in ref.~\cite{Hang:2024xas}, which was previously verified at low loop orders.\
Our derivation identifies the lattice and chain structures underlying the relation between the shifted-tree formula and the graph-tubing representation.

For the bubble-like family, the shifted-tree sum reduces to a finite difference of a single inverse total-energy divisor.\
The alternating subset expansion on the Boolean lattice $\PP(\YY)$ and the maximal-chain expansion \eqrefe{eq:subset-findiff-maxchain} are therefore two representations of the same finite-difference identity.\
The former is organized by subset choices with M\"obius signs, while the latter is organized by complete filtrations and their associated products of inverse shifted divisors.\
This chain structure also explains the permutohedral organization of the examples: a maximal chain labels a vertex, and its intermediate subsets label the incident facets, whose divisors give the chain-dependent factor, as in \eqrefe{eq:permutohedron-chain-facets}.

We have established the central identity by two complementary proofs.\
The algebraic proof proceeds by induction in the number of internal lines, with each step implemented by the finite-difference Leibniz rule, and the resulting maximal-chain identity has also been formalized in {\sc Lean 4}.\
The geometric proof represents the chain contributions as Laplace integrals over ordered simplices, whose sum gives the cubical integral associated with the finite-difference expression.\
It thus interprets the chain expansion as a simplex decomposition of the cube, with each chain selecting one ordering of the auxiliary Euclidean times.\
This cubical construction provides a higher-dimensional counterpart of the two-dimensional hyperplane arrangements in the plane of twist variables studied in ref.~\cite{Hang:2024xas}.

Tadpole decorations extend this structure to products of inverse affine-linear divisors.\
The discrete Leibniz rule distributes the tadpole shifts among the divisor factors, giving ordered partitions of $\bT$, with empty blocks allowed, and a local maximal chain for each assigned block.\
The partitions specify the assignment of shifts, while the local chains determine the associated divisor products.\
For a fixed partition, the lifted local chains can be concatenated into a distinguished global maximal chain within the product sublattice.\
Since each divisor is shifted precisely along its lifted chain, the general tadpole expansion \eqrefe{eq:tadpole-chain-expansion} becomes a combinatorial ``path integral'' over central maximal chains in $\PP(\YY)$ and global maximal chains in $\PP(\bT)$ passing through the cuts of the partition, as in \eqrefe{eq:chain-path-sum}.

The identification with the physical wavefunction coefficient follows from the matching to the graph-tubing representation of the all-loop recursion in section~\ref{sec:6}.\
With the natural divisor ordering, the non-vanishing chain weights are in one-to-one correspondence with the graph-tubing contributions.\
Subset complementation reverses the cumulative-shift filtration to match the graphical filtration, reflecting the dependence of tubing divisors on the internal lines intersecting the tubes.\
Matching the divisor products together with the common factor $\Vol[\YY,\bT]$, fixed by residue normalization, completes the proof for the tadpole-decorated family.\
The examples in section~\ref{sec:7} and appendix~\ref{app:A2} illustrate this correspondence for graphs with and without tadpoles.\
This correspondence also clarifies the role of divisor ordering.\
Changing the ordering of the commuting divisor factors changes the cumulative translations in the asymmetric Leibniz expansion and can modify individual chain weights.\
In the geometric picture of section~\ref{sec:5.1}, the set of paths is unchanged, and a change of ordering only permutes the divisors carried by their segments.\
The admissible tadpole assignments remain unchanged up to relabeling of the blocks, and all orderings give the same complete coefficient, with some orderings yielding identical weights.\
The natural ordering is distinguished only because its individual weights reproduce the conventional graph-tubing rule, so the particular term-wise matching is a technical choice of construction.\
The alternative ordering in section~\ref{sec:7.2} shows explicitly how different rational contributions combine to give the same result, while the shifted-tree formula remains independent of this choice.

A direct consequence is the reduction of the two-site FRW wavefunction coefficients to known tree coefficients.\
For the conformally coupled scalar models considered here, linearity of the integral transform from flat space and its commutation with the relevant energy translations carry the rational shifted-tree identity to power-law FRW backgrounds, and more generally to any FRW background whose couplings admit a spectral representation, with each shifted tree integrated with the measure of the target graph.\
Once the appropriate FRW tree is known, the finite shifted-tree sum gives the two-site coefficients without requiring a separate differential system to be solved at each loop order.\
The de Sitter formula \eqref{eq:wfc-generic-loop-dS} provides an explicit example in terms of shifted dilogarithmic tree contributions.\
These results concern the energy-dependent graph coefficients before spatial loop-momentum integration.

Several questions remain open within the two-site family.\
It would be useful to classify the divisor orderings that give identical weights and to determine the rational identities relating the remaining decompositions.\
This could help identify representations adapted to particular poles or factorization limits.\
It would also be interesting to investigate how the finite shifted-tree representation can be used in spatial loop-momentum integration, and how the relations among graph coefficients constrain the analytic structure of the integrated result.\
Since the decomposition does not rely on the twisted-cohomology structure of power-law backgrounds, it would also be interesting to apply it to more general FRW cosmologies, where differential-equation methods are less readily available.

A broader question is whether analogous shifted-tree formulas exist for graphs with more than two sites.\
The affine-divisor identities and the discrete Leibniz expansion remain applicable, but the physical correspondence must account for the compatibility of energy divisors associated with overlapping connected subgraphs.\
Identifying the partitions, filtrations, or cell decompositions appropriate to these constraints would be a step towards extending the construction.\
Such an extension would clarify to what extent the relation between shifted trees, chain expansions, and graph tubings persists beyond the two-site family.

\begin{acknowledgments}
YH acknowledges support from the Department of Physics at the University of Wisconsin--Madison and thanks Professor Gary Shiu for providing the opportunity to pursue this research.\ 
CS acknowledges the Northwestern University Amplitudes and Insight group, the Department of Physics and Astronomy, and Weinberg College for support.
\end{acknowledgments}

\clearpage
\appendix
\section{Additional Examples}
\label{app:A}

\begin{figure}[t]
\centering
\begin{subfigure}{0.48\textwidth}
\centering
\hspace*{-2mm}
\includegraphics[scale=1.1]{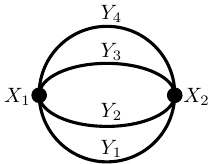}
\vs
\caption{}
\label{Afig:banana}
\end{subfigure}
\hspace*{-0.5cm}
\begin{subfigure}{0.48\textwidth}
\centering
\hspace*{-5mm}
\includegraphics[scale=1.1]{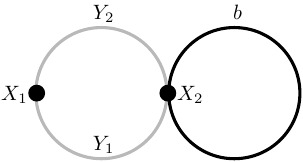}
\vs
\caption{}
\label{Afig:gourd}
\end{subfigure}
\caption{(a)\,Two-site three-loop banana diagram.\
(b)\,Central one-loop bubble decorated with one right tadpole.}
\label{Afig:add-egs}
\end{figure}

In this appendix, we provide two additional examples that extend the constructions in sections~\ref{sec:3} and~\ref{sec:5}.
The first is the three-loop banana, which makes the $P_4$ maximal-chain organization explicit; the second is a mixed bubble-tadpole graph, which illustrates how a single right-tadpole shift is distributed among several eligible divisors of each one-loop central-chain contribution.
The two diagrams are shown in figure~\ref{Afig:add-egs}.

\subsection{Three-Loop Banana}
\label{app:A1}

For the two-site three-loop banana, we have $\ell=3$; the diagram is shown in figure~\ref{Afig:banana}. The set of internal energies is $\YY=\{Y_1,Y_2,Y_3,Y_4\}$. The Boolean lattice $\PP(\YY)$ has four rank-one nodes, six rank-two nodes, and four rank-three nodes. It has $4!=24$ maximal chains, and its Hasse diagram is given by
\begin{equation}
\label{Aeq:bool-banana}
\begin{tikzpicture}[
scale=1.2,
baseline=(current bounding box.center),
every node/.style={inner sep=2pt, font=\small},
edge/.style={line width=1.1pt}
]
\node (e) at (0,0) {$\emp$};
\node (s1) at (-4,1.3) {$\{Y_1\}$};
\node (s2) at (-1.3,1.3) {$\{Y_2\}$};
\node (s3) at (1.3,1.3) {$\{Y_3\}$};
\node (s4) at (4,1.3) {$\{Y_4\}$};
\node (s12) at (-5,3) {$\{Y_1,Y_2\}$};
\node (s13) at (-3,3) {$\{Y_1,Y_3\}$};
\node (s14) at (-1,3) {$\{Y_1,Y_4\}$};
\node (s23) at (1,3) {$\{Y_2,Y_3\}$};
\node (s24) at (3,3) {$\{Y_2,Y_4\}$};
\node (s34) at (5,3) {$\{Y_3,Y_4\}$};
\node (s123) at (-4,4.7) {$\{Y_1,Y_2,Y_3\}$};
\node (s124) at (-1.3,4.7) {$\{Y_1,Y_2,Y_4\}$};
\node (s134) at (1.3,4.7) {$\{Y_1,Y_3,Y_4\}$};
\node (s234) at (4,4.7) {$\{Y_2,Y_3,Y_4\}$};
\node (Y) at (0,6) {$\YY$};
\draw[edge] (e) -- (s1);
\draw[edge] (e) -- (s2);
\draw[edge] (e) -- (s3);
\draw[edge] (e) -- (s4);
\draw[edge] (s1) -- (s12);
\draw[edge] (s1) -- (s13);
\draw[edge] (s1) -- (s14);
\draw[edge] (s2) -- (s12);
\draw[edge] (s2) -- (s23);
\draw[edge] (s2) -- (s24);
\draw[edge] (s3) -- (s13);
\draw[edge] (s3) -- (s23);
\draw[edge] (s3) -- (s34);
\draw[edge] (s4) -- (s14);
\draw[edge] (s4) -- (s24);
\draw[edge] (s4) -- (s34);
\draw[edge] (s12) -- (s123);
\draw[edge] (s12) -- (s124);
\draw[edge] (s13) -- (s123);
\draw[edge] (s13) -- (s134);
\draw[edge] (s14) -- (s124);
\draw[edge] (s14) -- (s134);
\draw[edge] (s23) -- (s123);
\draw[edge] (s23) -- (s234);
\draw[edge] (s24) -- (s124);
\draw[edge] (s24) -- (s234);
\draw[edge] (s34) -- (s134);
\draw[edge] (s34) -- (s234);
\draw[edge] (s123) -- (Y);
\draw[edge] (s124) -- (Y);
\draw[edge] (s134) -- (Y);
\draw[edge] (s234) -- (Y);
\end{tikzpicture}
\end{equation}
The corresponding maximal chains are
\begin{align}
\label{Aeq:P4-max-chains}
C[(1,2,3,4)]:&~~ \emp\subsetneq\{Y_1\}\subsetneq\{Y_1,Y_2\}\subsetneq\{Y_1,Y_2,Y_3\}\subsetneq\YY\,,
\nn\\
C[(1,2,4,3)]:&~~ \emp\subsetneq\{Y_1\}\subsetneq\{Y_1,Y_2\}\subsetneq\{Y_1,Y_2,Y_4\}\subsetneq\YY\,,
\nn\\
C[(1,3,2,4)]:&~~ \emp\subsetneq\{Y_1\}\subsetneq\{Y_1,Y_3\}\subsetneq\{Y_1,Y_2,Y_3\}\subsetneq\YY\,,
\nn\\
C[(1,3,4,2)]:&~~ \emp\subsetneq\{Y_1\}\subsetneq\{Y_1,Y_3\}\subsetneq\{Y_1,Y_3,Y_4\}\subsetneq\YY\,,
\nn\\
C[(1,4,2,3)]:&~~ \emp\subsetneq\{Y_1\}\subsetneq\{Y_1,Y_4\}\subsetneq\{Y_1,Y_2,Y_4\}\subsetneq\YY\,,
\nn\\
C[(1,4,3,2)]:&~~ \emp\subsetneq\{Y_1\}\subsetneq\{Y_1,Y_4\}\subsetneq\{Y_1,Y_3,Y_4\}\subsetneq\YY\,,
\nn\\
C[(2,1,3,4)]:&~~ \emp\subsetneq\{Y_2\}\subsetneq\{Y_1,Y_2\}\subsetneq\{Y_1,Y_2,Y_3\}\subsetneq\YY\,,
\nn\\
C[(2,1,4,3)]:&~~ \emp\subsetneq\{Y_2\}\subsetneq\{Y_1,Y_2\}\subsetneq\{Y_1,Y_2,Y_4\}\subsetneq\YY\,,
\nn\\
C[(2,3,1,4)]:&~~ \emp\subsetneq\{Y_2\}\subsetneq\{Y_2,Y_3\}\subsetneq\{Y_1,Y_2,Y_3\}\subsetneq\YY\,,
\nn\\
C[(2,3,4,1)]:&~~ \emp\subsetneq\{Y_2\}\subsetneq\{Y_2,Y_3\}\subsetneq\{Y_2,Y_3,Y_4\}\subsetneq\YY\,,
\nn\\
C[(2,4,1,3)]:&~~ \emp\subsetneq\{Y_2\}\subsetneq\{Y_2,Y_4\}\subsetneq\{Y_1,Y_2,Y_4\}\subsetneq\YY\,,
\nn\\
C[(2,4,3,1)]:&~~ \emp\subsetneq\{Y_2\}\subsetneq\{Y_2,Y_4\}\subsetneq\{Y_2,Y_3,Y_4\}\subsetneq\YY\,,
\nn\\
C[(3,1,2,4)]:&~~ \emp\subsetneq\{Y_3\}\subsetneq\{Y_1,Y_3\}\subsetneq\{Y_1,Y_2,Y_3\}\subsetneq\YY\,,
\nn\\
C[(3,1,4,2)]:&~~ \emp\subsetneq\{Y_3\}\subsetneq\{Y_1,Y_3\}\subsetneq\{Y_1,Y_3,Y_4\}\subsetneq\YY\,,
\nn\\
C[(3,2,1,4)]:&~~ \emp\subsetneq\{Y_3\}\subsetneq\{Y_2,Y_3\}\subsetneq\{Y_1,Y_2,Y_3\}\subsetneq\YY\,,
\nn\\
C[(3,2,4,1)]:&~~ \emp\subsetneq\{Y_3\}\subsetneq\{Y_2,Y_3\}\subsetneq\{Y_2,Y_3,Y_4\}\subsetneq\YY\,,
\nn\\
C[(3,4,1,2)]:&~~ \emp\subsetneq\{Y_3\}\subsetneq\{Y_3,Y_4\}\subsetneq\{Y_1,Y_3,Y_4\}\subsetneq\YY\,,
\nn\\
C[(3,4,2,1)]:&~~ \emp\subsetneq\{Y_3\}\subsetneq\{Y_3,Y_4\}\subsetneq\{Y_2,Y_3,Y_4\}\subsetneq\YY\,,
\nn\\
C[(4,1,2,3)]:&~~ \emp\subsetneq\{Y_4\}\subsetneq\{Y_1,Y_4\}\subsetneq\{Y_1,Y_2,Y_4\}\subsetneq\YY\,,
\nn\\
C[(4,1,3,2)]:&~~ \emp\subsetneq\{Y_4\}\subsetneq\{Y_1,Y_4\}\subsetneq\{Y_1,Y_3,Y_4\}\subsetneq\YY\,,
\nn\\
C[(4,2,1,3)]:&~~ \emp\subsetneq\{Y_4\}\subsetneq\{Y_2,Y_4\}\subsetneq\{Y_1,Y_2,Y_4\}\subsetneq\YY\,,
\nn\\
C[(4,2,3,1)]:&~~ \emp\subsetneq\{Y_4\}\subsetneq\{Y_2,Y_4\}\subsetneq\{Y_2,Y_3,Y_4\}\subsetneq\YY\,,
\nn\\
C[(4,3,1,2)]:&~~ \emp\subsetneq\{Y_4\}\subsetneq\{Y_3,Y_4\}\subsetneq\{Y_1,Y_3,Y_4\}\subsetneq\YY\,,
\nn\\
C[(4,3,2,1)]:&~~ \emp\subsetneq\{Y_4\}\subsetneq\{Y_3,Y_4\}\subsetneq\{Y_2,Y_3,Y_4\}\subsetneq\YY\,.
\end{align}
Applying \eqrefe{eq:fin-diff-sum-max-chain}, the finite-difference product becomes
\begin{align}
\label{Aeq:fin-diff-banana}
&\Delta_{2\YY}\bigg(\frac1{D}\bigg)
=\frac{16Y_1Y_2Y_3Y_4}{D_{\emp}D_{\YY}}\Bigg(
\frac1{D_{\{Y_1\}}D_{\{Y_1,Y_2\}}D_{\{Y_1,Y_2,Y_3\}}}
+\frac1{D_{\{Y_1\}}D_{\{Y_1,Y_2\}}D_{\{Y_1,Y_2,Y_4\}}}
\nn\\
&+\frac1{D_{\{Y_1\}}D_{\{Y_1,Y_3\}}D_{\{Y_1,Y_2,Y_3\}}}
+\frac1{D_{\{Y_1\}}D_{\{Y_1,Y_3\}}D_{\{Y_1,Y_3,Y_4\}}}
+\frac1{D_{\{Y_1\}}D_{\{Y_1,Y_4\}}D_{\{Y_1,Y_2,Y_4\}}}
\nn\\
&+\frac1{D_{\{Y_1\}}D_{\{Y_1,Y_4\}}D_{\{Y_1,Y_3,Y_4\}}}
+\frac1{D_{\{Y_2\}}D_{\{Y_1,Y_2\}}D_{\{Y_1,Y_2,Y_3\}}}
+\frac1{D_{\{Y_2\}}D_{\{Y_1,Y_2\}}D_{\{Y_1,Y_2,Y_4\}}}
\nn\\
&+\frac1{D_{\{Y_2\}}D_{\{Y_2,Y_3\}}D_{\{Y_1,Y_2,Y_3\}}}
+\frac1{D_{\{Y_2\}}D_{\{Y_2,Y_3\}}D_{\{Y_2,Y_3,Y_4\}}}
+\frac1{D_{\{Y_2\}}D_{\{Y_2,Y_4\}}D_{\{Y_1,Y_2,Y_4\}}}
\nn\\
&+\frac1{D_{\{Y_2\}}D_{\{Y_2,Y_4\}}D_{\{Y_2,Y_3,Y_4\}}}
+\frac1{D_{\{Y_3\}}D_{\{Y_1,Y_3\}}D_{\{Y_1,Y_2,Y_3\}}}
+\frac1{D_{\{Y_3\}}D_{\{Y_1,Y_3\}}D_{\{Y_1,Y_3,Y_4\}}}
\nn\\
&+\frac1{D_{\{Y_3\}}D_{\{Y_2,Y_3\}}D_{\{Y_1,Y_2,Y_3\}}}
+\frac1{D_{\{Y_3\}}D_{\{Y_2,Y_3\}}D_{\{Y_2,Y_3,Y_4\}}}
+\frac1{D_{\{Y_3\}}D_{\{Y_3,Y_4\}}D_{\{Y_1,Y_3,Y_4\}}}
\nn\\
&+\frac1{D_{\{Y_3\}}D_{\{Y_3,Y_4\}}D_{\{Y_2,Y_3,Y_4\}}}
+\frac1{D_{\{Y_4\}}D_{\{Y_1,Y_4\}}D_{\{Y_1,Y_2,Y_4\}}}
+\frac1{D_{\{Y_4\}}D_{\{Y_1,Y_4\}}D_{\{Y_1,Y_3,Y_4\}}}
\nn\\
&+\frac1{D_{\{Y_4\}}D_{\{Y_2,Y_4\}}D_{\{Y_1,Y_2,Y_4\}}}
+\frac1{D_{\{Y_4\}}D_{\{Y_2,Y_4\}}D_{\{Y_2,Y_3,Y_4\}}}
+\frac1{D_{\{Y_4\}}D_{\{Y_3,Y_4\}}D_{\{Y_1,Y_3,Y_4\}}}
\nn\\
&+\frac1{D_{\{Y_4\}}D_{\{Y_3,Y_4\}}D_{\{Y_2,Y_3,Y_4\}}}
\Bigg)\,.
\end{align}
Then, we order the proper subsets of $\YY$ first by cardinality and then lexicographically. The ordered subsets and the corresponding shifted divisors are 
\begin{equation}
\(V_0,V_1,\ldots,V_{14}\) =
\(\emp,\{Y_1\},\ldots,\{Y_2,Y_3,Y_4\}\),\qquad
D_j\equiv D_{V_j}=D+2\sum_{y\in V_j}y\,.
\end{equation}
The common shifted local divisors are
\begin{equation}
L_1=X_1+\sum_{j=1}^{4}Y_j\,,\qquad~~~
L_2=X_2+\sum_{j=1}^{4}Y_j\,.
\end{equation}
Multiplying \eqrefe{Aeq:fin-diff-banana} by $D_{\YY}/(L_1L_2)$, we obtain the maximal-chain form of the three-loop banana wavefunction coefficient,
\begin{align}
\label{Aeq:max-chain-banana}
\psi^{(2;3)}
=&\ \frac{16Y_1Y_2Y_3Y_4}{L_1L_2D_0}\Bigg(
\frac{1}{D_{1}D_{5}D_{11}}
+\frac{1}{D_{1}D_{5}D_{12}}
+\frac{1}{D_{1}D_{6}D_{11}}
+\frac{1}{D_{1}D_{6}D_{13}}
+\frac{1}{D_{1}D_{7}D_{12}}
\nn\\
&+\frac{1}{D_{1}D_{7}D_{13}}
+\frac{1}{D_{2}D_{5}D_{11}}
+\frac{1}{D_{2}D_{5}D_{12}}
+\frac{1}{D_{2}D_{8}D_{11}}
+\frac{1}{D_{2}D_{8}D_{14}}
\nn\\
&+\frac{1}{D_{2}D_{9}D_{12}}
+\frac{1}{D_{2}D_{9}D_{14}}
+\frac{1}{D_{3}D_{6}D_{11}}
+\frac{1}{D_{3}D_{6}D_{13}}
+\frac{1}{D_{3}D_{8}D_{11}}
\nn\\
&+\frac{1}{D_{3}D_{8}D_{14}}
+\frac{1}{D_{3}D_{10}D_{13}}
+\frac{1}{D_{3}D_{10}D_{14}}
+\frac{1}{D_{4}D_{7}D_{12}}
+\frac{1}{D_{4}D_{7}D_{13}}
\nn\\
&+\frac{1}{D_{4}D_{9}D_{12}}
+\frac{1}{D_{4}D_{9}D_{14}}
+\frac{1}{D_{4}D_{10}D_{13}}
+\frac{1}{D_{4}D_{10}D_{14}}
\Bigg)\,.
\end{align}
With the same labels, the shifted-tree expression \eqref{eq:wfc-loop} for the three-loop banana case is
\begin{align}
\label{Aeq:wfc-3loop}
\psi^{(2;3)}=&~ \frac{(L_1+L_2)-D_0}{L_1L_2D_0}
-\sum_{j=1}^{4}\frac{(L_1+L_2)-D_j}{L_1L_2D_j} 
+\sum_{j=5}^{10}\frac{(L_1+L_2)-D_j}{L_1L_2D_j}
\nn\\
&~-\sum_{j=11}^{14} \frac{(L_1+L_2)-D_j}{L_1L_2D_j} \,,
\end{align}
which is equal to \eqrefe{Aeq:max-chain-banana}.

\begin{figure}[tp!]
\centering
\includegraphics[scale=0.65]{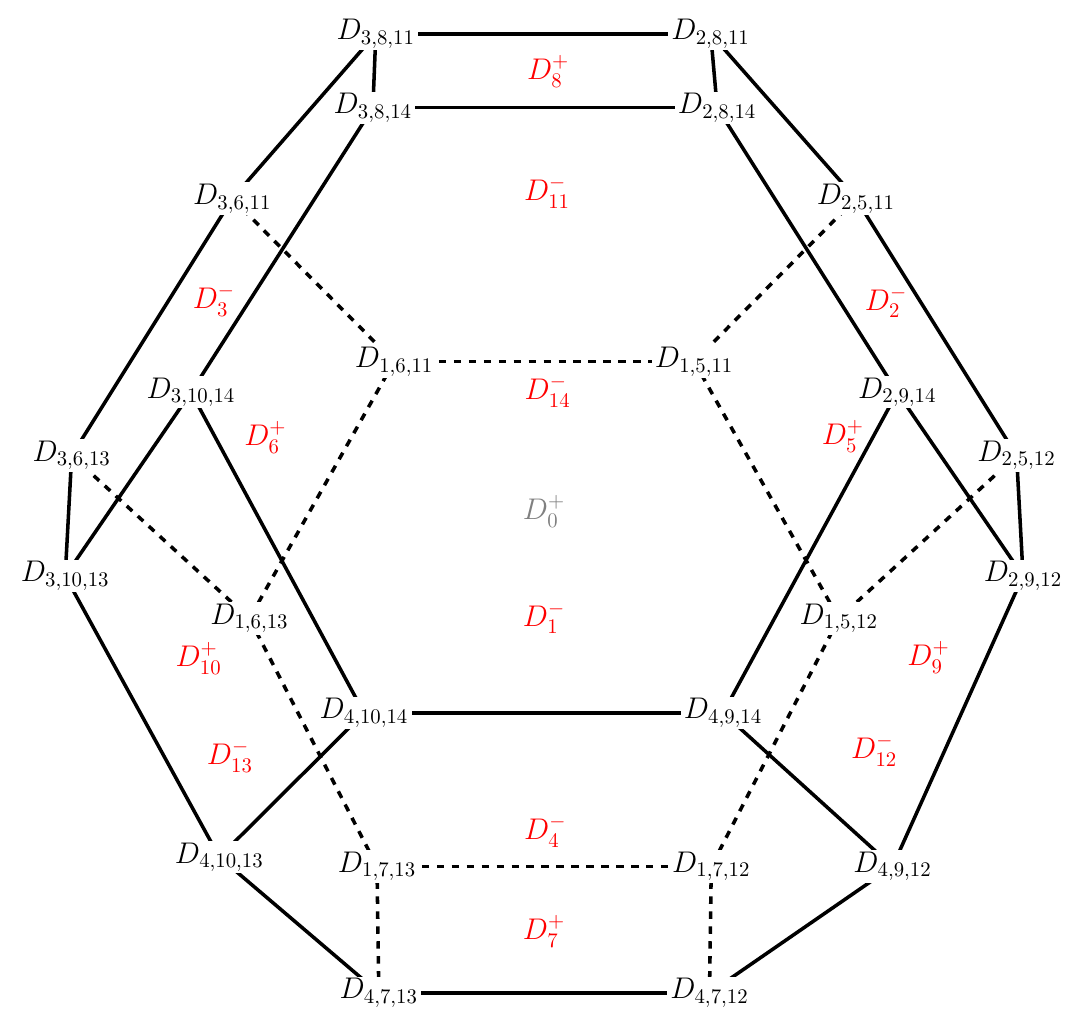}
\caption{The permutohedron $P_4$ as a truncated octahedron.\ 
Its twenty-four vertices correspond to the maximal chains in \eqrefe{Aeq:P4-max-chains}, with divisor labels given in \eqrefe{Aeq:P4-vertex-labels}.\ 
Its fourteen facets are indexed by the nonempty proper subsets of $\YY$, with vertex incidences and divisor labels given in \eqrefe{Aeq:P4-facet-labels}.}
\label{Afig:P4}
\end{figure}

For $|\YY|=4$, the permutohedron $P_4$ is a truncated octahedron with twenty-four vertices and fourteen facets, shown in figure~\ref{Afig:P4}.\ 
Each maximal chain in \eqrefe{Aeq:P4-max-chains} has three intermediate subsets, which index its three incident facets.\ 
By \eqrefe{eq:permutohedron-chain-facets}, their intersection is the corresponding vertex, whose divisor label is the product of their labels.\
The twenty-four chain--vertex correspondences and their divisor labels are:
\begin{alignat}{2}
\label{Aeq:P4-vertex-labels}
C[(1,2,3,4)]&\longleftrightarrow v_{(1,2,3,4)}\longmapsto D_{1,5,11}\,,\quad &C[(1,2,4,3)]&\longleftrightarrow v_{(1,2,4,3)}\longmapsto D_{1,5,12}\,,
\nn\\
C[(1,3,2,4)]&\longleftrightarrow v_{(1,3,2,4)}\longmapsto D_{1,6,11}\,,\quad &C[(1,3,4,2)]&\longleftrightarrow v_{(1,3,4,2)}\longmapsto D_{1,6,13}\,,
\nn\\
C[(1,4,2,3)]&\longleftrightarrow v_{(1,4,2,3)}\longmapsto D_{1,7,12}\,,\quad &C[(1,4,3,2)]&\longleftrightarrow v_{(1,4,3,2)}\longmapsto D_{1,7,13}\,,
\nn\\
C[(2,1,3,4)]&\longleftrightarrow v_{(2,1,3,4)}\longmapsto D_{2,5,11}\,,\quad &C[(2,1,4,3)]&\longleftrightarrow v_{(2,1,4,3)}\longmapsto D_{2,5,12}\,,
\nn\\
C[(2,3,1,4)]&\longleftrightarrow v_{(2,3,1,4)}\longmapsto D_{2,8,11}\,,\quad &C[(2,3,4,1)]&\longleftrightarrow v_{(2,3,4,1)}\longmapsto D_{2,8,14}\,,
\nn\\
C[(2,4,1,3)]&\longleftrightarrow v_{(2,4,1,3)}\longmapsto D_{2,9,12}\,,\quad &C[(2,4,3,1)]&\longleftrightarrow v_{(2,4,3,1)}\longmapsto D_{2,9,14}\,,
\nn\\
C[(3,1,2,4)]&\longleftrightarrow v_{(3,1,2,4)}\longmapsto D_{3,6,11}\,,\quad &C[(3,1,4,2)]&\longleftrightarrow v_{(3,1,4,2)}\longmapsto D_{3,6,13}\,,
\nn\\
C[(3,2,1,4)]&\longleftrightarrow v_{(3,2,1,4)}\longmapsto D_{3,8,11}\,,\quad &C[(3,2,4,1)]&\longleftrightarrow v_{(3,2,4,1)}\longmapsto D_{3,8,14}\,,
\nn\\
C[(3,4,1,2)]&\longleftrightarrow v_{(3,4,1,2)}\longmapsto D_{3,10,13}\,,\quad &C[(3,4,2,1)]&\longleftrightarrow v_{(3,4,2,1)}\longmapsto D_{3,10,14}\,,
\nn\\
C[(4,1,2,3)]&\longleftrightarrow v_{(4,1,2,3)}\longmapsto D_{4,7,12}\,,\quad &C[(4,1,3,2)]&\longleftrightarrow v_{(4,1,3,2)}\longmapsto D_{4,7,13}\,,
\nn\\
C[(4,2,1,3)]&\longleftrightarrow v_{(4,2,1,3)}\longmapsto D_{4,9,12}\,,\quad &C[(4,2,3,1)]&\longleftrightarrow v_{(4,2,3,1)}\longmapsto D_{4,9,14}\,,
\nn\\
C[(4,3,1,2)]&\longleftrightarrow v_{(4,3,1,2)}\longmapsto D_{4,10,13}\,,\quad &C[(4,3,2,1)]&\longleftrightarrow v_{(4,3,2,1)}\longmapsto D_{4,10,14}\,,
\end{alignat}
where $D_{i,j,k}\equiv D_iD_jD_k$.\
Multiplying the reciprocal vertex labels by the common factor $16Y_1Y_2Y_3Y_4/(L_1L_2D_0)$ gives the twenty-four chain contributions in \eqrefe{Aeq:max-chain-banana}. The bottom and top elements $\emp$ and $\YY$ do not index facets: $D_{\YY}$ has canceled, while $D_\emp=D_0$ remains in the common factor.

Conversely, each facet $F_V$ contains precisely the vertices whose chains pass through $V$. The four singleton subsets and the four three-element subsets index hexagonal facets, while the six two-element subsets index square facets. The following lists give all vertices of each facet together with its divisor label:
\begin{alignat}{2}
\label{Aeq:P4-facet-labels}
v_{(1,2,3,4)}\,,v_{(1,2,4,3)}\,,v_{(1,3,2,4)}\,,v_{(1,3,4,2)}\,,v_{(1,4,2,3)}\,,v_{(1,4,3,2)}\in{}&F_{\{Y_1\}}&&\longmapsto D_1\,,
\nn\\
v_{(2,1,3,4)}\,,v_{(2,1,4,3)}\,,v_{(2,3,1,4)}\,,v_{(2,3,4,1)}\,,v_{(2,4,1,3)}\,,v_{(2,4,3,1)}\in{}&F_{\{Y_2\}}&&\longmapsto D_2\,,
\nn\\
v_{(3,1,2,4)}\,,v_{(3,1,4,2)}\,,v_{(3,2,1,4)}\,,v_{(3,2,4,1)}\,,v_{(3,4,1,2)}\,,v_{(3,4,2,1)}\in{}&F_{\{Y_3\}}&&\longmapsto D_3\,,
\nn\\
v_{(4,1,2,3)}\,,v_{(4,1,3,2)}\,,v_{(4,2,1,3)}\,,v_{(4,2,3,1)}\,,v_{(4,3,1,2)}\,,v_{(4,3,2,1)}\in{}&F_{\{Y_4\}}&&\longmapsto D_4\,,
\nn\\
v_{(1,2,3,4)}\,,v_{(1,2,4,3)}\,,v_{(2,1,3,4)}\,,v_{(2,1,4,3)}\in{}&F_{\{Y_1,Y_2\}}&&\longmapsto D_5\,,
\nn\\
v_{(1,3,2,4)}\,,v_{(1,3,4,2)}\,,v_{(3,1,2,4)}\,,v_{(3,1,4,2)}\in{}&F_{\{Y_1,Y_3\}}&&\longmapsto D_6\,,
\nn\\
v_{(1,4,2,3)}\,,v_{(1,4,3,2)}\,,v_{(4,1,2,3)}\,,v_{(4,1,3,2)}\in{}&F_{\{Y_1,Y_4\}}&&\longmapsto D_7\,,
\nn\\
v_{(2,3,1,4)}\,,v_{(2,3,4,1)}\,,v_{(3,2,1,4)}\,,v_{(3,2,4,1)}\in{}&F_{\{Y_2,Y_3\}}&&\longmapsto D_8\,,
\nn\\
v_{(2,4,1,3)}\,,v_{(2,4,3,1)}\,,v_{(4,2,1,3)}\,,v_{(4,2,3,1)}\in{}&F_{\{Y_2,Y_4\}}&&\longmapsto D_9\,,
\nn\\
v_{(3,4,1,2)}\,,v_{(3,4,2,1)}\,,v_{(4,3,1,2)}\,,v_{(4,3,2,1)}\in{}&F_{\{Y_3,Y_4\}}&&\longmapsto D_{10}\,,
\nn\\
v_{(1,2,3,4)}\,,v_{(1,3,2,4)}\,,v_{(2,1,3,4)}\,,v_{(2,3,1,4)}\,,v_{(3,1,2,4)}\,,v_{(3,2,1,4)}\in{}&F_{\{Y_1,Y_2,Y_3\}}&&\longmapsto D_{11}\,,
\nn\\
v_{(1,2,4,3)}\,,v_{(1,4,2,3)}\,,v_{(2,1,4,3)}\,,v_{(2,4,1,3)}\,,v_{(4,1,2,3)}\,,v_{(4,2,1,3)}\in{}&F_{\{Y_1,Y_2,Y_4\}}&&\longmapsto D_{12}\,,
\nn\\
v_{(1,3,4,2)}\,,v_{(1,4,3,2)}\,,v_{(3,1,4,2)}\,,v_{(3,4,1,2)}\,,v_{(4,1,3,2)}\,,v_{(4,3,1,2)}\in{}&F_{\{Y_1,Y_3,Y_4\}}&&\longmapsto D_{13}\,,
\nn\\
v_{(2,3,4,1)}\,,v_{(2,4,3,1)}\,,v_{(3,2,4,1)}\,,v_{(3,4,2,1)}\,,v_{(4,2,3,1)}\,,v_{(4,3,2,1)}\in{}&F_{\{Y_2,Y_3,Y_4\}}&&\longmapsto D_{14}\,.
\end{alignat}
For example, the six vertices of $F_{\{Y_1\}}$ share the factor $D_1$ in their divisor labels. 
The divisor $D_0$ does not label a facet of the permutohedron and is placed at the center to indicate the denominator factor common to all maximal-chain contributions.\
The superscripts $\pm$ on the divisors in figure~\ref{Afig:P4} record the overall signs of each term in \eqrefe{Aeq:wfc-3loop}.

\subsection{One-Loop Bubble with One Right Tadpole}
\label{app:A2}

As a complementary example for maximal-chain decomposition and matching proof, consider the one-loop central bubble decorated with a single right tadpole of energy $b$, as shown in figure~\ref{Afig:gourd}.\ 
In this case, we have $\YY=\{Y_1,Y_2\}$ and $\bmY_\R=\bT=\{b\}$.

Organizing the central one-loop bubble \eqref{eq:max-chain-l1} via \eqrefe{eq:tadpole-central-block} gives two contributions:
\begin{equation}
\psi^{(2;1)}=\sum_{j=1}^{2}W[\pi_j]\,, \qquad
W[\pi_j]=4Y_1Y_2\frac{1}{L_1L_2D_0D_j}\,,
\end{equation}
where $\pi_1=(1,2)$ and $\pi_2=(2,1)$ label the two central maximal chains in $\PP(\{Y_1,Y_2\})$.\
The right-tadpole shift $\delta_{2b}$ only acts nontrivially on the divisors that depend on $X_2$:
\begin{equation}
\delta_{2b}L_2=\delta_{2b}D_0=\delta_{2b}D_j=2b\,.
\end{equation}
For each fixed central contribution $W[\pi_j]$, there are $Q_{\pi_j}=4$ divisor factors.\ 
Following \eqrefe{eq:natural-divisor-ordering}, we adopt the natural ordering
\begin{equation}
\(H[\pi_j]_{1},H[\pi_j]_{2},H[\pi_j]_{3},H[\pi_j]_{4}\)=\(D_0,D_j, L_1,L_2\).
\end{equation}
Since $L_1$ is invariant under the right-tadpole shift, the non-vanishing condition \eqref{eq:nonvanishing-weight-condition} allows $b$ to be assigned only to $L_2$, $D_j$, or $D_0$. Therefore, the three possible nontrivial ordered partitions are given by
\begin{align}
\hTT_{(1)}&=\big(\,\emp\,,\,\emp\,,\,\emp\,,\,\{b\}\,\big)\,, \qquad~~
\hTT_{(2)}=\big(\,\emp\,,\,\{b\}\,,\,\emp\,,\,\emp\,\big)\,,
\nn\\[1mm]
\hTT_{(3)}&=\big(\,\{b\}\,,\,\emp\,,\,\emp\,,\,\emp\,\big)\,.
\end{align}
In each partition, the unique nonempty block is a singleton, so the local ordering is unique and will be suppressed. Substituting these assignments into \eqrefe{eq:all-loop-wfc-weight} gives three weights for each central chain
\beqs
\label{Aeq:gourd-chain-weights}
\begin{align}
\label{Aeq:Wj1}
\W_{\pi_j;\hTT_{(1)}}
&=\Vol\frac1{L_1L_2(L_2+2b)D_0D_j}\,,
\\[1mm]
\W_{\pi_j;\hTT_{(2)}}
&=\Vol\frac1{L_1(L_2+2b)D_0D_j(D_j+2b)}\,,
\\[1mm]
\W_{\pi_j;\hTT_{(3)}}
&=\Vol\frac1{L_1(L_2+2b)D_0(D_0+2b)(D_j+2b)}\,,
\end{align}
\eeqs
where $j=1,2$ and $\Vol\equiv\Vol[\{Y_1,Y_2\},\{b\}]=8Y_1Y_2b$.\
Summing the six weights yields the full wavefunction coefficient, which can be checked explicitly to coincide with the signed shifted-bubble sum:
\begin{align}
\psi^{(2;1,0,1)}&=\sum_{j=1}^{2}\sum_{k=1}^{3}\W_{\pi_j;\hTT_{(k)}}
\nn\\[1mm]
&=\Delta_{2b}\!\(\psi^{(2;1)}\)
\nn\\[1mm]
&=\frac{4Y_1Y_2}{L_1L_2D_0}\(\!\frac{1}{D_1}\!+\!\frac{1}{D_2}\!\)-\frac{4Y_1Y_2}{L_1(L_2+2b)(D_0+2b)}\(\!\frac{1}{D_1+2b}\!+\!\frac{1}{D_2+2b}\!\).
\end{align}

The explicit correspondence with the graph tubings is given, for example, by considering the first central chain $\pi_1$:
\beqs
\begin{align}
\W_{\pi_1;\hTT_{(1)}}
&\,\iff~\letterfig{0mm}{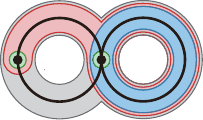}{0.75}{0cm} \hspace*{2.75cm}=~
{\lgre\A^\L_\emp}\,
{\lgre\A^\R_\emp}\,
{\lblu\A^\R_{\{b\}}}\,
{\red\B_{\{Y_2\},\{b\}}}\,
{\lgray\B_{\YY,\{b\}}}\,,
\\
\W_{\pi_1;\hTT_{(2)}}
&\,\iff~\letterfig{0mm}{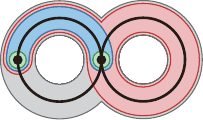}{0.75}{0cm} \hspace*{2.75cm}=~
{\lgre\A^\L_\emp}\,
{\lgre\A^\R_\emp}\,
{\lblu\B_{\{Y_2\},\emp}}\,
{\red\B_{\{Y_2\},\{b\}}}\,
{\lgray\B_{\YY,\{b\}}}\,,
\\
\W_{\pi_1;\hTT_{(3)}}
&\,\iff~\letterfig{0mm}{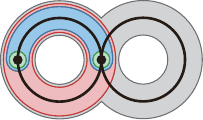}{0.75}{0cm} \hspace*{2.75cm}=~
{\lgre\A^\L_\emp}\,
{\lgre\A^\R_\emp}\,
{\lblu\B_{\{Y_2\},\emp}}\,
{\red\B_{\YY,\emp}}\,
{\lgray\B_{\YY,\{b\}}}\,.
\end{align}
\eeqs
These configurations arise from three types of graphic filtrations, which are defined according to eqs.~\eqref{eq:1-site-tube-ordering}, \eqref{eq:2-site-tube-ordering} and \eqref{eq:V-U} in terms of complement chains as
\beqs
\begin{align}
\W_{\pi_1;\hTT_{(1)}}\!\iff~~&
\left.\begin{array}{r}
\A_\emp^\L ~ \\ 
\A_\emp^\R\,\subsetneq\,\A_{\{b\}}^\R
\end{array}
\right\}\,\subsetneq\,
\underbrace{\B_{\{Y_2\},\{b\}}}_{\B_{\min}}
\,\subsetneq\,
\underbrace{\B_{\YY,\{b\}}}_{\B_{\max}}\,,
\\
\W_{\pi_1;\hTT_{(2)}}\!\iff~~&
\A_\emp^{\L/\R}
\,\subsetneq\,
\underbrace{\B_{\{Y_2\},\emp}}_{\B_{\min}}
\,\subsetneq\,
\B_{\{Y_2\},\{b\}}
\,\subsetneq\,
\underbrace{\B_{\YY,\{b\}}}_{\B_{\max}}\,,
\\
\W_{\pi_1;\hTT_{(3)}}\!\iff~~&
\A_\emp^{\L/\R}\,\subsetneq\,
\underbrace{\B_{\{Y_2\},\emp}}_{\B_{\min}}
\,\subsetneq\,
\B_{\YY,\emp}
\,\subsetneq\,
\underbrace{\B_{\YY,\{b\}}}_{\B_{\max}}\,.
\end{align}
\eeqs
The only non-identity complement map acts on the central chain, as reflected by the $\{Y_2\}$ labels of the 2-site tubes above, rather than $\{Y_1\}$.\ 
The other three filtrations associated with $\pi_2$ follow by exchanging $Y_1\leftrightarrow Y_2$ in the expressions above.\ 
Together, the six graphic filtrations reproduce the complete graph-tubing configurations. 

The associated factors for the 1-site and 2-site tubes are 
\begin{align}
\A_\emp^\L&=\frac1{L_1}\,,\qquad~~
\A_\emp^\R=\frac1{L_2\!+\!2b}\,,\qquad~~
\A_{\{b\}}^\R=\frac1{L_2}\,,
\nn\\
\B_{\{Y_2\},\emp}&=\frac1{D_1\!+\!2b}\,,\quad~~~
\B_{\{Y_2\},\{b\}}=\frac1{D_1}\,, \quad~~~
\B_{\YY,\emp}=\frac1{D_0\!+\!2b}\,, \quad~~~
\B_{\YY,\{b\}}=\frac1{D_0}\,.
\end{align}
Multiplying each graph-tubing contribution by the volume factor $8Y_1Y_2b$ reproduces the corresponding algebraic weight, whose sum yields the gourd coefficient given in ref.~\cite{Hang:2024xas}.

\section{Basic Lattice Theory}
\label{sec:B}

This appendix collects the elementary lattice-theoretic notions used in the main text. The goal is to set up the terminology needed to discuss direct products of lattices and chains. We mostly follow the order-theoretic approach of Birkhoff~\cite{birkhoff1967lattice}, defining lattices and chains as posets with certain properties.

\subsection{Posets, chains, and diagrams}
\label{app:B1}

\begin{definition}[Partially ordered set]
A \emph{partially ordered set}, or \emph{poset}, is a set $P$ equipped with a binary relation $\leqq$ satisfying, for all $x,y,z\in P$,
\begin{align}
&\text{Reflexivity:}\quad &&x\leqq x \,,
\nn\\
&\text{Antisymmetry:} \quad &&x\leqq y\ \text{and}\ y\leqq x
\quad\Longrightarrow\quad x=y \,,
\nn\\
&\text{Transitivity:} \quad &&x\leqq y\ \text{and}\ y\leqq z
\quad\Longrightarrow\quad x\leqq z \,.
\end{align}
Two elements $x,y\in P$ are \emph{comparable} if either $x\leqq y$ or $y\leqq x$.
\end{definition}

\begin{definition}[Chain]
A \emph{chain} $C$ is a totally ordered poset, namely every pair of elements of
$C$ is comparable. In other words, of any two distinct elements in a chain, one is less and the other greater.
\end{definition}

\begin{theorem}[Subsets]
Any subset of a poset is a poset under the same binary relation. Any subset of a chain is a chain.
\end{theorem}

\begin{definition}[Covering relation and Hasse diagram]
For $x,y\in P$, if $x<y$ and there is no $z\in P$ such that $x<z<y$, one says that $y$ \emph{covers} $x$ or $x$ is \emph{covered by} $y$, written as $x\lessdot y$ or $y\gtrdot x$, with the notation adopted from Stanley~\cite{Stanley_2011}.
For a finite poset, its \emph{Hasse diagram} has one vertex for each element of $P$ and one edge for each covering relation, with the larger element conventionally drawn above the smaller element.
\end{definition}

\begin{definition}[Least and greatest elements]
Let $P$ be a poset. An element $0\in P$ is the \emph{least element} of $P$ if $0\leqq x$ for all $ x\in P$. An element $1\in P$ is the \emph{greatest element} of $P$ if $x\leqq 1$ for all $x\in P$. The least and greatest elements, whenever they exist, are unique.
\end{definition}

\begin{definition}[Bounded poset]
A poset $P$ is \emph{bounded} if it has both a least element $0$ and a greatest element $1$, so that $0\leqq x\leqq1$ for all $x\in P$.
\end{definition}

\begin{definition}[Maximal chain]
A chain $C\subseteq P$ is \emph{maximal} if there is no chain $D\subseteq P$ such that $C\subsetneq D$.
\end{definition}

\begin{definition}[Saturated chain]
\label{def-saturated-chain}
For a finite chain $C$, two elements $x,y\in C$ with $x<y$ are \emph{consecutive} if there is no $z\in C$ satisfying $x<z<y$.
The chain $C\subseteq P$ is \emph{saturated} if every consecutive pair
$x<y$ in $C$ is a covering pair $x\lessdot y$ in $P$. 
\end{definition}

\begin{theorem}[Maximal chains in a finite bounded poset]
\label{Maximal-chains-in-a-finite-bounded-poset}

Let $P$ be a finite bounded poset. A chain $C\subseteq P$ is maximal if and only if it is a saturated chain from the least element $0$ to the greatest element $1$:
\begin{equation}
0=x_0 \lessdot x_1 \lessdot\cdots \lessdot x_n=1\,.
\end{equation}
\end{theorem}

\begin{definition}[Isomorphisms and duality of posets]
Let $P$ and $Q$ be posets. A bijection $f:P\to Q$ is an
\emph{(order-preserving) isomorphism} if it preserves the ordering:
$x\leqq_P y$ if and only if $f(x)\leqq_Q f(y)$. The \emph{dual poset}
$P^{\rm{op}}$ is the same set equipped with the reversed ordering.
A \emph{dual isomorphism} is a
bijection that reverses the ordering; equivalently, it is an isomorphism from $P$ to $Q^{\rm{op}}$. 
A (dual) isomorphism from a poset to itself is called a \emph{(dual) automorphism}. A poset is \emph{self-dual} if it is isomorphic to its dual.
\end{definition}

\subsection{Lattices, sublattices, and direct products}
\label{app:B2}

\begin{definition}[Meet, join, and lattice]
Let $P$ be a poset and let $x,y\in P$. An element $z\in P$ is a lower bound of $x$ and $y$ if $z\leqq x$ and $z\leqq y$. If the set of lower bounds has a greatest element, it is called the \emph{meet} of $x$ and $y$, denoted by $x\wedge y$. Dually, if the set of upper bounds of $x$ and $y$ has a least element, it is called their \emph{join}, denoted by $x\vee y$.

A nonempty poset $L$ is a \emph{lattice} if every pair $x,y\in L$ has both a meet and a join.
\end{definition}

\begin{definition}[Bounded and complete lattices]
A lattice $L$ is \emph{bounded} if its underlying poset is bounded. A lattice $L$ is \emph{complete} if every subset $S\subseteq L$ has both a greatest lower bound $\bigwedge S $ and a least upper bound $\bigvee S$ in $L$.
\end{definition}

\begin{theorem}[Completeness and boundedness]
Every nonempty complete lattice is bounded, with $
0=\bigvee\emp=\bigwedge L$ and $ 1=\bigwedge\emp=\bigvee L$. Every finite lattice is complete, thus bounded.
\end{theorem}

Since the lattices considered in the main text are all finite, completeness and the existence of the elements $0$ and $1$ will be understood there without further comment. Here finiteness is not assumed a priori.

\begin{corollary}[Chains as lattices]
\label{chain-algebra}
Every nonempty chain is a lattice, with meet and join given by $x\wedge y=\min\{x,y\}$ and $x\vee y=\max\{x,y\}$.
\end{corollary}

\begin{definition}[Sublattice]
A \emph{sublattice} of $L$ is a nonempty subset $S\subseteq L$ closed under the meet and join operations of $L$, namely $x,y\in S$ implies $x\wedge y\in S$ and $x\vee y\in S$. The meet and join in $S$ are therefore the restrictions of the meet and
join in $L$.
\end{definition}

\begin{remark}
A subset of a lattice may itself form a lattice under the induced ordering without being a sublattice of the ambient lattice. In other words, to be a lattice, its internal joins and meets only need to exist. To be a sublattice, the meets and joins must agree with the corresponding ambient operations.
\end{remark}

\begin{corollary}[A chain is a sublattice]
Every nonempty chain $C$ contained in a lattice $L$ is a sublattice of $L$.
\end{corollary}

\begin{definition}[Direct product of posets]
The \emph{direct product} of posets $P_1,\ldots,P_Q$ is the set-theoretic Cartesian product $\prod_{q=1}^{Q}P_q$ equipped with the componentwise ordering
\begin{equation}
\label{Aeq:product-poset-order}
(x_1,\ldots,x_Q)\leqq(y_1,\ldots,y_Q)
\quad\iff\quad
x_q\leqq y_q \,,
\quad\forall q\,.
\end{equation}
\end{definition}

\begin{theorem}[Direct product of lattices]
\label{direct-product-lattice-algebra}

The direct product of lattices $L_1,\ldots,L_Q$, understood as that of posets ordered as \eqrefe{Aeq:product-poset-order}, is a lattice. Its meet and join are given componentwise by
\begin{align}
\label{Aeq:product-lattice-operations}
&(x_1,\ldots,x_Q)\wedge(y_1,\ldots,y_Q)
= (x_1\wedge y_1,\ldots,x_Q\wedge y_Q)\,,
\nn\\
&(x_1,\ldots,x_Q)\vee(y_1,\ldots,y_Q)
= (x_1\vee y_1,\ldots,x_Q\vee y_Q)\,.
\end{align}
If $S_q\subseteq L_q$ is a sublattice for every $q$, then $\prod_{q=1}^{Q}S_q\subseteq \prod_{q=1}^{Q}L_q$ is a sublattice.
\end{theorem}

\begin{corollary}[Direct products of chains]
\label{product-of-chains}

Let $C_1,\ldots,C_Q$ be nonempty chains. Their direct product $C_1\times\cdots\times C_Q$, understood as that of posets, is a lattice. It is a chain if and only if all but at most one of the factors consist of a single element.

If $C_q$ is a nonempty chain in a lattice $L_q$ for every $q$, then
\begin{equation}
C_1\times\cdots\times C_Q \subseteq L_1\times\cdots\times L_Q
\end{equation}
is a sublattice.
\end{corollary}

\begin{remark}
A direct product of nontrivial chains is generally a grid-like lattice rather than a single chain, since componentwise order is only a partial order.
\end{remark}

\begin{definition}[Isomorphisms and duality of lattices]
Lattice (dual) isomorphisms and duality are defined through the underlying posets.
\end{definition}

\begin{theorem}[Preservation of lattice operations]
A lattice isomorphism preserves meet and join, while a dual isomorphism exchanges them, since meet and join are characterized order-theoretically as greatest lower bounds and least upper bounds.
\end{theorem}

\subsection{Distributive, modular, and Boolean lattices}
\label{app:B3}

\begin{theorem}[Algebraic characterization of lattices]
The meet and join operations in every lattice satisfy the following identities:
\begin{align}
\label{Aeq:lattice-algebra-identities}
&\text{\emph{Idempotence:}}
&& x\wedge x=x\,, \quad 
x\vee x=x\,,
\nn\\
&\text{\emph{Commutativity:}}
&& x\wedge y=y\wedge x\,, \quad 
x\vee y=y\vee x\,,
\nn\\
&\text{\emph{Associativity:}}
&& x\wedge(y\wedge z)=(x\wedge y)\wedge z\,,\quad
x\vee(y\vee z)=(x\vee y)\vee z\,,
\nn\\
&\text{\emph{Absorption:}}
&& x\wedge(x\vee y)=x\,,\quad 
x\vee(x\wedge y)=x\,.
\end{align}
Conversely, any nonempty set equipped with two binary operations satisfying these four pairs of identities becomes a lattice under the ordering defined by
\begin{equation}
\label{Aeq:order-from-algebra}
x\leqq y \quad \Longleftrightarrow\quad
x\wedge y=x \quad\Longleftrightarrow\quad
x\vee y=y\,.
\end{equation}
Thus, the order-theoretic and algebraic definitions of a lattice are equivalent.
\end{theorem}

\begin{theorem}[Universal lattice inequalities]
In every lattice, the following inequalities hold:
\begin{align}
\label{Aeq:universal-inequalities}
&\text{\emph{Distributive inequalities:}}
\quad&&x\wedge(y\vee z)\geqq(x\wedge y)\vee(x\wedge z)\,,
\nn\\
& \ &&x\vee(y\wedge z)\leqq(x\vee y)\wedge(x\vee z)\,,
\nn\\[1mm]
&\text{\emph{Modular inequality:}} \quad &&x\leqq z \implies x\vee(y\wedge z)
\leqq(x\vee y)\wedge z\,.
\end{align}
\end{theorem}

\begin{definition}[Distributive lattice]
A lattice $L$ is \emph{distributive} if the two distributive inequalities become equalities for all $x,y,z\in L$.
\end{definition}

\begin{theorem}[Inheritance of distributivity]
\label{Inheritance-of-distributivity}
Every nonempty chain is a distributive lattice. Every sublattice of a distributive lattice is distributive. Every direct product of distributive lattices is distributive. In particular, a direct product of nonempty chains is a distributive lattice, although it is generally not itself a chain.
\end{theorem}

\begin{definition}[Modular lattice]
A lattice $L$ is \emph{modular} if the modular inequality becomes an equality for $x\leqq z$.
\end{definition}

\begin{theorem}[Distributivity implies modularity]
Every distributive lattice is modular. The converse is not true.
\end{theorem}

\begin{definition}[Complement and complemented lattice]
Let $L$ be a bounded lattice. An element $y\in L$ is a \emph{complement} of $x\in L$ if $x\wedge y=0$ and $x\vee y=1$. A bounded lattice is \emph{complemented} if every element has at least one complement.
\end{definition}

\begin{remark}
Complements need not be unique in a general complemented lattice.
\end{remark}

\begin{theorem}[Uniqueness of complements]
An element of a bounded distributive lattice has at most one complement. Consequently, complementation is a well-defined unary operation in every complemented distributive lattice.
\end{theorem}

\begin{definition}[Boolean lattice and Boolean algebra]
A \emph{Boolean lattice} is a complemented distributive lattice. Since complements are unique, the complement of $x$ is denoted by $x^c$. With the underlying set $B$, the corresponding algebraic structure
\begin{equation}
(B,\wedge,\vee,{}^c,0,1)
\end{equation}
is called a \emph{Boolean algebra}.
\end{definition}

\begin{definition}[Boolean subalgebra]
A (Boolean) \emph{subalgebra} of a Boolean algebra $(B,\wedge,\vee,{}^c,0,1)$ is a nonempty subset $S\subseteq B$ which is closed under all operations in $B$ when restricted to $S$. Therefore, the subalgebra $(S,\wedge,\vee,{}^c,0,1)$ must inherit the Boolean operations and distinguished elements from $B$.
\end{definition}

\begin{remark}
We distinguish the notion of \emph{Boolean sublattice} as a subset $S\subseteq B$ which is closed under the operations $(\wedge,\vee)$ in $B$, and is itself a Boolean lattice with induced lattice structure. In other words, $S$ is a sublattice of $B$, but may have its own $0,1$, and complement operation. Finally, if a Boolean sublattice contains the ambient $0$ and $1$, then its internal complement necessarily agrees with the ambient complement, because complements are unique in a distributive lattice. Hence in that case, it is equivalent to a Boolean subalgebra. See also \cite[Ch.~11]{givant2009introduction}.
\end{remark}

\begin{theorem}[Complementation and self-duality]
Every Boolean lattice $B$ satisfies
\begin{align}
\label{Aeq:boolean-complement-identities}
(x^c)^c&=x\,, & 0^c&=1\,,
& 1^c&=0\,,
\nn\\
(x\vee y)^c&=x^c\wedge y^c\,,
& (x\wedge y)^c&=x^c\vee y^c\,.
\end{align}
Complementation reverses the ordering
\begin{equation}
\label{Aeq:boolean-order-reversal}
x\leqq y \quad\Longleftrightarrow\quad
y^c\leqq x^c \,,
\end{equation}
and defines an isomorphism
\begin{equation}
B\xrightarrow{\;\sim\;}B^{\mathrm{op}}\,, \qquad
x\longmapsto x^c\,,
\end{equation}
where $B^{\rm{op}}$ is the same lattice with the ordering reversed. Equivalently, complementation is an involutive lattice dual automorphism. In this sense, every Boolean lattice is self-dual.
\end{theorem}

\begin{corollary}[Complementation of chains]
\label{coro-complementation-of-chains}

Let $\ch(B)$ denote the set of chains in a Boolean lattice $B$. For $C\in\ch(B)$, define its complement $C^c$ as the set
\begin{equation}
C^c \equiv \{x^c:x\in C\}\,,
\end{equation}
where the complement $x^c$ is taken in $B$. Then $C^c$ is again a chain, and
\begin{equation}
\kappa: \ch(B) \longrightarrow \ch(B)\,, \qquad \kappa(C)=C^c
\end{equation}
is an involutive bijection. If $\ch(B)$ is ordered by set inclusion, then $\kappa$ is an order-preserving automorphism of $\ch(B)$. The element-wise complement map is, however, a dual isomorphism from the chain $C$ to the chain $C^c$, understood as a map between posets.

In particular, if a finite maximal chain is written as
\begin{equation}
0=c_0 \lessdot c_1 \lessdot\cdots \lessdot c_n=1\,,
\end{equation}
then its complement chain, written in ascending order, is
\begin{equation}
\label{Aeq:abstract-complement-chain}
0=c_n^c \lessdot c_{n-1}^c \lessdot\cdots \lessdot c_0^c=1\,.
\end{equation}
Equivalently, its rank-$r$ element is
\begin{equation}
(C^c)_r=(c_{n-r})^c\,.
\label{Aeq:abstract-complement-chain-rank}
\end{equation}
\end{corollary}

\begin{theorem}[Direct products of Boolean lattices]
A finite direct product of Boolean lattices is a Boolean lattice. Its meet, join, least element, greatest element, and complement are all componentwise:
\begin{align}
(x_1,\ldots,x_Q)\wedge(y_1,\ldots,y_Q)
&= (x_1\wedge y_1,\ldots,x_Q\wedge y_Q)\,,
\nn\\[1mm]
(x_1,\ldots,x_Q)\vee(y_1,\ldots,y_Q)
&= (x_1\vee y_1,\ldots,x_Q\vee y_Q)\,,
\nn\\[1mm]
(x_1,\ldots,x_Q)^c = (x_1^c,\ldots,x_Q^c)\,, &\quad 0=(0_1,\ldots,0_Q)\,,\quad 1=(1_1,\ldots,1_Q)\,.
\end{align}
\end{theorem}

\subsection{Power-set Boolean lattices and direct products}
\label{app:B4}

We now specialize the general theory of Boolean lattices to the power-set lattices used in the main text.

\begin{theorem}[Power-set Boolean lattice]
Let $U$ be a set. Its power set $\PP(U)$, ordered by inclusion, is a complete Boolean lattice with
\begin{align}
A\wedge B &= A\cap B\,, \quad 
A\vee B=A\cup B\,,
\nn\\
A^c&=U\setminus A\,, \quad 0=\emp\,, \quad 1=U\,.
\end{align}
The order relation defined by the algebra from \eqrefe{Aeq:order-from-algebra} is set inclusion. The covering relation is strict inclusion by exactly one element:
\begin{equation}
\label{Aeq:powerset-cover-relation}
A\lessdot B \quad\Longleftrightarrow\quad
A\subsetneq B \quad\text{and}\quad |B\setminus A|=1\,.
\end{equation}
\end{theorem}

\begin{remark}
A complemented lattice need not be complete. The distinction only matters for infinite lattices. That is, having $0$ and $1$ does not imply completeness for infinite lattices. Therefore, a Boolean lattice need not be complete. But the full power-set lattice $\PP(U)$ is always complete, even when $U$ is infinite, since union and intersection of arbitrary subsets are still subsets.
\end{remark}

Thus, between consecutive elements of a maximal chain, the cover relation may be displayed simply by the strict-inclusion symbol $\subsetneq$.

\begin{corollary}[Maximal chains in a power-set lattice]
For finite $U$, every maximal chain in $\PP(U)$ has the form
\begin{equation}
\label{Aeq:maximal-chain-powerset}
\emp = A_0 \subsetneq A_1 \subsetneq \cdots \subsetneq A_{|U|} = U\,, \qquad |A_r|=r\,.
\end{equation}
Each strict inclusion in \eqrefe{Aeq:maximal-chain-powerset} is a covering relation. Equivalently, a maximal chain is obtained by adjoining the elements of $U$ one at a time and therefore determines an ordering of $U$.
\end{corollary}

In the main text, a maximal chain with finite underlying set $U$ associated with an ordering $\lam=(\lam_1,\ldots,\lam_{|U|})$ is denoted by $C[\lam|U]$. By corollary~\ref{coro-complementation-of-chains}, its complement chain corresponds to the reverse ordering $\lam^c=(\lam_{|U|},\ldots,\lam_1)$, and its rank-$r$ element is
\begin{equation}
\label{Aeq:complement-chain-appendix}
C[\lam^c|U]_r = U\setminus C[\lam|U]_{|U|-r}\,, \qquad 0\leqq r\leqq |U|\,.
\end{equation}

\begin{theorem}[Product decomposition of a power-set lattice]
Let $U=\bigsqcup_{q=1}^{Q}U_q$ be a partition of $U$ into possibly empty pairwise disjoint subsets. Disjoint union defines a Boolean-lattice isomorphism between the direct-product lattice and total lattice
\begin{equation}
\label{Aeq:powerset-product-isomorphism}
\Phi: \prod_{q=1}^{Q}\PP(U_q) \xrightarrow{\;\sim\;}
\PP(U)\,, \qquad
\Phi(A_1,\ldots,A_Q) = \bigsqcup_{q=1}^{Q}A_q\,.
\end{equation}
Its inverse is given by
\begin{equation}
\label{Aeq:powerset-product-isomorphism-inverse}
\Phi^{-1}(A) = (A\cap U_1,\ldots,A\cap U_Q)\,.
\end{equation}
\end{theorem}

\begin{corollary}[Product of local maximal chains]
\label{product-of-local-maximal-chains}
Assume that each $U_q$ is finite. For each block $U_q$, let
\begin{equation}
\label{Aeq:local-maximal-chain-appendix}
C_q: \qquad \emp = A_{q,0} \subsetneq A_{q,1} \subsetneq \cdots \subsetneq A_{q,|U_q|} = U_q
\end{equation}
be a local maximal chain in $\PP(U_q)$. By corollary~\ref{product-of-chains} and theorem~\ref{Inheritance-of-distributivity}, the direct product of the local maximal chains,
\begin{equation}
\LL \,\equiv \prod_{q=1}^{Q}C_q
= \LB (A_{1,r_1},\ldots,A_{Q,r_Q})
\,\middle|\, 0\leqq r_q\leqq |U_q|\,,\, \forall q
\RB \,\subseteq\,
\prod_{q=1}^{Q}\PP(U_q)\,,
\label{Aeq:product-local-maximal-chains}
\end{equation}
is a distributive sublattice of the direct-product lattice. Its meet and join are given componentwise by theorem~\ref{direct-product-lattice-algebra} and corollary~\ref{chain-algebra}.

Under the disjoint-union isomorphism \eqref{Aeq:powerset-product-isomorphism}, the restriction $\Phi|_{\LL}$ is an isomorphism onto its image
\begin{equation}
\label{Aeq:image-product-local-chains}
\Phi(\LL) = \LB \bigsqcup_{q=1}^{Q}A_{q,r_q}
\,\middle|\, 0\leqq r_q\leqq |U_q|\,,\, \forall q
\RB \,\subseteq\, \PP(U)\,.
\end{equation}
Consequently, the lattice $\LL$ may be identified with the distributive sublattice $\Phi(\LL)$ of the total Boolean lattice $\PP(U)$. Moreover, $\LL$ contains the least and greatest elements of
$\prod_{q=1}^{Q}\PP(U_q)$,
\begin{equation}
(\emp,\ldots,\emp)\in \LL \,, \qquad
(U_1,\ldots,U_Q)\in \LL \,,
\end{equation}
which are mapped by $\Phi$ to $\emp$ and $U$, respectively. Hence $\Phi(\LL)$ is a bounded distributive sublattice of $\PP(U)$. It is not, in general, a Boolean subalgebra, since it need not be closed under the ambient complementation.
\end{corollary}

\begin{remark}
If at least two blocks $U_q$ are nonempty, then neither $\LL$ nor $\Phi(\LL)$ is itself a chain.
The Cartesian product of the local maximal chains should therefore be viewed as a product sublattice containing mutually incomparable elements, rather than as a single maximal chain in the total Boolean lattice, since there is no implicit choice of an ordering between steps belonging to different blocks.
\end{remark}

\bibliographystyle{utphys}
\bibliography{refs.bib}
\end{document}